\documentclass[epj,usenames,dvipsnames]{svjour}
\pdfoutput=1
\usepackage{amsmath}
\usepackage{colortbl}
\usepackage{xcolor}

\newcommand{\beq}{\begin{equation}}
\newcommand{\eeq}{\end{equation}}

\newcommand{\carbon}{\rm $_6^{12}$C}

\newcommand{\neut}{$\textsc{neut}$}
\newcommand{\genie}{$\textsc{genie}$}

\newcommand{\minerva}{\textsc{miner}$\nu$\textsc{a}}

\usepackage{array, graphicx, subfigure}
\usepackage{graphics}
\begin{document}
\title{ Removal Energies and Final State Interaction in Lepton Nucleus Scattering
}
%\subtitle{ none }
\author{Arie Bodek %\inst{1}
 and Tejin Cai%\inst{1}  -
 }
\institute{Department of Physics and Astronomy, University of
Rochester, Rochester, NY  14627-0171}
\date{Received: 29 January 2018/ Accepted 4 March 2019  published in European Physics Journal C (2019), ~~~~         
%    Version 3.3:  Wed March 11, 2019}  
DOI: 10.1140/epjc/s10052-019-6750-3, ~~~~~ 	arXiv:1801.07975 [nucl-th]} 
\abstract{
We investigate the  binding energy  parameters that should be used in modeling  electron and neutrino scattering from nucleons bound in a nucleus within the framework of the impulse approximation. We discuss the relation between binding energy, missing energy,  removal energy ($\epsilon$),   spectral functions and  shell model energy levels and extract updated removal energy parameters from  ee$^{\prime}$p spectral function data. We address the difference in parameters for scattering from bound protons and neutrons. We also use inclusive e-A data to extract an  empirical parameter 
$U_{FSI}( (\vec q_3+\vec k)^2)$  to account for the interaction of final state nucleons (FSI) with the optical potential of the nucleus. Similarly we use  $V_{eff}$ to account for the  Coulomb potential  of the nucleus.  With  three parameters  $\epsilon$,   $U_{FSI}( (\vec q_3+\vec k)^2)$ and $V_{eff}$ we can describe the energy of final state electrons for all available electron QE scattering data. The use of the updated parameters in neutrino Monte Carlo generators reduces the systematic uncertainty in the combined removal energy (with FSI corrections) from $\pm$ 20 MeV to $\pm$ 5 MeV.
 % end of PACS codes
} %end of abstract
\maketitle
%

%%%%%%%         SECTION   1      %%%%%%%%%
\section{Introduction}
%section 1
\label{intro}
 The  modeling of neutrino cross sections on nuclear targets is of great interest to neutrino oscillations experiments. 
 Neutrino  Monte Carlo (MC) generators
  include  $\textsc{genie}$\cite{genie}, $\textsc{neugen}$\cite{neugen}, $\textsc{neut}$\cite{neut}, 
 $\textsc{nuwro}$\cite{nuwro} and GiBUU\cite{gibuu}.
   
  Although more sophisticated
 % two dimensional spectral function 
  models are available\cite{benhar,spectral-theory2,Sakuda,optical,optical1}, calculations using a one-dimensional momentum distribution and an average removal energy parameter are still widely used.  One example is the  simple relativistic Fermi gas (RFG)  model. 
  %Here, we re-extract the  parameters without  approximations.  In addition, when available,  more precise parameters are extracted from recent ee$^{\prime}$p spectral function measurements.
 
The RFG model does not describe the tails in the energy distribution of the final state lepton very well\cite{electron,neutrino}. Improvements to the RFG model  such as  a better momentum distribution are usually made within the existing Monte Carlo (MC)   frameworks.  All RFG-like models  with one dimensional nucleon momentum distributions  require in addition removal energy parameters ($\epsilon^{P,N}$) to account for the average removal energy of a proton or neutron from the nucleus.  These parameters  should be approximately the same for all one-dimensional momentum distributions. 

Alternatively two dimensional spectral functions (as a function of nucleon momentum and missing energy) can be used.  However, even in this case,  MC generators currently used in neutrino oscillations experiments do not account for the final state interaction (FSI) of the final state lepton and nucleon in the optical and Coulomb potentials of the nucleus. 
 % We model the effects of FWI in the optical and Coulomb potentials of the nucleus  with two parameters ($U_{FSI}$, and $V_{eff}^{P,N}$)

In this paper we extract empirical average removal energy parameters from spectral function measured in exclusive ee$^\prime$p electron scattering experiments on several nuclei. We use $V_{eff}$ (see Appendix A) to account for the  Coulomb potential of the nucleus, and extract empirical nucleon final state interaction parameter   $U_{FSI}( (\vec q_3+\vec k)^2)$ from all available inclusive e-A electron scattering data. With these three parameters  
$\epsilon$,   $U_{FSI}( (\vec q_3+\vec k)^2)$ and $V_{eff}$ we can describe the energy of final state electrons for all  available electron QE scattering data. These parameters  can  be used to improve the predictions of current neutrino MC event generators such as \genie~and \neut~for the final state muon and nucleon energies in QE events.

A large amount of computer time has been used by various experiments to generate and reconstruct simulated neutrino interactions using MC generators such as \genie ~2.  We show how approximate  post-facto corrections could be applied to these existing MC samples to improve the modeling of the reconstructed muon, final state proton, and unobserved energy in quasielastic (QE) events.
\subsection{Relevance to neutrino oscillations experiments}
\label{parameters}
%----   subsection  1.1
 In a two neutrinos oscillations framework the oscillation parameters which are extracted from long baseline experiments are the  mixing  angle $\vartheta$  and the square
of the difference in mass between the two neutrino mass eigenstates  $\Delta{m}^2$.  A correct modeling of the reconstructed neutrino energy is very important in the measurement of  $\Delta{m}^2$.  In general, the resolution in the measurement of energy in  neutrino experiments is much worse than the resolution in electron scattering experiments. However,  a precise determination of  $\Delta{m}^2$ is possible if the MC prediction for  {\it average} value of the experimentally  reconstructed neutrino energy is unbiased.    At present the  uncertainty in the value of the removal energy parameters  is a the largest  source of systematic error in the extraction of the neutrino oscillation parameter $\Delta{m}^2$ (as shown below).  
%The measurement of the mixing angle  $\vartheta$ is sensitive to the both the experimental resolution and to the shape o

The two-neutrino transition probability can  be written as
\begin{equation}
P_{\nu_{\alpha}\to\nu_{\beta}}(L)
=
\sin^2 2\vartheta
\,
\sin^2\left(
1.27
\,
\frac
{ \left( \Delta{m}^2 / \text{eV}^2 \right) \left( L / \text{km} \right)}
{ \left( E_\nu / \text{GeV} \right) }
\right).
%\,.
\label{oscl}
\end{equation}
Here,   L (in $\text{km}$) is the distance between
the neutrino source and the detector and  $\Delta{m}^2$ is in $\text{eV}^2$.

The location of the first oscillation maximum in neutrino energy  ($E_\nu^{1st-min}$) is when the term in brackets is equal to $\pi/2$. 
  An estimate  of the extracted value of  $\Delta{m}^2$ is given by:
 \begin{equation}
\Delta{m}^2 = \frac{2 E_\nu^{1st-min}}{ 1.27\pi L}.
\label{Enumin}
\end{equation}
 % The uncertainty in the measurement of the neutrino 
 %energy $\Delta E_\nu$ results in a systematic error in the extracted value of $\Delta{m}^2$ of: 

For example,  for the \textsc{t2k}  experiment $L= 295~Km$, and $E_\nu$ is peaked around 0.6 GeV.
For the normal hierarchy the \textsc{t2k} experiment\cite{T2K}  reports a value of
$$\Delta{m_{32}^2} (\textsc{t2k}-2018) =(2.434  \pm 0.064) \times 10^{-3}~\rm eV^2.$$ 
$$\\sin^2\theta_{23} (\textsc{t2k}-2018) =0.536^{+0.031}_{-0.045}$$ 
 %In the Monte Carlo
%generator used by \ttok (\neut) a value of  the interaction energy
 % $\langle \epsilon^{\prime N}_{SM}\rangle$ 
%  of  27 MeV for $_8^{16}$O has been used.
%However, as we show in section \ref{extraction}  a value of  43 MeV
%should be used.
Using equation \ref{Enumin} and \ref{QEequ}
% in conjunction with equation \ref{QEequ}  of Appendix \ref{neutrinoE} 
we estimate
that a +20 MeV change in the removal energy used in the MC results in
a change in  $\Delta{m_{32}^2}$  of $+0.03 \times 10^{-3}~\rm eV^2$, which is the $largest$ contribution
to the total systematic error in  $\Delta{m_{32}^2}$.
  
The above estimate is consistent with the estimate of the  \textsc{t2k} collaboration.
The  \textsc{t2k} collaboration reports\cite{t2k-impact} that 
  ``for the statistics of the 2018 data set, a shift of 20 MeV in the binding energy parameter introduces a bias of 20\% for %sin2(th23)
 $\sin^2\theta_{23}$  and 40\% for $\Delta{m_{32}^2}$  with respect to the size of the systematics errors, assuming maximal  $\sin^2\theta_{23}$''.
 % sin2(th23).
 
 For the case of normal hierarchy a combined analysis\cite{combined} of the world's neutrino oscillations data in 2018 finds a best fit of 
 $$\Delta{m_{32}^2}(\textsc{combined}-2018) =(2.50  \pm 0.03) \times 10^{-3}~\rm eV^2,$$ 
 $$\\sin^2\theta_{23} (\textsc{combined}-2018) =0.547^{+0.020}_{-0.030},$$ 
 which illustrates the importance of using a common definition of removal energy parameters and
 the importance in handling the correlations in  the uncertainties  between
 various experiments when performing a combined analysis.
 
 For comparison, we find that a change of +20 MeV/c in the assumed value of the Fermi momentum $k_F$ yields a much smaller
change of  $+0.005\times 10^{-3}~eV^2$ in the extracted value of $\Delta{m_{32}^2}$.
%
%       subsection 1.2
  \subsection{ Neutrino near detectors}
In general,  neutrino oscillations experiments use data taken from a near detector to reduce the systematic error from uncertainties in the neutrino flux and in the modeling of neutrino interactions.  However, near detector data cannot constrain the absolute energy scale of final state muons and protons, or account for the energy that goes into the undetected nuclear final state.  These issues are addressed in this paper.
  \subsection{ Simulation of  QE events and reconstruction of neutrino energy}
  In order to simulate the reconstruction of neutrino  QE events within the framework of the impulse approximation
  the experimental empirical parameters that are used should describe:
  \begin{enumerate} 
  \item The momentum of the final state muon including the effect of  Coulomb corrections\cite{gueye}.
  \item The mass,  excitation energy, and recoil energy of the spectator nuclear state.
  \item The effect of the interaction of the final state nucleon (FSI) with the optical and Coulomb potential of the spectator nucleus.
      \end{enumerate} 
    %
 %  Within the impulse approximation,  the momentum of the final state muon can be extracted from the
%   shape of the distribution of QE events in  inclusive electron scattering. The mass and excitation energy of the spectator nuclear state
%   can be extracted from spectral function measurements in ee$\prime$p experiments and the effects of FSI on the energy of the final state lepton and nucleon
%   can be extracted from the location of the peak of the distribution of QE events in inclusive electron scattering.
%
 %     Subsection 1.3
  \subsection{ Nucleon momentum distributions}
\begin{figure}
\begin{center}
\includegraphics[width=3.5in,height=2.5in]{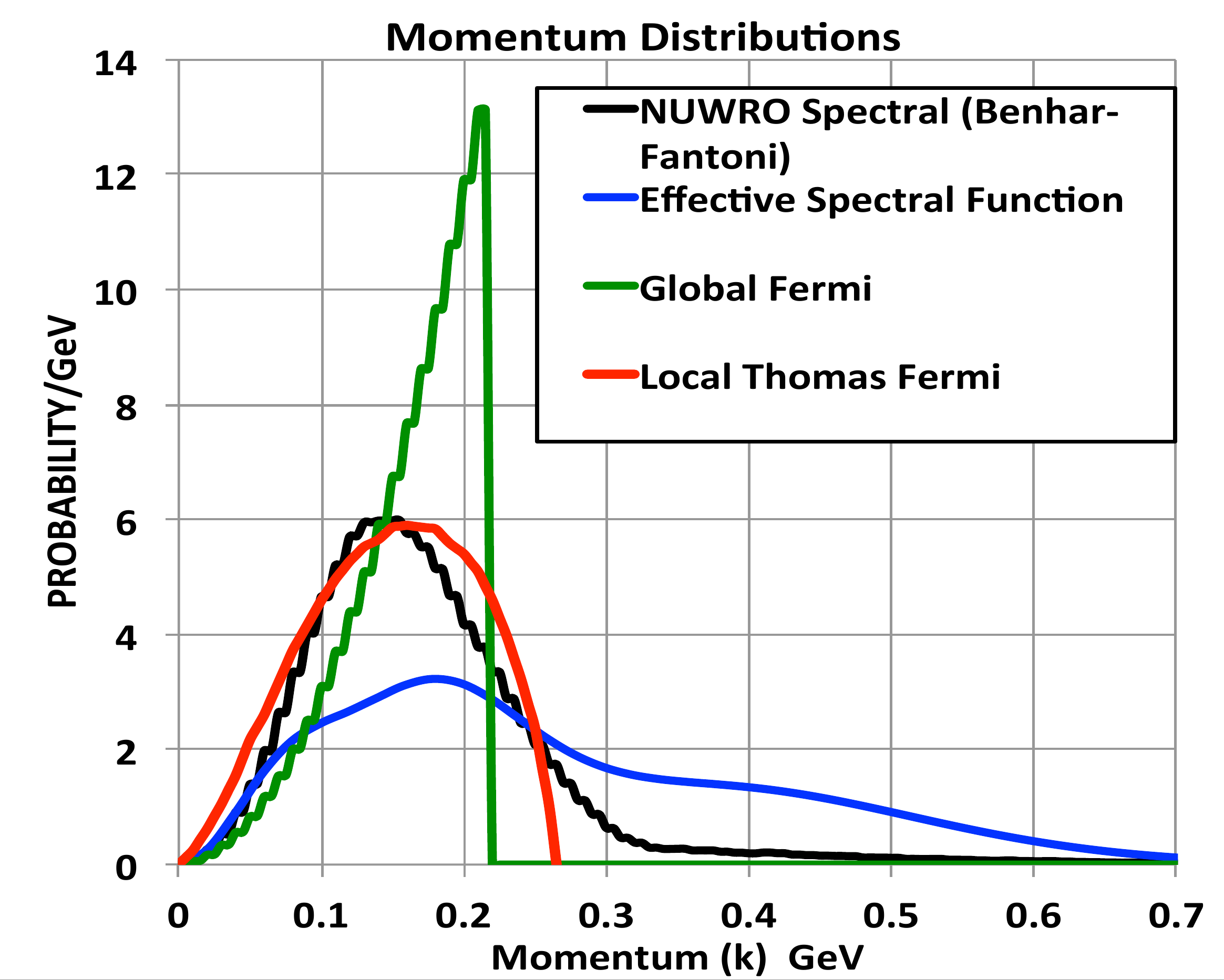}
\caption{ 
One-dimensional nucleon momentum distributions  in a \carbon~nucleus. The green curve (Global Fermi)  is 
the momentum distribution for the relativistic Fermi gas (RFG) model.
%which is related to global average density of nucleons.
 The red curve is the  Local-Thomas-Fermi (LTM) gas distribution.
 % which is related to the local density of nucleons in the nucleus.
 The black curve is the projected momentum distribution of the Benhar-Fantoni two dimensional spectral function.
 %  as implemented in  $\textsc{nuwro}$.
The blue line is the momentum distribution for the  {\it {effective spectral function}}  model, which approximates the 
 $\psi^\prime$ superscaling prediction for the final state muon in quasielastic scattering. 
 %The {\it {effective spectral function}}  is only valid for QE scattering 
 % because the $\psi^\prime$  superscaling model is only valid for QE scattering.
   }
\label{momentum}
\end{center}
\end{figure}

 Fig. \ref{momentum} shows  a few models for the nucleon momentum distributions in the \carbon~nucleus. The solid green line (labeled Global Fermi gas) is the nucleon momentum distribution for the  Fermi gas\cite{electron} model  which is currently implemented in all neutrino event generators and is related to global average density of nucleons. The solid black line is  the projected momentum distribution of the  Benhar-Fantoni\cite{benhar}  2D spectral function as implemented in $\textsc{nuwro}$.   The solid red line is the nucleon momentum distribution for  the Local-Thomas-Fermi (LTF) gas which is  is related to the local density of nucleons in the nucleus and is implemented in $\textsc{neut}$,  $\textsc{nuwro}$ and GiBUU.

A more sophisticated formalism is the $\psi^\prime$ superscaling model\cite{Donnelly}, {\it which is
only valid for QE scattering}.  It can be used to predict the kinematic distribution of the final state muon but does not describe the details of the hadronic final state.  Therefore, it has not been implemented in neutrino MC generators.  However,   the predictions  of the $\psi^\prime$ superscaling model can be approximated with an effective spectral function\cite{effective} which has been implemented in  $\textsc{genie}$. The momentum distribution of the effective spectral function for nucleons bound in \carbon~is  shown as the blue curve in Fig. \ref{momentum}.

\begin{figure}
\includegraphics[width=3.5in,height=2.5in]{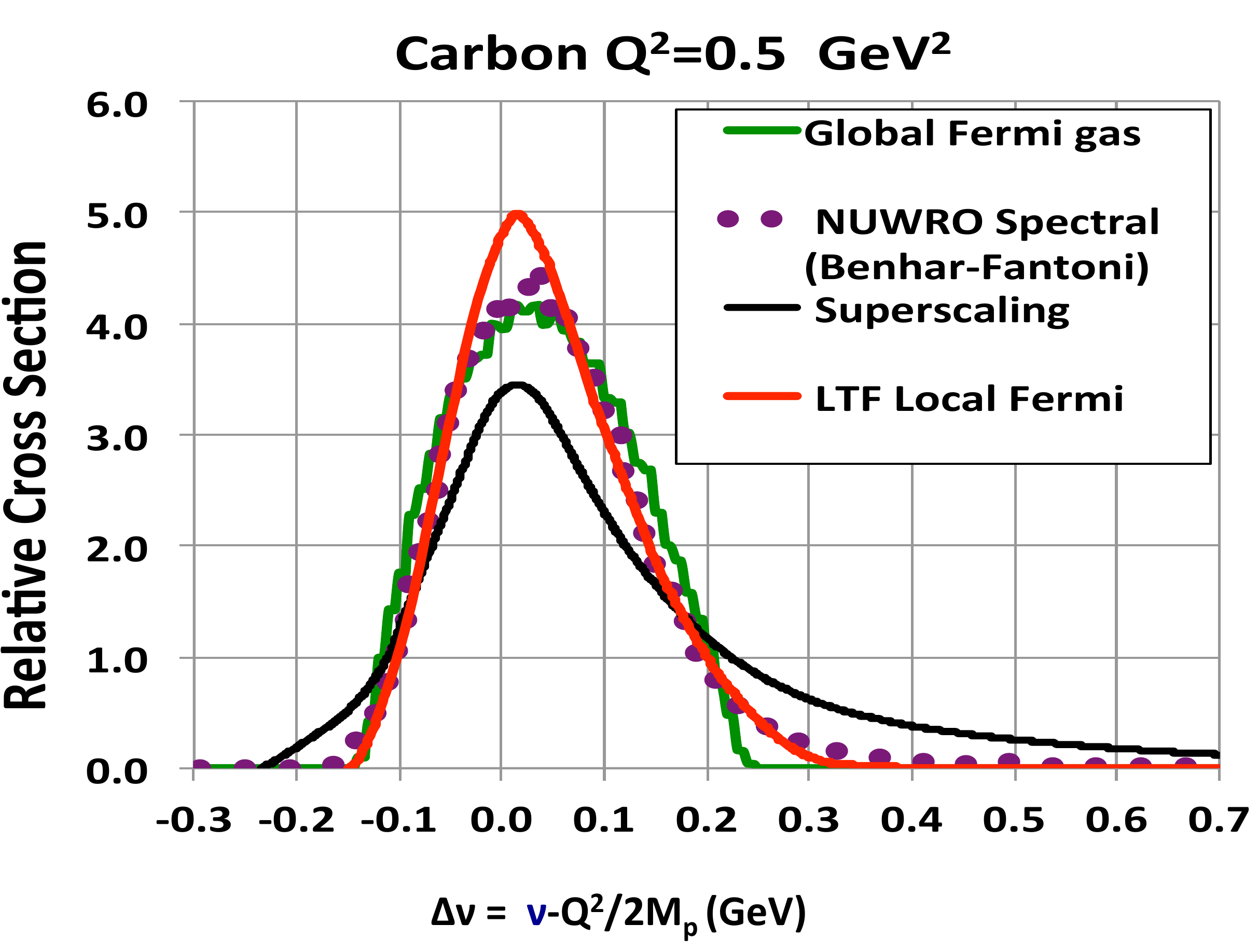}
\caption{ Comparison of the  $\psi^\prime$ superscaling prediction (solid black line)  for  the normalized  quasielastic  $\frac{1}{\sigma} \frac{d\sigma}{d\nu}(Q^2,\nu)$  at $Q^2$=0.5 GeV$^2$ for 10 GeV neutrinos on \carbon~to the predictions with several momentum distribution ($\nu = E_0-E^\prime$).
 Here the  solid green curve labeled ``Global Fermi'' gas  is the distribution for the Fermi gas model.
The red line is the prediction for the local Thomas Fermi (LTF) gas, and the purple dots are the prediction using the two dimensional Benhar-Fantoni
spectral function as implemented in $\textsc{nuwro}$. 
}
\label{nuwro-vs-scaling}
\end{figure}

Although the nucleon momentum distributions are very different for the various models, 
the predictions for the  normalized  quasielastic neutrino cross section  $\frac{1}{\sigma} \frac{d\sigma}{d\nu}(Q^2,\nu)$ are similar as shown in Fig. \ref{nuwro-vs-scaling}. 
 These predictions as a function of $\nu=E_\nu-E_\mu$  are calculated for  10 GeV neutrinos on \carbon~at $Q^2$=0.5 GeV$^2$.   
The prediction with the local Fermi gas distribution are similar to the prediction of the Benhar-Fantoni two dimensional spectral function  as implemented in  $\textsc{nuwro}$.   Note that the prediction of the  $\psi^\prime$ superscaling  model are based on fits to longitudinal QE differential cross sections. Subsequently, they includes 1p1h  and some  2p2h processes (discussed in section \ref{impulse}).

The following nuclear targets are (or were)  used in neutrino experiments:  
Carbon (scintillator)  used in the $\textsc{nova}$ and \minerva\;experiments.  Oxygen (water)  used in \textsc{t2k}  and  in  \minerva.   Argon used in the $\textsc{argoneut}$ and $\textsc{dune}$ experiments.  Calcium (marble)  used in $\textsc{charm}$.  Iron
 used in \minerva, $\textsc{minos}$, $\textsc{cdhs}$, $\textsc{nutev}$, and $\textsc{ccfr}$.
 Lead used in $\textsc{chorus}$ and  \minerva.
%
%
%%FIG 1  momentum.pdf
%FIGURE 2   nuwro-vs-scaling.pdf
%

%FIG 3 -------------------------------- -------------------------------- -------------------------------- --------------------------------
\begin{figure}[ht]
\begin{center}
\includegraphics[width=3.5in,height=2.5in]{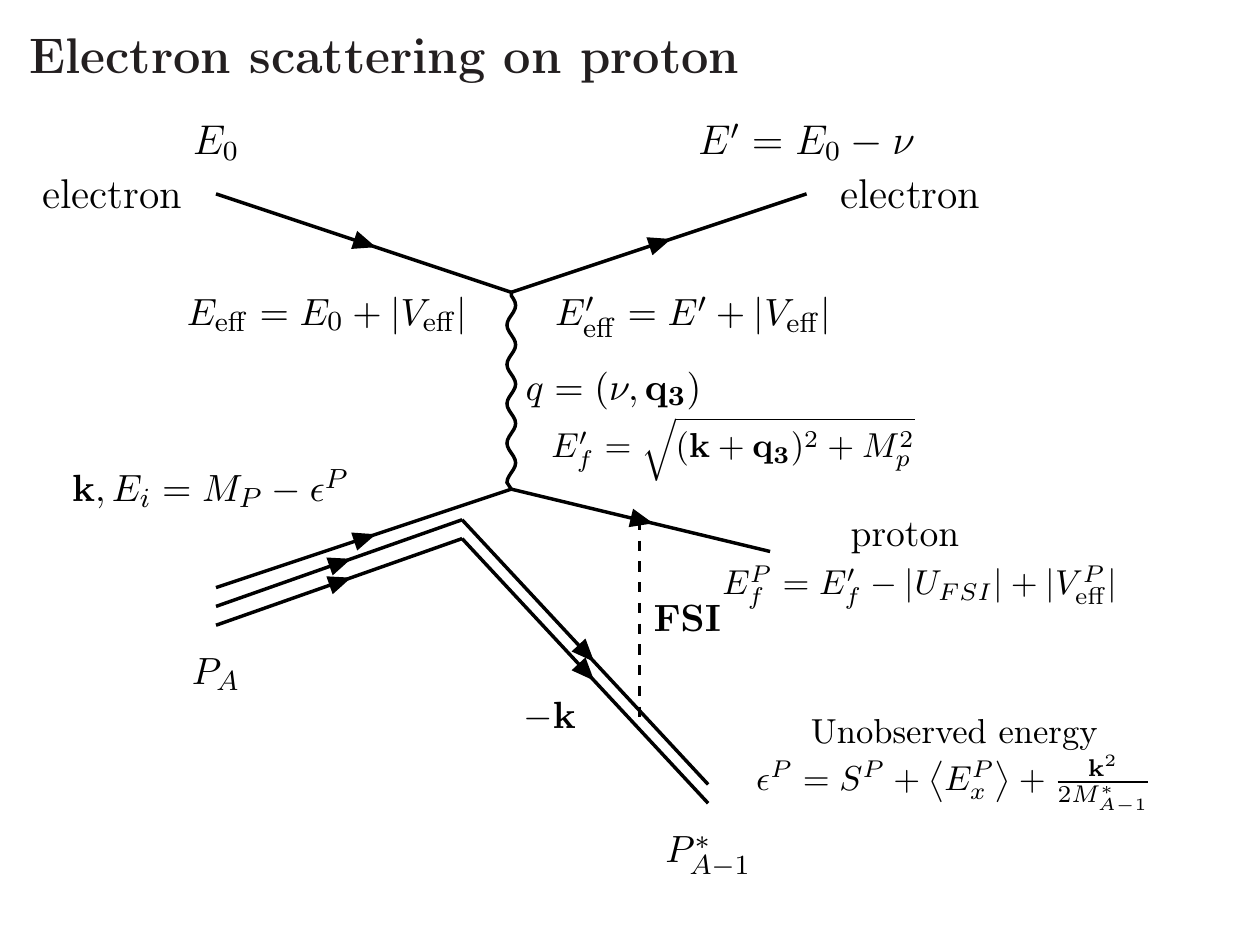}
\caption{ 
1p1h process: Electron scattering from an off-shell bound proton of momentum $\vec {p_i}$=$\vec k$ in 
a nucleus of mass A.  Here, the nucleon is moving in the mean field (MF) of all  the other nucleons in the nucleus.  The on-shell recoil excited   $[A-1]^*$ spectator nucleus has a momentum  $\vec p_{ (A-1) *}=-\vec k$ and  a mean excitation energy  $\langle {E_x^P} \rangle$.
The off-shell energy of the interacting nucleon is 
$E_i  =  M_A - \sqrt{ (M_{A-1}*) ^2+\vec k^2}=  M_A - \sqrt{ (M_{A-1}+{{E_x}})^2+\vec k^2} = M_P -\epsilon^P $, where 
$\epsilon^P = S^P +\langle E_x  \rangle+\frac{\vec k^2}{2M^*_{A-1}}$. 
  We model the effect of FSI (strong and EM interactions) by setting $E_f=\sqrt{(\vec {k}+\vec {q_3})^2+M_P^2} -|U_{FSI}|$+$|V_{eff}^P|$,
  where  $U_{FSI}= U_{FSI}( (\vec q_3+\vec k)^2)$.
  For electron QE scattering  on bound protons  $|V_{eff}^P|=\frac{Z}{Z-1}|V_{eff}|$, 
 $E_{eff}=E_0+V_{eff}$,   $E^\prime_{eff}=E^\prime+V_{eff}$.
%$~~\vec \vec q_3, \nu~~$ $~~ E_0+V_{eff}~~$  $E^\prime+V_{eff}$  $\vec k, E_i= M_P -\epsilon^P $
}
\label{Aoff-shell}
\end{center}
\end{figure}
%FIG 3 -------------------------------- -------------------------------- -------------------------------- --------------------------------
%
%%
%TABLE 1 -------------------------------- -------------------------------- -------------------------------- --------------------------------
 \begin{table}[ht]
\begin{center}
\begin{tabular}{|c|c|c|c|c|c|} \hline
$_{Z}^ANucl$ &  remove  &  & remove& &   \\
 & proton& $S^P$ &neutron& $S^N$ & $S^{N+P}$\\ \hline
%  \arrayrulecolor{red}\hline
%   \arrayrulecolor{black}\hline
   & Spectator & &Spectator&  &\\
 \hline \hline
  $_1^2$H                          & N  &2.2 &  P & 2.2 &2.2   \\ \arrayrulecolor{gray} \hline
$_3^6$Li  1+                            &  $_2^5$He $\frac{3}{2}$- & 4.4    &  $_3^5$Li $\frac{3}{2}$- &5.7 &4.0  \\  \arrayrulecolor{gray}\hline
$_6^{12}$C 0+       &          $_5^{11}$B $\frac{3}{2}$-  &   16.0     & $_6^{11}$C $ \frac{3}{2}$- & 18.7&27.4\\  \arrayrulecolor{gray} \hline
$_8^{16}$O 0+                     &$_7^{15}$N $ \frac{1}{2}$-&12.1    & $_8^{15}$O $\frac{1}{2}$- & 15.7&23.0 \\  \arrayrulecolor{gray} \hline
$_{12}^{24}$Mg 0+            &   $_{11}^{23}Na$  $\frac{3}{2}$+&11.7& $_{12}^{23}$ Mg $\frac{3}{2}$+ &16.5&24.1  \\  \arrayrulecolor{gray}\hline
$_{13}^{27}$Al$\frac{5}{2}$+            &   $_{12}^{26}Mg$  0+ &8.3 & $_{12}^{23}$ Al 5+ &13.1&19.4  \\ \arrayrulecolor{gray} \hline
$_{14}^{28}$Si 0+           &  $_{13}^{27}$Al$\frac{5}{2}$+ &11.6 & $_{14}^{27}$Si $\frac{5}{2}$+  &17.2 & 24.7 \\ \arrayrulecolor{gray} \hline
$_{18}^{40}$Ar$\frac{3}{2}$+      & $_{17}^{39}$CL  $\frac{3}{2}$+ &  12.5 & $_{18}^{39}$Ar $\frac{7}{2}$- &   9.9 & 20.6\\  \arrayrulecolor{gray}\hline 
$ _{20}^{40}$Ca 0+            &  $_{19}^{39}$K $\frac{3}{2}$ +& 8.3   &$_{20}^{39}$Ca $\frac{3}{2}$+&15.6     &21.4\\    \arrayrulecolor{gray} \hline
$ _{23}^{51}$V $\frac{7}{2}$-             &  $_{22}^{50}$Ti  0+& 8.1   &$_{23}^{50}$V 6+&11.1     &19.0\\    \arrayrulecolor{gray} \hline
 $_{26}^{56}$Fe 0+              &  $_{25}^{55}$Mn $\frac{5}{2}$ -&  10.2  &$_{26}^{55}$Fe $\frac{3}{2}$-&11.2    &20.4\\  \arrayrulecolor{gray}   \hline
 $_{28}^{58}$Ni $\frac{3}{2}$ -  &    $_{27}^{58}$Co 2+&  8.2 &       $_{87}^{58}$Ni 0+ & 12.2&19.5  \\ \arrayrulecolor{gray} \hline
$_{39}^{89}$Y $\frac{1}{2}$ -        &  $_{38}^{88}$Sr $\frac{1}{2}$-& 7.1  & $_{39}^{88}$Y  4- & 11.5 &18.2   \\  \arrayrulecolor{gray}\hline
$_{40}^{90}$Zr  0+       &  $_{39}^{89}$Y  $\frac{1}{2}$ -  & 8.4  & $_{40}^{88}$Zr$\frac{9}{2}$+   & 12.0 &17.8   \\  \arrayrulecolor{gray}\hline
$_{50}^{120}$Sn 0+     &  $_{49}^{119}$In $\frac{9}{2}$ +&  10.1 &  $_{50}^{119}$Sn $\frac{1}{2}$+&8.5&17.3    \\  \arrayrulecolor{gray}\hline
$_{73}^{181}$Ta$\frac{7}{2}$ - &  $_{72}^{180}$Hf  0+& 5.9  & $_{73}^{180}$Ta 1+  &  7.6&13.5   \\ \arrayrulecolor{gray} \hline
$_{79}^{197}$Au$\frac{3}{2}$+     & $_{78}^{196}$Pt  0+& 5.8  & $_{79}^{196}$Au  2-& 8.1&13.7 \\ \arrayrulecolor{gray} \hline
$_{82}^{208}$Pb 0+    &  $_{81}^{207}$TI $\frac{1}{2}$+&  8.0  &$_{82}^{207}$Pb $\frac{1}{2}$-&7.4&14.9  \\  \arrayrulecolor{black}\hline   
\end{tabular}
\caption{ The spin parity transitions and separation energies $S^P$, $S^N$  and $S^{N+P}$ when a proton or a neutron  or both are removed from various nuclei.
All energies are in MeV.
}
\label{spin-parity} 
\end{center}
\end{table}
%TABLE 1 -------------------------------- -------------------------------- -------------------------------- --------------------------------
%
%       Section 1
\section {The Impulse Approximation}
\label{impulse}
%
%             subsection 2.1
\subsection{1p1h process}
 Fig. \ref{Aoff-shell} is a descriptive diagram for QE {\it electron scattering} on an off-shell proton which is bound in a nucleus of mass $M_A$, and is
moving in the mean field (MF) of all other nucleons in the nucleus.  The on-shell recoil excited   $[A-1]^*$ spectator nucleus has a momentum  $\vec p_{ (A-1) *}=-\vec k$ and  a mean excitation energy  $\langle {E_x^P} \rangle$.
The off-shell energy of the interacting nucleon is 
$E_i  =  M_A - \sqrt{ (M_{A-1}*) ^2+\vec k^2}=  M_A - \sqrt{ (M_{A-1}+{{E_x}})^2+\vec k^2} = M_P -\epsilon^P $, where 
$\epsilon^P = S^P +\langle E_x  \rangle+\frac{\vec k^2}{2M^*_{A-1}}$. 
As discussed in section   \ref{OpticalC} we  model the effect of FSI (strong and EM interactions) by setting $E_f=\sqrt{(\vec {k}+\vec {q_3})^2+M_P^2} -|U_{FSI}|$+$|V_{eff}^P|$, where  $U_{FSI}=U_{FSI}((\vec q_3+\vec k)^2)$.

  Table \ref{spin-parity} shows the spin and parity of the initial state nucleus, and the spin parity of the ground state of the spectator nucleus when a  bound proton or a bound neutron is removed via the 1p1h process. 
 
 The four-momentum transfer to the nuclear target is defined as $q = (\vec q_3,\nu) $.  Here $\vec q_3^2$ is the 3-momentum transfer,  $\nu$ is the energy transfer, and $Q^2= -q^2 = \vec{q}^2- \nu^2$  is the square of the four-momentum transfer.  For QE electron scattering on unbound protons (or neutrons)  the energy transfer $\nu$ is equal to $Q^2/2M_{P,N}$ where $M_P$ is  mass of the proton and $M_N$ is the mass of the  neutron, respectively.
 \subsection{Nuclear Density corrections to $k_F^P$ and $k_F^N$}
 %
 %section 3.10
 The values of the Fermi momentum $k_F$ that are currently used in neutrino Monte Carlo generators are usually taken from an analysis of e-A data by Moniz et al.\cite{electron}.
 The Moniz published values of $k_F$ were extracted using the RFG model under the assumption that the Fermi momenta for protons and neutrons
are different and are related to $k_F$ via the relations  $k_F^N = k_F (2N/A) ^{1/3}$ and $k_F^P = k_F (2Z/A) ^{1/3}$, respectively. What is  actually measured is $k_F^P$, and what is published is $k_F$.  Moniz assumes that the nuclear density (nucleons per unit volume)   is constant.  Therefore, in the same nuclear radius R,  $k_F^N$  for neutrons is larger if N is greater than Z.  Moniz used these expressions to extract the published value of  $k_F$ from the measured value of  $k_F^P$.  

 We undo this correction and re-extract the measured values of  $k_F^P$ for nuclei which have a different number of neutrons and protons. In order to obtain the values of $k_F^N$  from the measured values of $k^P_F$ we use the fact that the Fermi momentum is proportional  to  the cube root of the nuclear density. Consequently $k_F^N=C\frac{N^{1/3}}{R_N}$, and
  $k_F^P=C\frac{Z^{1/3}}{R_P}$, and  $k_F^N=k_F^P\frac{N^{1/3} R_P} {Z^{1/3} R_N}$.  For the proton and neutron radii, we use  the fits for the half density radii  of nuclei (in units of femtometer)  given in ref.\cite{radii}.
  \begin{eqnarray}
R_P&=&1.322 Z^{1/3}+0.007 N +0.022\\
R_N&=&0.953 N^{1/3}+0.015 Z +0.774.
\end{eqnarray}
  We only these fits for nuclei which do not have an equal number of protons and neutrons.   For nuclei which have an equal number of neutrons and protons we assume that  $k_F^N=k_F^P$=$k_F (Moniz)$. 
  
 However for the $_{82}^{208}$Pb nucleus only we use $k_F^P$=0.275 GeV which we obtain from our own fits to inclusive e-A scattering data. For all  other nuclei, our values are consistent with the values extracted by Moniz et. al. 
\subsection{Separation energy}
 The separation energy for a proton ($S^P$) or neutron $S^N$ is defined as follows: 
 \begin{align}
\label{separation}
M_A= M_{A-1} + M_{N,P} - S^{N,P}
\end{align}
 The energy  to  separate both a proton and neutron ($S^{P+N}$) is defined as follows: 
 \begin{align}
M_A= M_{A-2} + M_P+M_N - S^{N+P}
\end{align}
The  proton and neutron  separation energies S$^P$ and S$^N$
are available in nuclear data tables. The values of 
%If both a neutron and a proton  are removed than we define $S^{N+P}$ as follows:
%\begin{eqnarray}
%M_{A-2}=M_A - M_{P} -M_{N} +S^{N+P}
%\end{eqnarray}
%S$^{N+P}$= ($_Z^{A}$BE$_N$) - ($_{Z-1}^{A-2}$BE$_{N-1}$),
%The values of 
S$^P$, S$^N$ and $S^{N+P}$ 
for various nuclei\cite{TUNL,nuclear-data}
are given in Table \ref{spin-parity} 
 %   -- subsection 2.4
 \subsection{Two nucleon correlations}
\begin{figure}
\begin{center}
\includegraphics[width=3.in,height=2.5in]{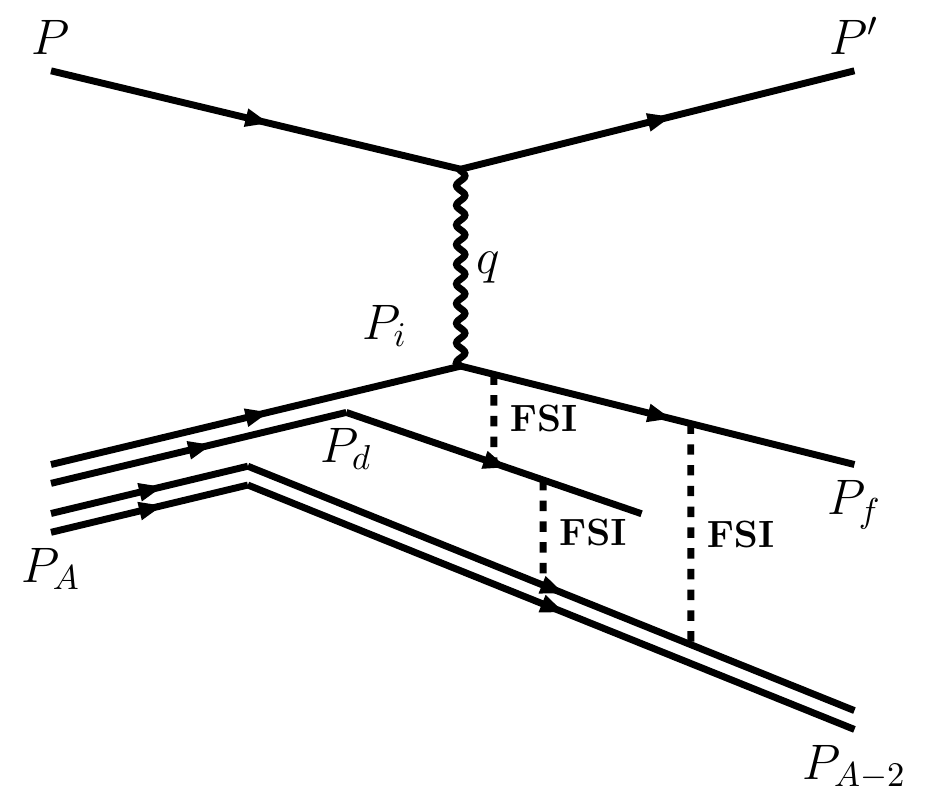}
\caption{ 2p2h process:  Electron scattering from an off-shell bound proton of momentum $\vec p_i$=$\vec k$ from two nucleon short range
correlations (quasi-deuteron).   There is an on-shell spectator (A-2) *  nucleus
and an on-shell spectator recoil neutron with momentum  $-\vec k$.
The off-shell energy of the interacting  bound proton is  $E_i^P (src) =  M_D - \sqrt{M_N + {\vec k}^2}  -S^{P+N}$.
%We model the effect of FSI by setting $E_f^P=\sqrt{\vec k^2+\vec q_3^2} -|U_{FSI}(\vec q_3^2)|$ +$|V_{eff}^P|$.
}
\label{Doff-shell}
\end{center}
\end{figure}

  Fig. \ref{Doff-shell} illustrates the  2p2h process originating  from  both long range and short range two nucleon correlations ($\textsc{src}$). Here the  scattering  is from an off-shell bound proton of momentum $\vec p_i$=$\vec k$. The momentum of the initial state off-shell interacting nucleon   is balanced by a single on-shell correlated recoil  neutron  which has momentum $-\vec k$. The $[A-2]^*$ spectator nucleus is left with two holes.   Short range nucleon-proton correlations  occur $\approx 20$\% of the time\cite{miller}. % As discussed in section \ref{2p2h}. 
   The off-shell energy of the interacting bound proton in a quasi-deuteron is  $ (E_i^P) _{src}=  M_D - \sqrt{M_N + {\vec k}^2}  -S^{P+N}$,  where
  % $\epsilon^{N+P}_{src}$ is the missing energy of the two nucleons and
   $M_D$ is the mass of the deuteron.
  For QE scattering there is an additional 2p2h transverse cross section from ``Meson Exchange Currents'' ($\textsc{mec}$)  and ``Isobar Excitation'' ($\textsc{ie}$).
 
 In this paper we only focus on the extraction of the  {\it average}  removal energy parameters for 1p1h processes. 
  Processes leading to 2p2h final states ($\textsc{src}$, $\textsc{mec}$ and $\textsc{ie}$) result in larger missing energy and  should be modeled separately.
   %
 %FIG 4------------------------------- -------------------------------- -------------------------------- --------------------------------
%FIG 4 -------------------------------- -------------------------------- -------------------------------- --------------------------------
%
%   Table 2
\begin{table}[]
\begin{center}
\begin{tabular}{|c|c||} \hline
Symbol   & \\ \hline \hline
 & Spectator Nucleus Excitation \\ 
$E_x^{P,N}$&Used in spectral functions\\
&implemented in $\textsc{genie}$\cite{genie} \\  \hline
$S^{P,N}$ & Separation Energy \\ 
  = $M_{A-1}+M_{P,N}-M_A$& Nuclear Data Tables \\ 
& (measured) \cite{TUNL,nuclear-data}\\\hline  
% &   \\  
   & missing  energy \\
$E_m^{P,N}$=$S^{P,N}$+$E_x^{P,N}$ & used in spectral functions     \\ \hline
 & removal energy is $\epsilon^{P,N}$  \\
$ {\epsilon^{P,N}}$=$E_m^{P,N}$+$T_{A-1}$& $E_i=M- {\epsilon^{P,N}}$\\ 
   $T_{A-1}=$&used in $E_\nu^{QE-\mu}$,  $Q^2_{QE-\mu}$,\\ 
$=\sqrt{\vec k^2+M_{A-1}^2} -M_{A-1}$    & and  $Q^2_{QE-P}$, also used in\\ 
 $\approx \frac{\vec k^2}{2M_{A-1}}$    & effective spectral functions\cite{effective}\\ \hline 
$\epsilon_{SM}^{\prime (P,N) }$=$  {\epsilon^{P,N}}+T^{P,N}_{av}$ & $\epsilon_{SM}^{\prime (P,N) }$ is Smith Moniz\cite{neutrino} \\ 
$T=\sqrt{\vec k^2+M^2} -M$&  Interaction energy \\
& $E_i=M + T-\epsilon_{SM}^{\prime (P,N) }$\\ 
$\langle k^2 \rangle=0.6k_F^2$ & used in $\textsc{old-neut}$\cite{neut}\\ \hline 
$x^{\nu}$= $\epsilon^N- |U_{FSI}|$ & we use $x^{\nu ,\bar \nu}$ to include \\
$+|V_{eff}^P|$ & the effects of FSI.\\ 
$x^{ \bar \nu}$=$\epsilon^P- |U_{FSI}|$& $U_{FSI} =U_{FSI}((\vec q_3+\vec k)^2)$ \\
%$\textsc{neut}$&\\ \hline 
%$T_{av} \approx\frac{\langle\vec k^2\rangle}{2M}$
%$T_{A-1}$= $\frac{\vec k^2}{2M_{A-1}}  
 \hline
\end{tabular}
\caption{Summary of the relationships between excitation energy $E_x^{P,N}$ (used in $\textsc{genie}$), separation energy $S^{P,N}$, missing  (missing)  energy $E_m^{P,N}$ (used in spectral function measurements), removal energy ${\epsilon^{P,N}}$ (used in the reconstruction of neutrino energy from muon kinematics only), the Smith-Moniz removal energy $\epsilon_{SM}^{\prime P,N}$ (that should be used in $\textsc{old-neut}$) and the  parameters $x$ and  $x^{\nu ,\bar \nu}(\vec q_3+\vec k)^2)$  which we use to include the effects of FSI in electron and neutrino/antineutrino scattering.
For QE neutrino scattering on bound neutrons $|V_{eff}^P|=|V_{eff}|$}.
\label{summary} 
\end{center}
\end{table}
% Table 2  
%
%%TABLE 3 ------------------------------- -------------------------------- -------------------------------- --------------------------------
%	
\begin{table}[]
 %[ht]					
 \begin{center}					
\begin{tabular}{|c|c|c|c|}					
\hline					
Target	&	Q2	&$\langle T^P \rangle$	&$\langle E_m^P \rangle$ \\ 
	&		&$E_m^P <80$	&$E_m^P <80$ \\ 
\hline \hline
$\bf_6^{12}C$	&	0.6	&15.9	&26.0	\\  
Jlab Hall C \cite{C12}	&	1.2	& 16.3 	&25.8	\\
	&	1.8	&16.0	&26.6	\\
	&	3.2	&17.3	&26.2	\\ \arrayrulecolor{gray} \hline \arrayrulecolor{black}
Jlab $ \langle T^P\rangle^{SF}$, $\langle E_m^P\rangle^{SF}$	&	Ave.	&16.4$\pm$0.6	&{\bf 26.1$\pm$0.4}	\\ 
%error	&	RMS	&	0.6&0.4	\\ \arrayrulecolor{black} \hline 
	Saclay  $\langle T^P\rangle^{SF}$,$\langle E_m^P\rangle^{SF}$		&		&16.9$\pm$0.5	&23.4$\pm$0.5	\\ 
	Saclay  $\langle E_m^P\rangle^{levels}$ &		&	&24.4$\pm$2	\\ 
$k_F^P$=221$\pm$5 & %	$\langle K^2 \rangle$=0.6$k_F^2$ 
	 &15.5 $\pm$1.2	&	\\ 	\hline\hline
Target	&	Q2	&$\langle T^P \rangle$	&$\langle E_m^P \rangle$ \\  \hline
$\bf_{14}^{28}Si$	&		&	&	\\  
	Saclay  $\langle T^P\rangle^{SF}$,  $\langle E_m^P\rangle^{SF}$ 	&		&17.0$\pm$0.6	&{\bf 24.0$\pm$0.6}	\\ 
	Saclay $\langle E_m^P\rangle^{levels}$ 	&		&	&27.6$\pm$2	\\ 
$k_F^P$=239$\pm$5 &		%$\langle K^2 \rangle$=0.6$k_F^2$  
&18.1$\pm$1.3 	&	\\ \hline  \hline
Target	&	Q2	&$\langle T^P \rangle$	&$\langle E_m^P \rangle$\\ \hline
$\bf_{20}^{40}Ca$	&		&	&	\\  
	Saclay $\langle T^P\rangle^{SF}$, $\langle E_m^P\rangle^{SF}$ 		&		&16.6$\pm$0.5	&{\bf 27.8$\pm$0.5}	\\ 
		Saclay$\langle E_m^P\rangle^{levels}$	&		&	&26.5$\pm$2	\\ 
$k_F^P$=239$\pm$5 &		%$\langle K^2 \rangle$=0.6$k_F^2$  
&18.1$\pm$1.3 	&	\\ \hline  \hline
Target	&	Q2	&$\langle T^P \rangle$	&$\langle E_m^P \rangle$\\ \hline 
$\bf_{26}^{56}Fe$	&	0.6	&20.4	&30.7	\\
Jlab Hall C \cite{C12}	&	1.2	&18.1	&29.4	\\
	&	1.8	&17.8	&27.8	\\
	&	3.2	&19.1	&28.8	\\  \arrayrulecolor{gray}\hline \arrayrulecolor{black}
Jlab   $\langle T^P\rangle^{SF}$, $ \langle E_m^P\rangle^{SF}$ 	&	Ave.	&18.8$\pm$1.0	&{\bf 29.2$\pm$1.1}	\\
%error  	&	RMS	&1.0	&1.1	\\ 
$k_F^P$=254$\pm$5 &		%$\langle K^2 \rangle$=0.6$k_F^2$ 
 &20.4$\pm$1.4	&	\\ \hline \hline
Target	&	Q2	&$\langle T^P \rangle$	&$\langle E_m^P \rangle$ \\ \hline
$\bf_{28}^{58}Ni$	&		&	&	\\  
Saclay  $\langle T^P\rangle^{SF}$,$\langle E_m^P\rangle^{SF}$ 		&		&18.8$\pm$0.7	&{\bf 25.0$\pm$0.7}	\\ 
Saclay$\langle E_m^P\rangle^{levels}$	&		&	&25.3$\pm$2	\\
$k_F^P$=257$\pm$5 &		%$\langle K^2 \rangle$=0.6$k_F^2$ 
 &20.9$\pm$1.4		&	\\ \hline \hline
Target	&	Q2	&$\langle T^P \rangle$	&$\langle E_m^P \rangle$ \\ \hline
$\bf_{79}^{197}Au$	&	0.6	&20.2	&25.5	\\
Jlab Hall C \cite{C12}	&	1.2	&18.4	&25.7	\\
	&	1.8	&18.3	&24.1	\\
	&	3.2	&19.4	& 26.1	 \\ \hline
Jlab $ \langle T^P\rangle^{SF}$, $\langle E_m^P\rangle^{SF}$		&	Ave.		&	19.1$\pm$0.8 & { \bf 25.3$\pm$0.8} 	\\
%error 	&	RMS	&0.8	&0.8	\\  
$k_F^P$=24.5$\pm$5 &		%$\langle K^2 \rangle$=0.6$k_F^2$  
&19.0$\pm$1.3		&	\\ \hline  \hline
\end{tabular}
     %TABLE 7------------------------------- -------------------------------- -------------------------------- ------e--------------------------
\caption{ Average values of  the  proton kinetic energy $\langle T^P \rangle^{SF}$ and
 missing energy $\langle E_m^P \rangle^{SF}$ for 1p1h final  states ($E_m^P <80$)  extracted 
 from published tests of the Koltun sum rule using spectral function (SF)  measurements  at Jefferson lab Hall A\cite{C12} and Saclay\cite{Saclay}.  
 For a Fermi gas distribution $\langle T^P \rangle= \frac{3}{5} (k_F^P)^2$ All energies are in MeV.  The bolded numbers are the best estimates for each target.
 % We take the  RMS variation with $Q^2$ of the Jefferson Lab data ($\approx$ 0.5 MeV)  as the statistical uncertainty in the Jlab measurements
%of  $\langle E_m^P\rangle^{SF}$.  We use  the difference in $\langle E_m^P \rangle ^{SF}$ between the Jefferson Lab  and Saclay measurements on  $_6^{12}C$  ($\approx$ 3 MeV)  as the systematic error in  $\langle E_m^P\rangle^{SF}$.  For comparison, we also  present the  {\it average}  energy 
 %$\langle E_m^P \rangle ^{levels}$ extracted from the average  of the missing energies of all shell-model levels (weighted by the number of nucleons)    published by the Saclay experiments.
% 
%Note that if e2p2h processes ($E_m >80$)  from short range correlations 
%are included, the average $\langle E_m^P \rangle$  for all $E_m$ is much larger\cite{Benharsrc} (52.2 MeV for Cabon and 70.5 MeV for
%nuclear matter).
 }					
\label{HallC} 					
\end{center}					
\end{table}
	%
 %section 3
\section{Spectral functions and  $ee^\prime p$ experiments}
\label{eep}
 In  $ee^\prime p$ experiments the following process is investigated:
\begin{align}
\label{process}
e + A \to e^\prime + (A-1) ^\star + p_f. 
\end{align}
Here,  an electron beam is incident on a nuclear target of mass  $M_A$. 
The hadronic final state consists of a proton of four momentum $p_f\equiv (E_f,{\vec  p_f}) $ and an undetected nuclear remnant
$ (A-1) ^\star$. Both the final state electron and the final state proton are  measured.
 The $ (A-1) ^\star$  nuclear remnant can be a  $ (A-1, Z-1) $ spectator nucleus with excitation $E_x^P$, or a  nuclear remnant with additional 
 unbound nucleons. 

At high energies, within the plane wave impulse approximation (PWIA)
the  initial momentum $\vec k$ of the initial state off-shell interacting nucleon can be  identified  approximately with
the missing momentum   ${\vec  p}_m$.  Here we  define $p_m$  = $|{\vec  p}_m|$ and $k=|\vec k|$
\begin{align}
\label{def:pmiss}
{\vec  p}_m = {\vec  p_f}-{\vec  q_3} \approx \vec k.
\end{align}
The missing energy $E_m$   is defined by the following relativistic energy conservation expression,
  \begin{eqnarray}
%\label{def1:Emiss}
& &\nu + M_A = \sqrt{  (M_{A-1}^*)^2 + \vec {p_m}^2 } + E_f^P \\
& & E_f^P=\sqrt{{\vec  p_f^2}+ M_P^2},~~~~~
 % \nonumber \\  
 M_{A-1}^*= M_A - M + E_m. \nonumber 
\end{eqnarray}
 The missing energy $E_m$ can be expressed in term of the excitation
  energy  ($E_x$) of the spectator (A-1) nucleus and the  separation
 energy of the proton $S^P$ (or neutron $S^N$).
    \begin{eqnarray}
E_m^{P,N}&=&S^{P,N}+E_x^{P,N}
  \end{eqnarray}
The probability distribution of finding a nucleon with initial state momentum $ p_m\approx k$ and missing energy
 $E_m$ from the target nucleus is described by the spectral function, defined as $P_{SF} (p_m, E_m) $. 
Note that  for spectral functions both  $P (p_m,E_m) $ and $S (p_m,E_m) $ notation are used in some publications.  
   The spectral functions  $P_{SF}^P (p_m,E_m^P)$ and   $P_{SF}^N ({ p_m},E_m^N) $ 
   for protons and neutrons are two dimensional distributions which can be measured (or calculated theoretically).  Corrections for final state interactions of the outgoing nucleon are required in the extraction of $P_{SF}^P (p_m,E_m^P)$ from $ee^\prime p$ data.    The kinematical region corresponding to low missing momentum and energy is where shell model\cite{brown} states dominate\cite{review}. 
 In practice, only the spectral function for protons can be measured reliably. 
 
 In addition to the  1p1h contribution in which the residual nucleus is left in the ground or excited bound state, the measured spectral function includes contributions from  nucleon-nucleon correlations in the initial state (2p2h)  where there is one or more additional spectator nucleons.   Spectral function  measurements cannot differentiate between a spectator (A-1)  nucleus and a spectator (A-2)  nucleus from  $\textsc{src}$ because the 2nd final state $\textsc{src}$ spectator nucleon is  not detected.   
  
  Here,  we focus  on the  spectral function for the 1p1h process, which dominates for   $E_m$ less than 80 MeV,  and ignore the spectral function for the 2p2h process which dominates at higher values of $E_m$.   We  use  shell model calculations to obtain  the difference in the binding energy parameters for neutrons and protons.  
    %
 %%  TABLE 5  -------------------------------- -------------------------------- -------------------------------- --------------------------------

  \begin{table*}
\begin{center}
\begin{tabular}{|c|		c		c     c		c		c|  c    c|				c		c|} \hline  
Nucleus	&		&$_6^{12}C$	&  	& 	& &  &	$_{14}^{28}Si$	&				&	$_{28}^{58}Ni$\\ 
		$S^P$              &		&	16.0	&    &	&		&	&11.6	&		  &8.2\\  \hline
%	$T_{A-1}$                &		&	1.4	&  &	& &   	&	0.7	&		  &0.4\\  \hline
 	&		&	         shell        &	shell &	shell &	&                       &shell	                &		&	  shell        	\\ 

	ee$^{\prime}$p               	&		&	         missing    &  missing  &	missing&& 	&	missing                                    & &	 missing      	\\ 
	$\epsilon^P$=$E_m$+$T_{A-1}$	   	&		&	          energy &energy   &		energy &	width    &    & energy             &				&	  energy     	    	\\ 
		&		&	          $E_m^P$    &	 $E_m^P$    &  $E_m^P$	&$\textsc{fwhm}$	  &         &  $E_m^P$        &		    &$E_m^P$		\\ 
				        	               	&		&	        Saclay       & $\textsc{nikhef}$       &	Tokyo & Tokyo & &	Saclay                                      &		&	 Saclay       	\\ \hline
\hline 
1s$_{1/2}$	&	2	&38.1$\pm$1.0	&42.6$\pm$5&    36.9$\pm$0.3	&19.8$\pm$0.5    &  2	&	51.0          	&		      	2 &	62.0 \\   
%&		&	            &		&	&                   &  &		&	               	    	    	\\ 
1p$_{3/2,1/2}$	&	4	&	17.5$\pm$0.4	&17.3$\pm0.4$	& 15.5$\pm0.1$ & 6.9$\pm$0.1 &6	&	32.0	&	6			&	45.0	\\ \hline
 1d$_{5/2,3/2}$ 	&		&		&	& & &  4	&	16.1$\pm$0.8	&		  10	&	21.0\\
2s$_{1/2}$	&		&		&&	&  & 2	&	13.8$\pm$0.5	&	2			&	14.7$\pm$0.2\\
1f$_{7/2}$	&		&		&	&	&	&	&		&	8	&	9.3$\pm$0.3\\ \hline
 \arrayrulecolor{gray}\hline
	
$\langle E_m^P\rangle^{levels}$	&	6  &		$\bf{\langle24.4\pm2	\rangle}$	& $\bf{\langle25.7\pm2\rangle}$&$\bf{\langle22.6	\pm3\rangle}$  &  & 14	&	$\bf{\langle27.6\pm2	\rangle}$	&				28 &$\bf{{\langle25.3	\pm2\rangle}}$	\\
$T_{A-1}$                &		&	1.4	&1.4  &1.4	& &   	&	0.7	&		  &0.4\\ 	
$\langle \epsilon^P\rangle^{levels}$	&	6  &		$\bf{\langle25.8\pm2	\rangle}$	& $\bf{\langle27.1\pm2\rangle}$&$\bf{\langle24.0\pm3\rangle}$  &   & 14	&	$\bf{\langle28.3\pm2	\rangle}$	&				28 &$\bf{{\langle25.7\pm2	\rangle}}$	\\	
\hline 
\end{tabular}
\end{center}
\caption{Results of a DPWA analysis of the ``level missing energies'' for different shell-model levels done by the  Saclay\cite{Saclay} and Tokyo\cite{Tokyo1,Tokyo2,Tokyo3} $e e^{\prime}p$ experiments on  $_6^{12}C$, $_{14}^{28}Si$ and $_{28}^{58}Ni$.}
\label{Saclay}
\end{table*}
%TABLE 5  -------------------------------- -------------------------------- -------------------------------- --------------------------------
%
% TABLE 6 ------------------------------- -------------------------------- -------------------------------- --------------------------------
 \begin{table*}[]
\begin{center}
\begin{tabular}{|c|		c		c|		c		c|		c		c				c c |		c		c c|} \hline	
	Nucleus       &		&	$_3^6Li$	&		&	$_{13}^{27}Al$	&	&$_{20}^{40}Ca$			&	&		&	& $_{23}^{51}V$ & \\
$S^p$       	&		&	          4.4     &		&	    8.3 	                 &		&	         8.3		&	   	&		&	& 8.1&   	\\\hline
		        	               	&		&	         Shell        &		&	Shell                       &		&	Shell                  		&	  Shell       	&		&&	Shell     &	\\ 	
	ee$^{\prime}$p               	&		&	         missing       &		&	missing                       		&	              &	missing   	&	  missing      	&		& &	missing     &	\\ 
	$\epsilon^P$=	   	&		&	          energy    &		&	energy                      &		&	energy               		&	  energy     	&	width	&	&energy    & width	\\ 
	$E_m$+$T_{A-1}$	&		&	          $E_m^P$    &		&	$E_m^P$                    &		&	 $E_m^P$               		&	
		  $E_m^P$    & $\textsc{fwhm}$	&	 & $E_m^P$    & $\textsc{fwhm}$	\\ 	  	       	               	&		&	        Tokyo       &		&	Tokyo    
		                        & 	&	   Saclay &              			  Tokyo       		&  Tokyo	&	&Tokyo    &   Tokyo\\ \hline  \hline
1s$_{1/2}$ &	2	&	22.6$\pm$0.2	&	2	&	57$\pm$3	&	2	&	56.0	&		59$\pm$3	& 34$\pm$10&	2	&	60$\pm$3& 36$\pm$	11\\
1p$_{3/2,1/2}$	&	1	&	4.5$\pm$0.2	&	6	&	34$\pm$1		&	6	&41.0	&		35$\pm$1	&21$\pm$3	&6	&	40$\pm$1&25$\pm$4	\\ \hline
 1d$_{5/2}$	&		&		&			4	&	14.0$\pm$0.6	&	6	& *14.9$\pm$0.8	&	19.0$\pm$1.1	& 10$\pm$3	&6		&19.5$\pm$0.5& 19$\pm$2	\\
2s$_{1/2}$	&		&		&		1&	14.3$\pm$0.2	&	2	&	11.2$\pm$0.3	&		14.4$\pm$0.3	&13$\pm$1	&2		&15.1$\pm$0.2& 5$\pm$2\\
1d$_{3/2}$	&		&		&		&		&	4	&	*14.9$\pm$0.8	&		10.9$\pm$0.7&	9$\pm$1	& &	&\\	
1f$_{7/2}$	& 		&		&		&		&		&	*combined	&				&	&7		&10.3$\pm$1.1&	5$\pm$3\\ \hline
  \arrayrulecolor{gray}\hline
%%%protons	&	3	&		&	13	&		&	20	&		&	20	&		&	23	&		&\\  \hline
${ \langle E_m^P\rangle}^{levels}$ 	&	3	&	$\bf{\langle16.6\pm2\rangle}$	&	13	&	$\bf{\langle29.9\pm3\rangle}$	&20&		
	$\bf{\langle26.5\pm3\rangle}$	&	$\bf{\langle25.7\pm3\rangle}$	&	&23		&$\bf{\langle25.2\pm3	\rangle}$  & \\ 
 $T_{A-1}$      	&		&	                  1.8     &		&	     0.7   	                 &			          & 0.5&	     0.5    	&		&	& 0.4  &   	\\  
%	protons	&	3	&		&	13	&		&	20	&		&	20	&		&	23	&		&\\  \hline
	${\bf \langle \epsilon^P\rangle }^{levels}$ 	&	3	&	$\bf{\langle18.4\pm2\rangle}$	&	13	&	$\bf{\langle30.6\pm3\rangle}$	&20&		
	     $\bf{\langle27.0\pm3\rangle}$	&	$\bf{\langle26.3\pm3\rangle}$	&	&23		&$\bf{\langle25.6\pm3	\rangle}$  & \\ \hline
\hline
\end{tabular}
\end{center}
\caption{
Results of a DPWA analysis of the ``level missing energies'' for different shell-model levels done by the  Saclay\cite{Saclay} and Tokyo\cite{Tokyo1,Tokyo2,Tokyo3} $e e^{\prime}p$ experiments on  on $_3^6Li$, $_{13}^{27}Al$, $_{20}^{40}Ca$ and  $_{23}^{51}V$.}
\label{Tokyo}
\end{table*}		
%
    %  -   Section  4
  \section {Effects of the optical and Coulomb potentials (FSI)}
  \label{OpticalC}
 We use empirical parameter  $U_{FSI}((\vec q_3+\vec k)^2)$  to approximate the effect of the  
  interaction of the final state proton with the optical potential of the spectator nucleus.  This is important
  at low  values of $\vec q_3^2$.
 In addition, we include the effect of the interaction of the final state proton with the Coulomb field of
 the nucleus ($V_{eff}^P$). 
 
 In QE scattering of electrons  a  three momentum transfer $\vec q_3$ to a bound proton
with  initial momentum $\vec k$ results in the following 
 energy $E_f^P$  of the final state proton:
 \begin{eqnarray}
 \label{def1:Emiss}
 E_f^P&=&\sqrt{{\vec  ( k + \vec q_3)^2} + M_P^2} -|U_{FSI}
 ((\vec q_3+\vec k)^2)| +|V_{eff}^P| \nonumber
%\nu + M_A &=& \sqrt{M_{A-1}^{*2} + k^2 } + E_f ^P \nonumber \\
%&=& M_A - M_P + E_m^P + \frac{\vec k^2}{M_{A-1}^*}  + E_f ^P\\
%\nu &+&E_i =   E_f,   ~~~~~~E_i=(M_P-\epsilon^P).\nonumber 
%\nu +(M-\epsilon) &=.&   \sqrt{{\vec  ( k + \vec q_3)^2} + M_P^2} +U_{FSI} +V_{eff}^P
%&=& \sqrt{\vec {p_m}^2 + M_P^2}
 \end{eqnarray}
where  for electron scattering on bound protons $V_{eff}^P= \frac{Z-1}{Z}|V_{eff}|$.  
The Coulomb correction $|V_{eff}|$ is discussed in Appendix A. 

We define the average removal energy $\epsilon^{P,N}$ 
in terms of the  average momentum $\langle k^2\rangle^{P,N}$ of the bound nucleon as follows:
\begin{eqnarray}
\epsilon^{P,N}&=&  E_m^{P,N}+T^{N,P}_{A-1} \\
&=& S^{P,N}+E_x^{P,N}+  \frac{\langle k^2\rangle^{P,N}}{2M^*_{A-1}}\nonumber 
\end{eqnarray}
 In order to properly simulate neutrino interactions we extract  values of the
 average missing energy (or equivalently the average excitation energy)  from spectral functions measured in 
  ee$^\prime$p experiments.  We then use these values and extract  $U_{FSI}((\vec q_3+\vec k)^2)$   from inclusive e-A as discussed in section \ref{U-FSI}.
  %  -------------------------------- -------------------------------- -------------------------------- --------------------------------
 %
 % FIG  -8----------------------------- -------------------------------- -------------------------------- --------------------------------
\begin{figure}
\begin{center}
\includegraphics[width=3.3in,height=2.5in]{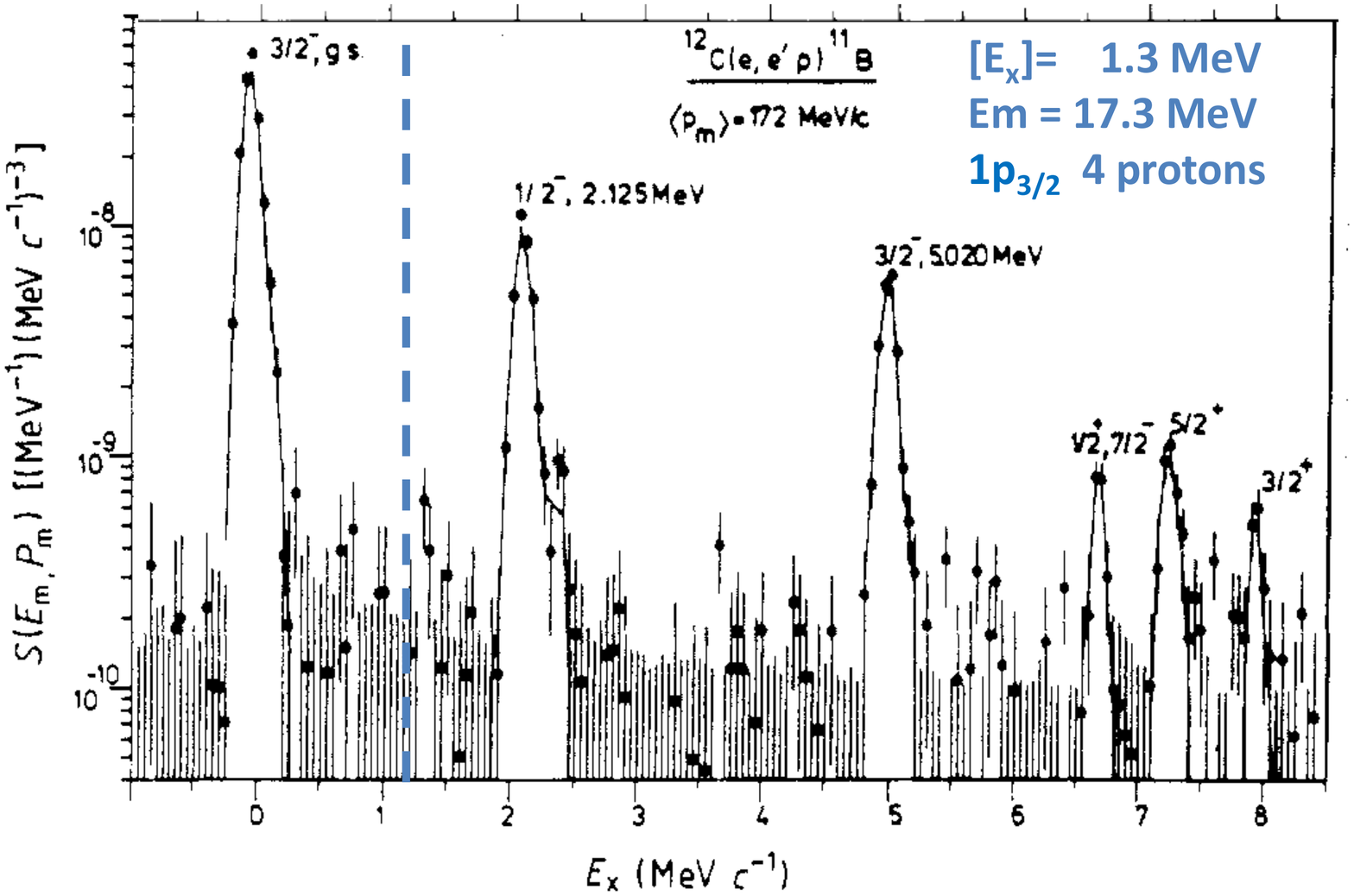}
\includegraphics[width=3.in,height=2.0in]{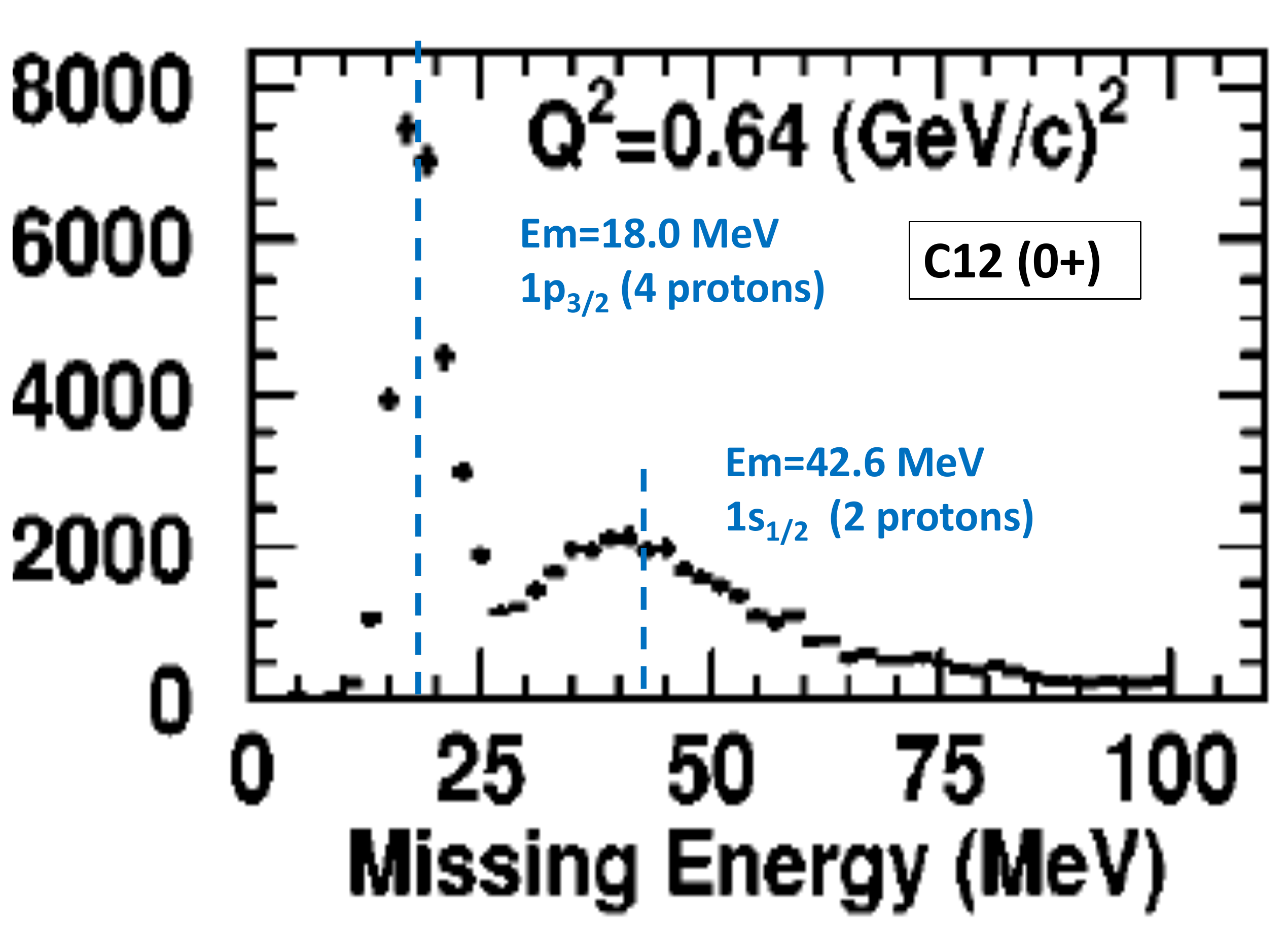}
\caption{ Top panel: The measured\cite{huberts} \textsc{NIKHEF} high resolution spectral function  for protons in  
the 1p level of  $_{6}^{12}C$ as a function of the spectator nucleus excitation energy  $E_x^P$  for ${\vec p_m}=\vec k$ = 172 MeV/c.  
 Bottom panel: The Jlab measurement\cite{C12} of the  one-dimensional spectral function for
 $_{6}^{12}C$ as a function of missing energy $E_m^P$ for $Q^2$= 0.64 GeV$^2$.  
  }
\label{carbon-spectral-detailed}
\end{center}
\end{figure}
%
%  FIG 8 ------------------------------ -------------------------------- -------------------------------- --------------------------------
%
% FIG 9 ------------------------------ -------------------------------- -------------------------------- --------------------------------
%  \textcolor{black}{text}
\begin{figure}
\begin{center}
\includegraphics[width=3.5in,height=2in]{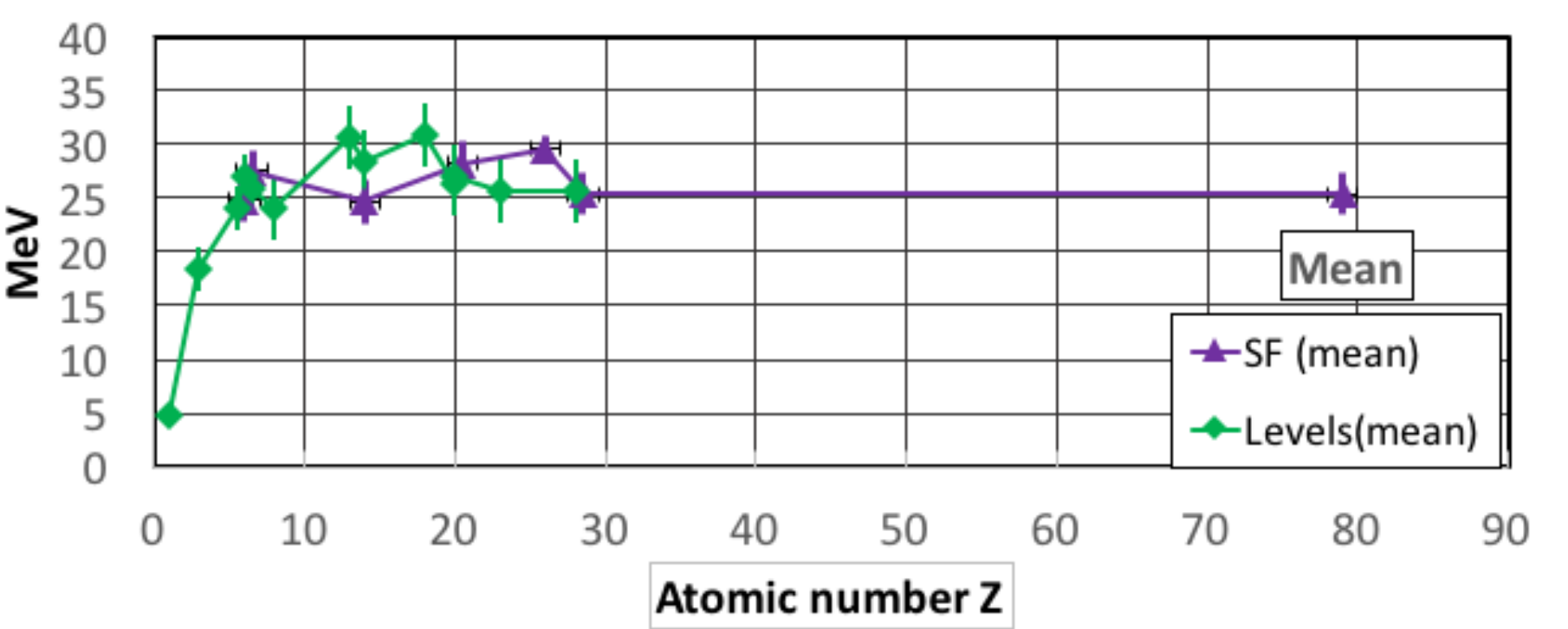}
\caption { Average removal energies versus atomic number Z. Values of  $\langle \epsilon^P\rangle^{SF}$  from tests of the Koltun sum rule in $ee^{\prime}p$  experiments (in purple)  are  compared to values of   $\langle \epsilon^P\rangle^{levels}$  extracted from $ee^{\prime}p$ measurements of  ``level missing energies'' (in green). }
\label{average-peak}
\end{center}
\end{figure}
% FIG 9 ------------------------------ -------------------------------- -------------------------------- --------------------------------
%
%\begin{figure}
%\begin{center}
%\includegraphics[width=3.5in,height=3.0in]{O16-Ar40.pdf}
%\caption{ 
%The measured (Frascati \cite{other})  energy loss spectra ($\nu=E_0-E^{\prime}$)   at $32^0$ for  0.7 GeV  incident electrons for oxygen $_{8}^{16}O$ (in squares)   and $_{18}^{40}Ar$ (in X's], respectively.  The  {\it peaks} of the QE distributions for oxygen and $_{18}^{40}Ar$ are  the same value of  ($\nu_{ {\it peak}}\approx$  105 MeV).  
% }
%\label{Fig-O16-Ar40}
%\end{center}
%\end{figure}
%---STOP  HERE $$$$$   
%section 5.7
 % TABLE A1 ------------------------------ -------------------------------- -------------------------------- --------------------------------
 \begin{table*}[]
\begin{center}
\begin{tabular}{|c||		c		c		c| c||		c	c|			c		c|c|}	\hline
	&	 ${\bf^{12}_{6}C}$	&	proton   &	neutron	&N-P&	 ${\bf ^{16}_{8}O}$	&proton	&proton		&	neutron	&N-P\\
$T_{A-1}$	&		&	1.4	&			1.4	& Diff &		&	1.1	&	1.1&		1.1&	Diff\\  \hline
&		&	 binding	&		 	binding&		&&	Jlab shell	&	binding&		binding&	\\ 
 	          &	          	&	energy				&	 energy	&  &	& missing energy  			& energy&	energy&	\\  \hline
	          $S^P$,$S^N$	&		&	16.0	&		18.7	&   {\bf 2.7}&		&	12.1	& 12.1&			15.7&  {\bf 3.6}	\\ \hline \hline
1s$_{1/2}$	&	2	&	42.6	&		43.9& 	1.3&	2	&42$\pm$2	 &45.0	&		47.0&	2.0\\ 
	&		&		&		&		&     &  & ( {\it40$\pm$8}) 	   	& &\\ \hline
1p$_{3/2}$	&	4	&	 {\it 16.0}		&	 {\it 18.7}	&2.7	&4	&18.9$\pm$0.5	& {\it 18.4}		&	 {\it 21.8}	&3.4 \\
1p$_{1/2}$	&		&		&			&	&	2	&12.1$\pm$0.5	& {\it 12.1}	&		 {\it15.7}&	3.6 \\ \hline \hline
$average~\langle E_m^P \rangle, BE$	&	6&	$ \langle 24.9 \rangle$&			$\langle 27.1 \rangle$	&{\bf  2.6} &	8&$\langle{23.0\pm2}\rangle$	 &$ \langle 23.5 \rangle$	&		$\langle 26.6 \rangle$	&$\bf 2.9\pm1$	\\ \hline
%levels not included	&		&	1s 	&			 1s 	&&		&1s	&  1s 	&1s	&	\\  
%$ peak~[ E_m],~BE$	&	4 &[16.0]	&	[18.7]	&	{2.7}	& 6&[16.6]	&[16.3]	&		[19.8]	& {3.5$\pm1$}\\  \hline
%$\bf average~\langle E_m^P \rangle$-$peak~[ E_m]$	&		&	8.9$\pm1$ 	&			 8.4$\pm$1 	&&		&6.4$\pm1$	&  7.5$\pm1$ 	&		6.8$\pm1$&	\\  \hline
%$\bf average~\langle E_m^P \rangle$-$peak~[ E_m]$	&	$^{12}_{6}C$	&	 	&		&&	 $^{16}_{8}O$	&{\bf 6.4$\pm1$}	&  {\bf 7.5$\pm1$} 	&	&		\\  \hline
%&	$^{12}_{6}C$	&		&			 	&&	 $^{16}_{8}O$		&	& 	&		&	\\  \hline
%\hline 
\end{tabular}
\end{center}
\caption{ Shell-model single particle binding energies for $^{12}_{8}C$ and  $^{16}_{8}O$  from ref.\cite{spectral-theory}.  When available, the experimental values shown in   {\it italics} are used.  %The single particle binding energies are assumed to be the same as the missing energies $E_m$. 
 The difference between  the  {\it average}  missing energies  $\langle E_m^{P,N} \rangle$ for  {\it neutrons} and  {\it protons} can be approximated by the difference in the weighted average of the single particle binding energies of all shell-model levels. We obtain N-P  differences of 2.6 and 2.9 MeV for $^{12}_{8}C$ and $^{16}_{8}O$, respectively.  These differences are close to the corresponding differences in separation energies of neutrons and protons ($S^{N}$-$S^{P}$)   of  2.7 and 3.6 MeV, for $^{12}_{8}C$ and $^{16}_{8}O$, respectively.  
  %The  {\it peak}  missing energies  $[E_m^{P,N}]$  can be approximated by a truncated  {\it average} (not including the 1s level).
  }   
 \label{Tcarbon}
\end{table*}
 
 %subsection 4.1
  \subsection {Smith-Moniz formalism}
The Smith-Moniz\cite{neutrino} formalism uses on-shell description of the initial state.
In the on-shell formalism, the energy conserving expression is 
 $$\nu+M -\epsilon=   E_f$$
  is replaced with
  $$\nu +\sqrt{k^2+M^2} -\epsilon_{SM}^{\prime P,N} =   E_f.$$
  Therefore, 
    $$\epsilon_{SM}^{\prime} = \epsilon +\langle T^{P,N} \rangle , $$
    where 
     $$\langle T^{P,N} \rangle=\sqrt{\langle k^2\rangle^{P,N}+M^2}-M\approx    \frac {3}{5} \frac {(k_F^{P,N})^2}{2M}$$
     
    A summary of the relationships between excitation energy $E_x^{P,N}$ used in $\textsc{genie}$ (which incorporates the Bodek-Ritchie  \cite{BodekRitchie} off-shell formalism), separation energy $S^{P,N}$, missing missing  energy $E_m^{P,N}$ (used in spectral function measurements),  removal energy ${\epsilon^{P,N}}$ (used in the reconstruction of neutrino energy from muon kinematics only)  and the Smith-Moniz removal energy $\epsilon_{SM}^{\prime P,N}$ (that should be used in $\textsc{old-neut}$)
is given in Table \ref{summary}. 
%
% section 5
\section{Extraction of average missing energy $\langle E_m \rangle$}
\label{section_5}
We extract the average  missing energy ${\langle E_m^P \rangle}$  and excitation energy ${\langle E_x^P \rangle}$  for the 1p1h process from $ee^\prime P$ electron scattering data using two methods.
\begin{enumerate}	
\item  $\langle E_m^P\rangle^{SF}$:  From direct measurements  of  the average missing energy ${\langle E_m^P \rangle}$ and  average proton kinetic energy ${\langle T^P \rangle}$.
These quantities have been extracted from spectral functions measured in ee$^{\prime}$p experiments for tests of the Koltun sum rule\cite{Koltun}. 
The contribution of two nucleon corrections is minimized by restricting the analysis to $E_m^P<$ 80 MeV. This is the most reliable determination of   $\langle E_m^P \rangle$.  We refer to this  {\it average} as  $\langle E_m^P\rangle^{SF}.$ $~~~~~~~~~~~~~$             

\item $\langle E_m^P\rangle^{levels}$:  By taking the average (weighted by shell model number of nucleons)  of the nucleon ``level missing energies'' of all shell model levels which are extracted from  spectral functions measured in ee$^{\prime}$p experiments.   We refer to this  {\it average} as   $\langle E_m^P\rangle^{levels}.$
\end{enumerate} 

  There could be bias in method 2 originating from the fact that  a fraction of the nucleons ($\approx 20\%$)  in each level are in a correlated state with other nucleons (leading to 2p2h final states).  The fraction of correlated nucleons is not necessarily the same for all shell-model levels.
As discussed in section \ref{Comparison-three} (and shown Fig. \ref{average-peak})we find that the values of  $\langle E_m^P\rangle^{levels}$ are consistent with  $\langle E_m^P\rangle^{SF}$ for nuclei for which both are available.

When available, we extract the removal energy parameters using  $\langle E_m^P\rangle^{SF}$ from method 1.  Otherwise  we use $\langle E_m^P\rangle^{levels}$ from method 2. 
%If complete spectral function measurements are not available we use  {\it peak} $[E_m^P]^{Moniz}$ from method 1, and apply the estimated difference between
%the  {\it average} and the  {\it peak} to obtain $\langle E_m ^P\rangle^{Moniz}$. 
For each of the two methods, we also use the nuclear shell model to estimate difference between  the missing energies for neutrons and protons.
%
%
 %Table 4   -------------------------------- -------------------------------- -------------------------------- --------------------------------
   \begin{table*}[ht]
\begin{center}
\begin{tabular}{|c|c|c|c|c|c||cc|} \hline
  \arrayrulecolor{black}\hline
  &  &  $k_F^P$,$k_F^N$&$k_F^P $ & $E_{\mathrm{shift}}$&$|V_{eff|}$ & &${ {\epsilon}}^P$ (MeV)    \\
   &   $_{Z}^ANucl$ &Moniz & $\psi^\prime$ fit      & $\psi^\prime$ fit   &Gueye   & &      \\
&   Nucl.   &  $\pm$5       &   ref.\cite{Donnelly} &   ref.\cite{Donnelly}  & ref.\cite{gueye}                   & &                    \\ 
Source&          & updated  &  &   &  & &   \\
Source&          &MeV/c  &MeV/c & MeV  & MeV  & &   \\ 
\hline 
  \arrayrulecolor{black}\hline
  \arrayrulecolor{black}\hline
  & $_1^2H$                           &{ 88,88}&  &   &      &  &      {\bf *4.7$\pm$1} \\ \hline 
ee$^{\prime}$p Tokyo\cite{Tokyo1,Tokyo2,Tokyo3}&   $_3^6Li$                         &  {169,169} &165    &  15.1      &  1.4  & $\epsilon^{levels}$   & 
{\bf *18.4$\pm$3}   \\ \hline \arrayrulecolor{black}\hline
%Tokyo\cite{Tokyo1,Tokyo2,Tokyo3}     
  \arrayrulecolor{black}\hline
% carbon Tokyo
ee$^{\prime}$p Tokyo\cite{Tokyo1,Tokyo2,Tokyo3} &        ${_6^{12}C}$  &  {221,221}           &  228  &20.0    &   3.1$\pm$0.25  &      $\epsilon^{levels}$ &   24.0$\pm$3   \\ 
  \arrayrulecolor{gray}\hline
  \arrayrulecolor{black}
  % carbon Nikhef
ee$^{\prime}$p  $\textsc{NIKHEF}$\cite{NIKHEFC12} &    ${_6^{12}C}$                   &    & &   &  &   $\epsilon^{levels}$   &    27.1$\pm$3\\
 \arrayrulecolor{gray}\hline
 % carbon Saclay
ee$^{\prime}$p Saclay\cite{Saclay} &     ${_6^{12}C}$               &  & &  & & $\epsilon^{levels}$  &    25.8$\pm$3      \\ 
   &                 & & &  & & $\langle \epsilon\rangle ^{SF}$ & ~~24.8$\pm$3.0    \\ 
    \arrayrulecolor{gray}\hline 
    % carbon Shell model
   Shell Model  binding E&     ${_6^{12}C}$               &  & &  & & $\epsilon_{shell-model} ^{levels}$  &     24.9 $\pm$5     \\   \arrayrulecolor{gray}\hline   
  % carbon jlab
ee$^{\prime}$p Jlab Hall C \cite{C12}&     ${_6^{12}C}$                &    &  &   &  &   $\langle \epsilon \rangle ^{SF}$  &   {\bf  *27.5$\pm$3} \\
 \arrayrulecolor{black}\hline  \arrayrulecolor{black}\hline 
% oxygen jlab Hall A
  ee$^{\prime}$p Jlab Hall A \cite{O16} &${_8^{16}O}$                     & {225,225} & & & 3.4& $\epsilon^{levels}$ &             {\bf *24.1$\pm$3}  \\ 
  \arrayrulecolor{gray}\hline 
    % Oxygen  Shell model
   Shell Model  binding E&     ${_8^{16}O}$              &  & &  & &$ \epsilon_{shell-model} ^{levels}$&   23.5$\pm$5     \\   \arrayrulecolor{gray}\hline   
 \arrayrulecolor{black}\hline
  \arrayrulecolor{black} \hline
%Aluminum
ee$^{\prime}$p Tokyo\cite{Tokyo1,Tokyo2,Tokyo3} &$_{13}^{27}Al$              &   {238,241}&   236&  18.0    &5.1  & $\epsilon^{levels}$   &  {\bf 30.6 $ \pm $3 }   \\ 
  \arrayrulecolor{black}\hline
   \arrayrulecolor{black}\hline
  %Silicon1
ee$^{\prime}$p Saclay\cite{Saclay} &$_{14}^{28}Si$               &    {239,241}  &  &     &  5.5& $\epsilon^{levels}$   &  28.3$\pm$2   \\
% Silicon2 
        &      & & &  & & $\langle \epsilon\rangle ^{SF}$  &     {\bf *24.7$\pm$3}      \\ 
   %Argon
 \arrayrulecolor{black}\hline
  \arrayrulecolor{black} \hline
${_{20}^{40}Ca}\rightarrow{_{18}^{40}Ar}$ Shell Model &  ${_{18}^{40}Ar}$               &{ 251,263}   &    &             &
    6.3 &$\epsilon^{levels}$ & {\bf  *30.9$\pm$4} \\      
% Calcium Tokyo
  \arrayrulecolor{black}\hline \arrayrulecolor{black}\hline
ee$^{\prime}$p Tokyo\cite{Tokyo1,Tokyo2,Tokyo3}&       ${_{20}^{40}Ca}$        & {251,251} &241  &   28.8     &7.4$\pm$0.6 & $\Delta^{levels}$  &   26.3$\pm$3   \\ 
 \arrayrulecolor{gray}\hline
 % Calcium Scalay
ee$^{\prime}$p Saclay\cite{Saclay} &      ${_{20}^{40}Ca}$        &   & &       & & $\epsilon^{levels}$ &    27.0$\pm$3       \\
  &                  &  & &  & & $\langle \epsilon \rangle ^{SF}$  &   {\bf *28.2$\pm$3}     \\ 
 \arrayrulecolor{gray}\hline
 % Calcium Shell model
Shell-model binding E &        ${_{20}^{40}Ca}$       &   &  &        & & $\epsilon_{shell-model}  ^{levels}$  &     23.6$\pm$5     \\ 
  \arrayrulecolor{black}\hline  \arrayrulecolor{black}\hline
%VV V V
ee$^{\prime}$p Tokyo\cite{Tokyo1,Tokyo2,Tokyo3}  &$_{23}^{50}V$               &  {253,266} &     &    &  8.1& $\epsilon ^{levels}$ &     {\bf *25.6$\pm$3}   \\ 
  \arrayrulecolor{black}\hline   \arrayrulecolor{black}\hline
ee$^{\prime}$p Jlab hall C \cite{C12}& $_{26}^{56} Fe$           &  {  254,268}   &   241 & 23.0  & 8.9$\pm$0.7&$\langle \epsilon \rangle ^{SF}$ &   {\bf *29.6$\pm$3}  \\   
   \arrayrulecolor{black}\hline     \arrayrulecolor{black}\hline
    % Calcium Ni
  ee$^{\prime}$p Saclay\cite{Saclay} &  $_{28}^{58.7} Ni$               &{  257,269} & 245  &   30.0     &9.8 & $\epsilon ^{levels}$  &     25.7$\pm$3      \\  
     &                  &    & &  && $\langle \epsilon \rangle ^{SF}$  & {\bf *25.4$\pm$3}      \\ 
  \arrayrulecolor{black}\hline \arrayrulecolor{black}\hline
  % Zicronium
 Shell-model binding E &$_{40}^{88}Zr$             &  &       &  & 11.9$\pm$0.9& $\epsilon_{shell-model}  ^{levels}$    &{\bf  25.1$\pm$5} \\  \hline
   \arrayrulecolor{black}\hline  \arrayrulecolor{black}\hline
   %Gold
   ee$^{\prime}$p Jlab Hall C \cite{C12}& $_{79}^{197}Au$   & {275,311} & 245   & 25.0 & 18.5& $\langle \epsilon \rangle ^{SF}$  & {\bf *25.4$\pm$3}    \\ 
    \arrayrulecolor{black}\hline \arrayrulecolor{black}\hline
    %$ Lead
   Shell Model   binding E           &$_{82}^{208}Pb$  &  {275,311}& 248  & 31.0    & 18.9$\pm$1.5 & $\epsilon_{shell-model} ^{levels}$  &  {\bf   22.8$\pm$5  }   \\ 
 \arrayrulecolor{black}\hline \arrayrulecolor{black}\hline 
\end{tabular}
\caption{ Comparison of   removal energies $\langle \epsilon \rangle ^{SF}$ from tests of the Koltun sum rule (by  Saclay\cite{Saclay}, and  Jlab Hall C \cite{C12})  to   $\langle \epsilon ^P\rangle^{levels}$ extracted from the missing
 energies of shell-model levels  measured in    
  $ee^{\prime}p$  experiments by Tokyo\cite{Tokyo1,Tokyo2,Tokyo3},
 Saclay\cite{Saclay},  Jlab Hall A\cite{O16},   Jlab Hall C \cite{C12},  and $\textsc{nikhef}$\cite{NIKHEFC12}. 
 In addition we show   $\langle \epsilon ^P\rangle^{levels}$ estimated from shell-model  binding energies.
   The value in {\bf *bold} is the best measurement for each nucleus. 
 }
\label{TMonizNew} 
\end{center}
\end{table*} 
%  TABLE 4  -------------------------------- -------------------------------- -------------------------------- --------------------------------

%--------subsection 5.1
 \subsection{Direct measurements  of  ${\langle E_m^P\rangle^{SF}}$ and ${\langle T^P \rangle^{SF}}$}
 \label{directEm}
The  best estimates of  the average missing energy ${\langle E_m^P \rangle}$ and
average nucleon kinetic energy ${\langle T^P \rangle}$  are those that are directly extracted  from  spectral function measurements in analyses that 
test the Koltun sum rule \cite{Koltun}.   The Koltun's sum rule states that
% {\bar T}
  \label{koltun}
  \begin{eqnarray}
\frac{E_B}{A} = \frac{1}{2} \, [ \,{\langle T^P\rangle^{SF} } \frac{A-2}{A-1} - {\langle E_m^P \rangle^{SF}}],
  \end{eqnarray}
where $E_B/A$ is the nuclear binding energy per particle obtained from nuclear
masses and includes a (small)  correction for the Coulomb energy,
  \begin{eqnarray}
\label{TandE}
{ \langle T^P\rangle ^{SF}  } = \int d^3k ~dE_m \ \frac{\vec k^2}{2M} P_{SF} (k,E_m)  \  , 
  \end{eqnarray}
and
  \begin{eqnarray}
{ \langle E_m\rangle ^{SF} } = \int d^3k~ dE_m \  E_m \  P_{SF} (k,E_m)  \ .
  \end{eqnarray}
For precise tests of the  Koltun sum rule a small contribution from three-nucleon processes
should  be taken into account.

Values of  $\langle E_m^P\rangle^{SF}$ and ${\langle T^P \rangle^{SF}}$ for the 1p1h  process ($E_m^P<$80 MeV)  published by  Jlab Hall C experiments\cite{C12} and by the Saclay group\cite{Saclay} are given in Table \ref{HallC}.
%	
%We refer to the  {\it peak} of the $E_m$ distribution as  $[E_m]$.
%

% FIG 6   -------------------------------- -------------------------------- -------------------------------- --------------------------------
%\begin{figure}[t]
%\begin{center}
%\includegraphics[width=3.in,height=2.in]{Adep.pdf}
%\caption{ Single ``level missing energies''   $ \langle E_m^P \rangle^{1s}$ and  $ \langle E_m^P \rangle^{1p}$ for the {\bf 1s} and {\bf 1p} states,
%respectively.  The data points are measurements done
%in $ee^\prime p$ experiments\cite{Tokyo3}.   The solid curves represent interpolations of the ``level missing energies'' observed
%in (p, 2p)  experiments\cite{p2p}. The  "level missing  energies"  for the {\bf 1s} and {\bf 1p} states measured in $ee^\prime p$ experiments are systematically %higher than those observed in (p, 2p)  experiments. 
% }
%\label{Adep}
%\end{center}
%\end{figure}
%FIG 6   -------------------------------- -------------------------------- -------------------------------- --------------------------------
%
%
%TABLE 7 ------------------------------- -------------------------------- -------------------------------- --------------------------------
%
%      subsection 5,2
\subsection{Spectral function  ``level missing energies''}
\label{E_levels}
Measured 2D spectral functions can be analyzed within the distorted plane wave approximation (DPWA)  to extract the   {\it peak}  and  {\it  width}  of the missing   energy distribution  $ E_m^P$ for protons for each shell model level. We refer to it as the ``level missing energy''.  In some publications it is  referred to as the ``shell separation energy''.  The  energies and widths of the  ``level missing energies''  for  $_3^6Li$, $_6^{12}C$, $_{3}^{17}Al$, $_{20}^{40} Ca$, $_{23}^{50}V,$  extracted  from data published by the  Tokyo group\cite{Tokyo1,Tokyo2,Tokyo3} are shown in Tables  \ref{Saclay} and \ref{Tokyo}. Also shown are  the  ``level missing energies''  for  $_6^{12}C$, $_{14}^{28}Si$, $_{20}^{40} Ca$,  and $_{28}^{58.7} Ni$, extracted  from the data published by the Saclay\cite{Saclay} group.

We obtain an estimate of the  {\it average}  missing energy $\langle E_m^P\rangle^{levels}$ for the 1p1h process by taking the average (weighted by the number of nucleons)  of the ``level missing energies''  of all shell model levels with $E_m^P<$80 MeV. The results of our analysis of the Saclay and Tokyo data are given in Tables  \ref{Saclay} and  \ref{Tokyo}.  As shown in Tables  \ref{Saclay}  and  \ref{Tokyo} for the deeply bound {\bf 1s} 
and {\bf 1p} levels  in heavy nuclei the averages and widths of the missing energy distributions are large. 
%
%      subsection 5.3
\subsection{ $_{6}^{12}C$ spectral function}
 \label{comp_spectral}
The measured\cite{huberts,NIKHEFC12} \textsc{NIKHEF} high resolution spectral function for the missing energy of a bound proton
in the 1p level of  $_{6}^{12}C$ as a function of the spectator nucleus excitation energy  $E_x^P$  for ${\vec p_m}=\vec k$ = 172 MeV/c
is shown in the top panel of Fig.  \ref{carbon-spectral-detailed}.
The Jlab measurement\cite{C12} of the  one-dimensional spectral function for the missing energy for bound proton from
 $_{6}^{12}C$ as a function of  $E_m^P$ for $Q^2$= 0.64 GeV$^2$ is shown in the bottom panel of  of Fig.  \ref{carbon-spectral-detailed}. 
The second  {\it peak}  at an average value of  $E_m^P$ $\approx$  42.6$\pm$5 MeV is for protons in the  1s level.
Combining the two results (weighted by the number of nucleons in each level)  we obtain  $\langle E_m^P \rangle^{levels}$=25.7 $\pm$2 MeV for $_{6}^{12}C$.
Additional details are given in Table \ref{Saclay}.
%or $\langle E_m^P \rangle^{levels}$=25.7 $\pm$2%
%-$[E_m^P]^{levels}$=  
%8.4$\pm$2 MeV.
%These data yield  $[\epsilon^P]^{levels}$= [$E_m^P$]+$T_{A-1}$=18.7$\pm$1 MeV,  and  $\langle \epsilon^P \rangle^{levels}$= 27.1 $\pm$2 MeV
% (where $T_{A-1}$=1.4 MeV).
%
%%  TABLE  8----------------------------- -------------------------------- -------------------------------- --------------------------------
\begin{table*}
\begin{center}
\begin{tabular}{|c|c|c||c||c|c|c||c|c|c|} \hline
 &  &     & ${\langle \epsilon ^{P,N}\rangle}$ &$\langle \epsilon ^{\prime P, N}_{SM}\rangle$ & &$\Delta S$   &$\langle E_x^{P,N} \rangle$&   & $\Delta  E_m$  \\
$_{Z}^AN$ & $\langle T^{P,N} \rangle$  & $T_{A-1}^{P,N}$ & Removal  &     $\textsc{smith-}$   &  &N-P &$\textsc{bodek-}$ &$\langle E_m^{P,N} \rangle$   &N-P  \\
&  {\it average} & {\it average} & energy  &  $\textsc{moniz}$ &        & &$\textsc{ritchie}$ &  {\it average}    &\\ \hline
&  & &      $E_m$+$T_{A-1}^{P,N}$          & $ { \epsilon ^{P,N}}$+$T$  &        & &$E_m^{P,N}$-$S^{P,N}$ &  &   \\ \hline
%&  & A-1 &use for &  $\bf{NEUT}$  &         &   &$\bf{GENIE}$  & &\\ 
&  & &use for &  $\bf{old}$  &         &   & & &  \\ 
&  & A-1  &$E_\nu^{QE-\mu}$ &     $\bf{NEUT}$  &     & &$\bf{GENIE}$      & &   \\ 
& nucleon & nucleus & $Q^2_{QE-\mu}$&     removal  &     & &excitation &&   \\ 
& $\langle KE \rangle$ & $\langle KE \rangle$ &  $Q^2_{QE-P}$&  energy   &        & &energy &  &diff\\ 
 & $T^P$,$T^N$  & P, N  &$ {\langle \epsilon ^{P}\rangle}$,$ {\langle \epsilon ^{N}\rangle}$& $\langle \epsilon ^{\prime P}_{SM}\rangle$,$\langle \epsilon ^{\prime N}_{SM}\rangle$  & $S^P$, $S^N $  &$ diff  $  &  $\langle E_x^{P} \rangle$,$\langle E_x^{N} \rangle$ &$E_m^P$, $E_m^N$   &$E_m^N$-$E_m^P$ \\ 
 \hline \hline
  $ (_1^2H) $                          & 2.5, 2.5 & 2.5, 2.5         & 4.7, 4.7          & 7.2, 7.2   & 2.2, 2.2    &  0.0    & 0.0, 0.0  & 2.2, 2.2     &0.0 \\ \hline
$_3^6Li$                               & 9.1, 9.1  &1.8, 1.8       & 18.4, 19.7   & 27.5, 28.8   & 4.4, 5.7 & 1.3  & 12.2, 12.2    &  16.6, 17.9     & (1.3) \\  \hline
%
  %  ${_6^{12}C}$                      & 15.5, 15.5  & 1.4, 1.4&     26.8, 29.4  &42.3, 44.9 &  16.0, 18.7& 2.7& 9.4, 9.3   &  25.4, 28.0     &2.6 \\  \hline
        ${_6^{12}C}$                      & 15.5, 15.5  & 1.4, 1.4&     27.5, 30.1  &43.0, 45.6 &  16.0, 18.7& 2.7& 10.1, 10.0   &  26.1, 28.7     &2.6 \\  \hline
${_8^{16}O}$    & 16.0, 16.0 &   1.1, 1.1    &               24.1, 27.0  &  40.1, 43.0 & 12.1, 15.7 & 3.6 & 10.9, 10.2   &  23.0, 25.9   &2.9 \\  \hline
%
%$_{12}^{24}Mg$         &  17.5, 17.5&   0.8, 0.8      & 27.0, 31.8  &   44.5, 49.3  &11.7, 16.5          &4.8   & 14.5, 14.5   &  26.2, 31.0    & (4.8) \\   \hline
%
$ _{13}^{27}Al$         & 17.9, 18.4  &     0.7, 0.7      & 30.6, 35.4          &  48.5, 53.3                  & 8.3, 13.1        &  4.8       &    21.6, 21.6       &  29.9, 34.7     & (4.8) \\  \hline
%$_{14}^{28}Si$         & 18.1, 18.4   &    0.7,  0.7    &  28.2,  33.8             &   46.3,  51.9               &  11.6, 17.2       &     5.6    &    15.9, 15.9       & 27.5, 33.1      &\\ \hline
$_{14}^{28}Si$         & 18.1, 18.4   &    0.7,  0.7    &  24.7,  30.3             &   42.8,  48.4               &  11.6, 17.2       &     5.6    &    12.4, 12.4       & 24.0, 29.6      & (5.6) \\ \hline
${_{18}^{40}Ar}$    &  19.9, 21.9 &    0.5, 0.6     &   30.9,  32.3  &  50.8,  52.2   & 12.5, 9.9 & -2.6 &   17.8, 21.8 &   30.2, 31.7    &1.4  \\ \hline
%
%${_{20}^{40}Ca}$               &  19.9,19.9  &    0.5, 0.5      &  26.2,  33.9    & 46.1, 53.8  & 8.3,  15.6 & 7.3 &   19.3, 19.7    & 27.6, 35.3    &   7.7 \\    \hline
${_{20}^{40}Ca}$               &  19.9,19.9  &    0.5, 0.5      &  28.2,  35.9    & 48.1, 55.8  & 8.3,  15.6 & 7.3 &   19.4, 19.8    & 27.7, 35.4    &   7.7 \\    \hline
$_{23}^{50}V$         & 20.2, 22.4    &    0.4, 0.5               & 25.6,  28.6            & 45.8,   48.8               & 8.1, 11.1        &  3.0       &   17.0, 17.0       &   25.1, 28.1    & (3.0) \\  \hline
${_{26}^{56} Fe}$          &  20.4, 22.6    &  0.4, 0.4        &29.6,  30.6    & 50.0, 51.0 & 10.2, 11.2& 1.0 & 19.0, 19.0&  29.2, 30.2    & (1.0) \\ \hline 
 $_{28}^{58.7} Ni$&  20.9, 22.8&      0.4, 0.4     &     25.4,  29.4  &46.3, 50.3  & 8.2, 12.2& 4.0 &  16.8, 16.8&    25.0, 29.0  & (4.0)        \\ \hline
%$_{39}^{89}Y$         &  18.7, 21.9 &    0.2, 0.3      &     31.0, 35.4   & 49.7,  54.1  &  7.1, 11.5 &4.4   &23.6, 23.6  &  30.7, 35.1   &    (4,4)   \\ \hline
$_{40}^{88}Zr$         &   &         &        &    &  8.4, 12.0 & 3.6 &    &    &  1.9    \\ \hline
%
%$_{50}^{118.7}Sn$       &  18.9, 23.1 &   0.2, 0.2          &  32.0, 30.4    & 50.9, 49.3   &  10.1, 8.5 &-1.6  & 21.7, 21.7 &  31.8, 30.2     &  (-1.6)      \\ \hline
%$_{73}^{181}Ta$ & 18.5, 23.2  &          0.1, 0.1        & 29.3, 31.0    & 47.8,   49.5   &   5.9, 7.6&1.7   & 23.3, 23.3 &   29.2, 39.9 & (1.7)      \\ \hline
${_{79}^{197}Au}$     & 23.9, 30.4    &    0.1, 0.1        & 25.4, 27.7     & 49.3,  57.6   & 5.8, 8.1 &2.3   & 19.5, 19.5  & 25.3, 27.6   &  (2.3)    \\ \hline
$ _{82}^{208}Pb$       & 23.9, 30.4    &   0.1, 0.1      & 22.8, 25.0     &    46.7, 55.4  &  8.0,  7.4&-0.6 & 14.7, 16.9    & 22.7, 24.9  & 2.2   \\ 
 \hline \hline   %%%%  {\it average} 
 % spreadsheet  DeltaR-average sheet 1 \rightarrow$ 
\end{tabular}
\caption{ Summary of the  parameters that enter into the extractions of  excitation $\langle E_m^{P,N} \rangle$,  removal energies $\langle \epsilon ^{P,N} \rangle$, and the Smith-Moniz removal energy  $\langle \epsilon ^{\prime N}_{SM}\rangle$.  All values are in MeV.}
\label{TMoniz2} 
\end{center}
\end{table*}
% %  end ABLE 8  ------------------------------ -------------------------------- -------------------------------- -------------------------------
%
%             subsection 5.4
\subsection{Comparison of the two methods}
\label{Comparison-three}
 Tests of the Koltun sum rule as a function of $Q^2$ were done by  $ee^\prime p$ experiments at Jlab Hall C\cite{C12} 
for $_6^{12}C$, $_{26}^{56}Fe$, and $_{79}^{197}Au$.  
Tests of the Koltun sum rule were also reported by  the Saclay\cite{Saclay} group for $_6^{12}C$, $_{14}^{28}Si$, $_{20}^{40}Ca$,  and $_{28}^{59}Ni$.
For both groups  values of   $\langle E_m^P\rangle^{SF}$ and $\langle T^P \rangle^{SF}$ were extracted from
the measured spectral functions. The results  from both groups are summarized in  Table \ref{HallC}. 
We take the  RMS variation with $Q^2$  of the Jefferson Lab Hall C data shown in Table \ref{HallC}  ($\approx$ 0.5 MeV)  as the random error in the Jlab Hall C measurements of  $\langle E_m^P\rangle^{SF}$.  

We use  the  2.7 MeV difference in the  measured values of  $\langle E_m^P\rangle^{SF}$ for  $_6^{12}C$ at Jefferson Lab  (26.1$\pm$0.4)  and Saclay  (23.4$\pm$0.5)  as the systematic error in measurements of   $\langle E_m^P\rangle^{SF}$.  Since  $\langle E_m^P\rangle^{SF}$  is the  most reliable measurement of  $\langle E_m^P\rangle$,   {\it we assign $\pm$3  MeV as the systematic uncertainty to all measurements of $\langle E_m^P\rangle$}. 

The  {\it average}  values of the removal energies $\langle \epsilon^P\rangle^{SF}$ = $\langle E_m^P \rangle^{SF}$ +$T_{A-1}$ versus atomic number from tests of the Koltun sum rule  are in agreement with the  {\it average}  values extracted from measurements of  ``level missing energies'' $\langle \epsilon^P\rangle^{levels}$= $\langle E_m^P \rangle^{levels}$ +$T_{A-1}$ as shown Fig. \ref{average-peak}.  
%The measurements of $\langle E_m^P\rangle^{SF}$ are consistent with the measurements of  $\langle E_m^P \rangle^{levels}$.
For example, for the Saclay data  shown in Table \ref{HallC}  the average of the difference  between  $\langle E_m^P \rangle^{levels}$ and  $\langle E_m^P \rangle^{SF}$  for $_6^{12}C$, $_{14}^{28}Si$, $_{20}^{40}Ca$,  and $_{28}^{59}Ni$ is 0.9$\pm$1.0 MeV.    
%
%
%    subsection 5.5
\subsection {Spectral function measurement of $_{8}^{16}O$}
\begin{figure}
\begin{center}
\includegraphics[width=3.5in,height=2in]{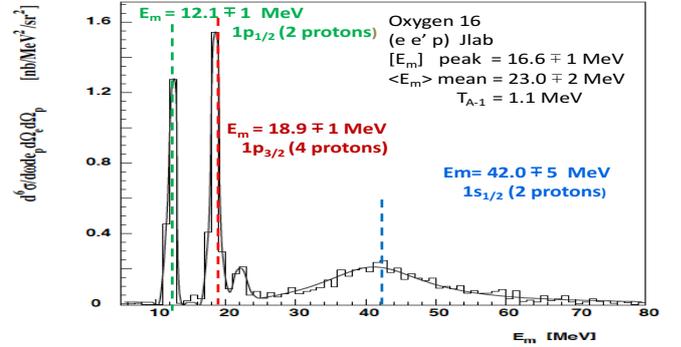}
\caption{ The spectral function for   $_{8}^{16}O$ as a function of  $E_m^P$ for ${\vec p_m}=\vec k$ = 60 MeV/c  measured in Jlab Hall A\cite{O16} with 2.4 GeV incident electrons.
%  Bottom:
%The measured  (Frascati \cite{other})  energy loss spectra ($\nu=E_0-E^{\prime}$)   at $32^0$ for  0.7 GeV  incident electrons on $_{8}^{16}O$ (in squares)   and $_{18}^{40}Ar$ (in X's], respectively.  The  {\it peaks} of the QE distributions for $_8^{16}$O and $_{18}^{40}Ar$ are  the same value of  ($\nu_{ {\it peak}}\approx$  105 MeV).  
 }
\label{oxygen-spectral}
\end{center}
\end{figure}

The spectral function for
 $_{8}^{16}O$ as a function of  $E_m^P$ for ${\vec p_m}=\vec k$ = 60 MeV/c  measured in Jlab Hall A\cite{O16} with 2.4 GeV incident electrons
 is shown in the top panel of Fig. \ref{oxygen-spectral}. 
  From this figure we extract a  {\it average}   $\langle E_m^P\rangle^{levels}$ = 23.0$\pm$2 MeV.
  % and a   {\it peak}  (average of the nucleons in the 1p level only)    $[E_m^P]^{levels}$ = 16.6$\pm$1 MeV. 
    Using  $T_{A-1}$= 1.1 MeV for $_{8}^{16}O$ we also obtain a   {\it average}   $\langle \epsilon ^P \rangle^{levels}$ = 24.1$\pm$2 MeV.
    % and a  {\it peak}  $[\epsilon ^P]^{levels}$ = 17.7$\pm$1 MeV for $_{8}^{16}O$.
%
%%FIG 7   -------------------------------- -------------------------------- -------------------------------- --------------------------------
%  %FIG 7   -------------------------------- -------------------------------- -------------------------------- --------------------------------
%
%         subsection 5.6
\subsection{Difference between neutrons and protons for $^{12}_{8}C$ and $^{16}_{8}O$}
For nuclei which have the same number of neutrons and protons
we expect that the average excitation energy $ E_x^{P,N}$ spectrum for protons and neutrons to be approximately the same 
($ \langle E_x^{P,N}\rangle \approx \langle E_x^{P,N}\rangle$).
Since  $E_m^{P,N}$=$S^{P,N}$+$E_x^{P,N}$
the difference in the average missing energies for neutrons and protons is approximately equal to the difference in separation energies $S^N-S^P$.
By definition, the single particle binding energy of the least bound state is equal to the separation energy.
The differences in the separation energies between neutrons and protons (S$^N$-S$^P$)   bound in
 $^{12}_{8}C$ and $^{16}_{8}O$ are  of 2.7 MeV and 3.6 MeV, respectively.

More generally, a better estimate of the difference between the average missing energies for neutrons
and protons can be obtained from the nuclear shell model.   The single nucleon missing energy $(E_m^{P,N})^{shell{\text-}model{\text-}level} $ for a nucleon
in a given shell-model level is close (somewhat larger)  to the single nucleon binding energy for that level. 
Consequently,  the difference in the average missing energies for neutrons and protons
for a nucleus  $\langle E_m^{P}\rangle$-$\langle E_m^{N}\rangle$  is also approximately equal to the difference in the  average binding energies. 

The binding energies of   different  shell-model levels\cite{spectral-theory} for $^{12}_{8}C$ and $^{16}_{8}O$ are shown in Table \ref{Tcarbon}. 
 When available, the experimental values shown in   {\it italics} are used. 
 % We set the binding energy for the least bound level  to be equal to the separation energies $S^P$ or $S^N$ for bound protons or neutrons, respectively.  The single particle binding energies are assumed to be close to the the missing energies $E_m$ for each level. 
 The differences between  the   {\it averages} of the 
 %  missing energies   for neutrons and protons $\langle E_m^{P,N} \rangle$ can be approximated by the difference in 
 %  the weighted average
 nucleon binding energies in all shell-model levels for neutrons and protons is  2.6 and 2.9 MeV for $^{12}_{8}C$ and $^{16}_{8}O$, respectively. 
 As expected these values are similar (within 1 MeV)  to  the differences in the separation energies for neutrons and protons (S$^N$-S$^P$)   bound in
 $^{12}_{8}C$ and $^{16}_{8}O$ of  of 2.7 and 3.6 MeV, respectively. 
 %
 % The   {\it peak} missing  energies   $[E_m^{P,N}]$  for neutrons and protons bound carbon and $^{16}_{8}O$ can be approximated by the corresponding truncated averages  (not including the deeply bound  {\bf 1s} level), as shown in Table \ref{Tcarbon}.
%\begin{figure}
%\begin{center}
%\includegraphics[width=2.4in,height=1.4in]{photon-proton.pdf}
%\vspace{-0.1in}
%\caption{ Scattering from an off-shell bound nucleon of momentum $ {k}$ which is perpendicular to the direction of the virtual photon. 
%This is the configuration at the   {\it peak} of the Fermi  motion smearing.   At the   {\it peak} of the distribution the $z$ component of the nucleon momentum ($k_z$)  is zero.}
%\label{photon-proton}
%\end{center}
%\end{figure}
%\vspace{-0.2in}
%
  %----------------------section  6---------- -------------------------------- -------------------------------- --------------------------------
\section {Inclusive e-A electron scattering}
\label{C12-Ufsi}
For  QE $electron$ scattering at low  $(\vec k+\vec q_3)^2$ we use an empirical parameter
$U_{FSI}((\vec k+\vec q_3)^2)$ to account for the effect of final state interactions.
The off-shell Bodek-Ritchie formalism (used by $\textsc{genie}$) for  the case of  QE $electron$ scattering from a bound $proton$,
should be implemented as follows:
 \begin{eqnarray}
\nu &+&(M_P-\epsilon^P) =   \sqrt{{\bf ( k + \vec q_3)^2} + M_P^2} -|U_{FSI}| +|V_{eff}^P|\nonumber \\
%&=& \sqrt{{\bf p_f}^2+M_P^2}\nonumber\\
\epsilon^P&=& S^P+\langle E_x^P \rangle + \frac { \langle \vec k^2 \rangle}{2M_{A-1}^*}  \nonumber\\ 
\nu& +&(M_P-x^{P}) =   \sqrt{{\bf (k + \vec q_3)^2} + M_P^2} \nonumber \\
x^P&=& S^P+\langle E_x^P \rangle+\frac { \langle k^2\rangle}{2M_{A-1}} -|U_{FSI}|+ |V_{eff}^P|  \nonumber \\
&&x^N= S^N+\langle E_x^N \rangle+\frac { \langle k^2\rangle}{2M_{A-1}} -|U_{FSI}| \nonumber \\
Q^2 &=& 4 (E_0+|V_{eff}|)(E^\prime+|V_{eff})|\sin^2(\theta/2) \nonumber\\
E^\prime &=& E_0 - \nu ,~~~~~E_f^P = \nu-\epsilon^P, ~~~~\vec q_3^2=Q^2+\nu^2,
 \end{eqnarray}
 and    $U_{FSI}=U_{FSI} ((\vec q_3+\vec k)^2)$. 
 For  $electron$ scattering from a bound $proton$  $|V_{eff}^P|=\frac{Z-1}{Z}|V_{eff}|$, where (Z-1) is the number
 of protons in the spectator final state nucleus.
% \begin{enumerate}
%  \item $\vec k$ is the initial  3-momentum of proton.
% \item $\vec q_3$ is the 3-momentum transfer to the nucleon.
% \item $\nu$ is the energy transfer to the nucleon.
%\item $S^P$ is the separation energy (energy it takes to remove a proton from nucleus A, leaving A-1 nucleus in the ground state).
%\item $\langle E_x \rangle$ is the average excitation energy of the  $(A-1)^*$ spectator nucleus.
%\item $ \frac {\vec k^2}{2M_{A-1}^*}$ is the kinetic energy of the recoil $(A-1)^*$ spectator nucleus.
%\item $U_{FSI}((\vec q_3+\vec k)^2)$is negative corresponding to the kinetic energy loss of the final state proton in the optical potential of the spectator $(A-1)^*$ nucleus.
%\item  $V_{eff}^P=\frac {z-1}{z} |V_{eff}|$ is positive and corresponds to gain in the kinetic energy of the final state proton on the way out as it is repelled by the Coulomb field of the $(A-1)^*$ spectator nucleus. 
%\end{enumerate}
 
 %----  subsection  6.1
 \subsection{Smith-Moniz on-shell formalism }
For QE $electron$ scattering on a bound proton in the Smith-Moniz on-shell formalism (used by $\textsc{old-neut}$) 
the following equations should be used:
  \begin{eqnarray}
 \nu &+&M_P+ T^P-\epsilon_{SM}^{\prime P} = \\ 
 && \sqrt{{\vec ( k + \vec q_3)^2} + M_P^2}  -|U_{FSI}| +|V_{eff}^P| \nonumber \\
T^P&=& \sqrt{\vec k^2+M^2_P}~~~~~~\epsilon^P= \epsilon_{SM}^{\prime P}- \langle T^P \rangle  \nonumber \\ 
\nu &+&[(M_P+T^P)-x^{SM})]=   \sqrt{{\bf (k + \vec q_3)^2} + M_P^2}\nonumber  \\
x^{SM}&=& \epsilon_{SM}^{\prime P}  - |U_{FSI}| +|V_{eff}^P| \nonumber \\
%&\approx&  \epsilon^P - |U_{FSI}({\vec q_3^2})| +|V_{eff}^P| \nonumber \\
Q^2 &=& 4 (E_0+|V_{eff}|)(E^\prime+|V_{eff}|)\sin^2(\theta/2) \nonumber \\
E^\prime &=& E_0 - \nu ,~~~~ E_f^P = \nu- \epsilon^P,~~~~~\vec q_3^2=Q^2+\nu^2,  \nonumber
 \end{eqnarray}
and  $U_{FSI}=U_{FSI} ((\vec q_3+\vec k)^2)$.
%\nu +(M_P-x) &=&   \sqrt{{\bf ( k + \vec q_3)^2} + M_P^2}\\
%x&=& S^P+E_x+\frac {k^2}{2M_{A-1}} -U_{FSI}((\vec q_3+\vec k)^2) +V_{eff}^P  \nonumber
%
%     subsection 6.2
%
\subsection{Extraction of $U_{FSI}$ from in inclusive e-A QE data} 
  We define  $k_z$ as the component of k along the direction of the  of the 3-momentum
transfer $\vec q_3$.
 \begin{eqnarray}
  \nu + (M_P-\epsilon^P) &=&
                           \sqrt{ \langle  \vec k^2(k_z) \rangle+ 2k_z {\vec q_3} +{\vec{q_3^2}}+M_P^2}\\
   &-&   
      |U_{FSI} ( \langle  \vec k^2(k_z) \rangle+ 2k_z {\vec q_3} +{\vec{q_3^2}})|
    +|V_{eff}^P|  \nonumber\\ 
    {\vec {q_3^2}}&=&Q^2+\nu^2 \nonumber \nonumber \\ 
  Q^2&=&4(E_0+|V_{eff})(E_0-\nu+|V_{eff}|)\sin^2\frac{\theta}{2} \nonumber
   \end{eqnarray}
 where  in the calculation of ${\vec q_3^2}$  we have applied Coulomb corrections to the initial and final electron energies
   as described in  Appendix \ref{Coulomb}.
 
In the peak region of the QE distribution  $ k_z \approx 0$.  Therefore, from the location of the peak in $\nu$ we extract
  $U_{FSI}( (\vec q_3+\vec k)^2)_{peak}$ for 
  \begin{eqnarray}
  (\vec q_3+\vec k)^2_{peak} &\approx& \langle  \vec k^2(k_z=0) \rangle + \vec{q_3^2} \\
  & =& \frac{1}{2}k_F^2 +  \vec{q_3^2}\nonumber\\
  & \approx&  0.02 ~GeV^2+ \vec{q_3^2}~~~~(for K_F=0.2)\nonumber
   \end{eqnarray}
where we have used equation \ref{kq32peak} for the Fermi gas distribution. If 
simplicity is needed
then $( \vec q_3+\vec k)^2 \approx   \vec q_3^2$ is a good approximation.

We fit  a large number of  electron scattering QE differential cross sections for various nuclei  and extract the values of   $U_{FSI}( (\vec q_3+\vec k)^2_{peak})$.  The data samples  include:   four  $_{3}^{6}$Li spectra,     33  $_{6}^{12}$C spectra, five $_{8}^{16}$O spectra,    seven  $_{18}^{27}$Al spectra,    29  $_{20}^{40}$Ca spectra, two $_{18}^{40}$Ar spectra,   30 $_{26}^{56}$Fe  spectra,   23 $_{82}^{208}$Pb  spectra and one  $_{79}^{197}$Au) spectra.  Most (but not all)  of the QE differential cross sections  given in references \cite{Heimlich:1973} to \cite{Zghiche:1993xg})
are available on the  QE electron scattering archive\cite{archive}.  
 Figures   \ref{Li6_fits}, \ref{C12_fits}, \ref{O16_fits}, \ref{Al27_fits}, \ref{Ca40_fits}, \ref{Ar40_fits},\ref{Fe56_fits}, \ref{Pb208_fits}    show examples of these fits to QE  differential cross sections for all these elements. The solid blue curve is the RFG fit with the best value of $U_{FSI}$. The black dashed curve is a simple parabolic fit used to estimate the systematic error.  The red dashed curve is the RFG model  with  $U_{FSI}=V_{eff}=0$.

\begin{figure*}
\centering
 \includegraphics[width=17.cm,height=7.cm]{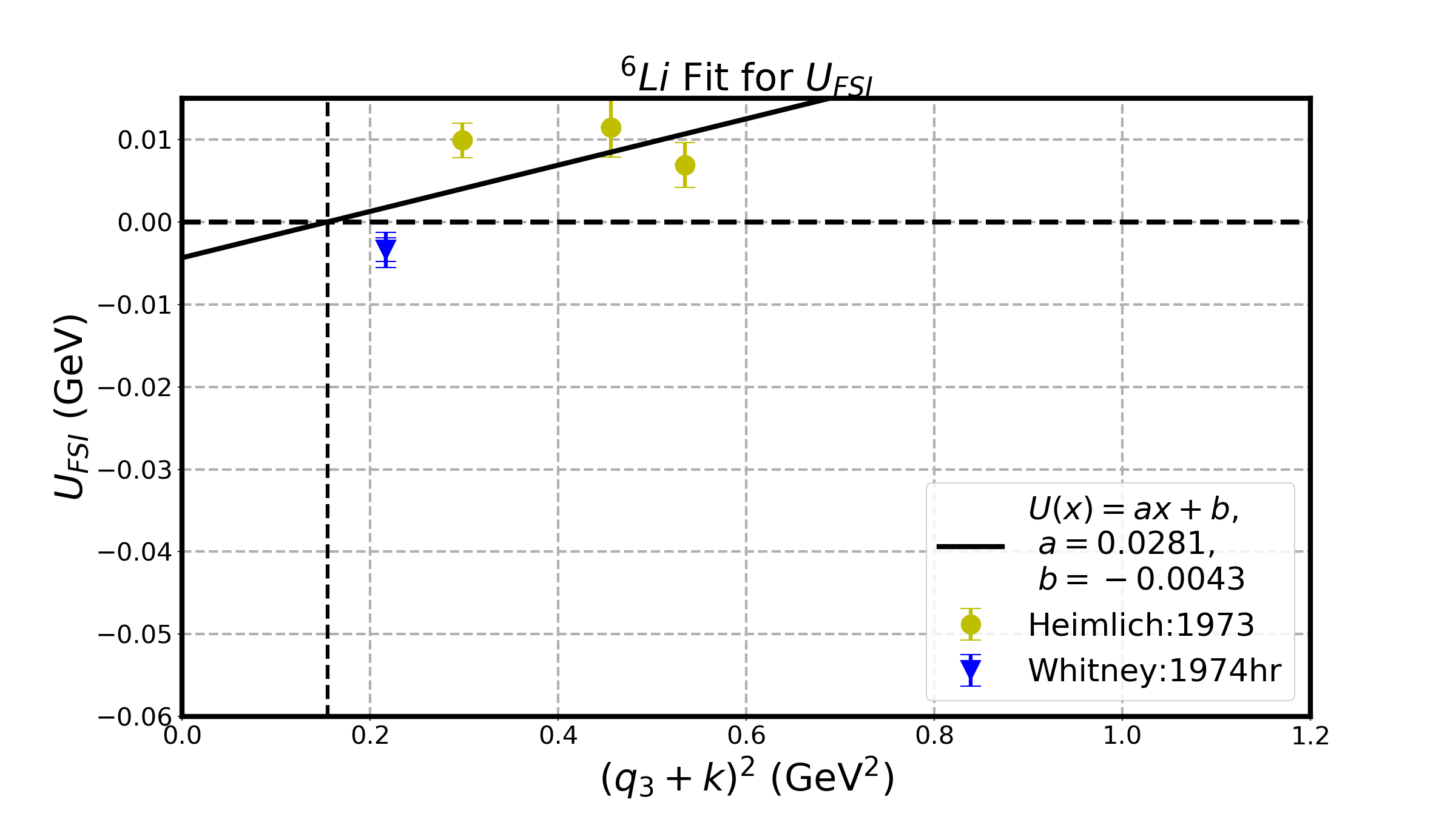}
\caption{
 Extracted values of $U_{FSI}$ versus $(\vec q_3+\vec k)^2$ for four Lithium ($_{3}^{6}$Li) spectra .}
  \label{Li6vskq32}
\end{figure*}

\begin{figure*}
\centering
  \includegraphics[width=17.cm,height=7.cm]{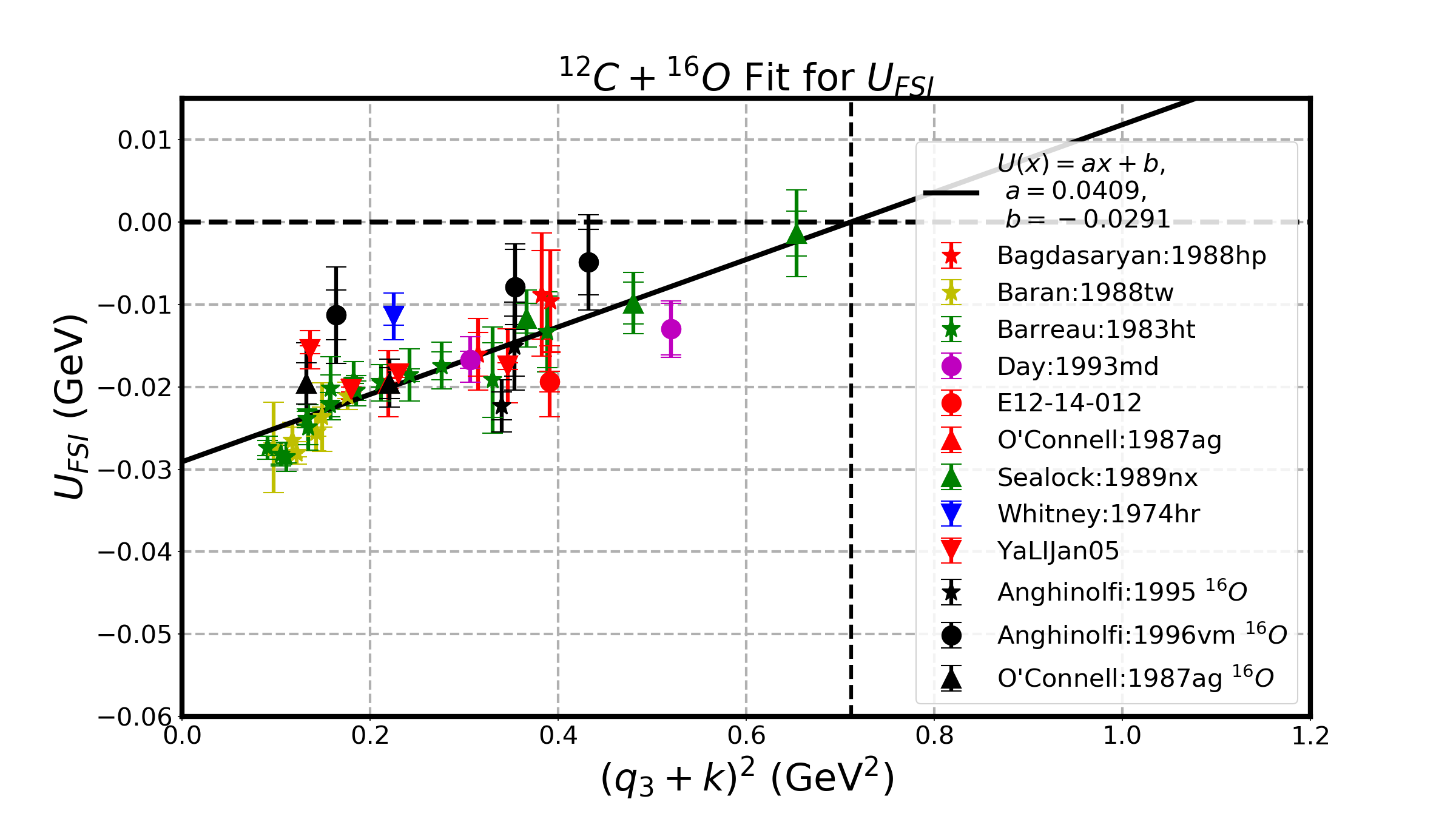}
\caption{
Extracted values of $U_{FSI}$ versus $(\vec q_3+\vec k)^2$ for 33  Carbon ($_{6}^{12}$C) and 8  Oxygen  ($_{8}^{16}$O) spectra. }
  \label{C12vskq32}
\end{figure*}

\begin{figure*}
    %[ht]
\centering
       \includegraphics[width=17.cm,height=7.cm]{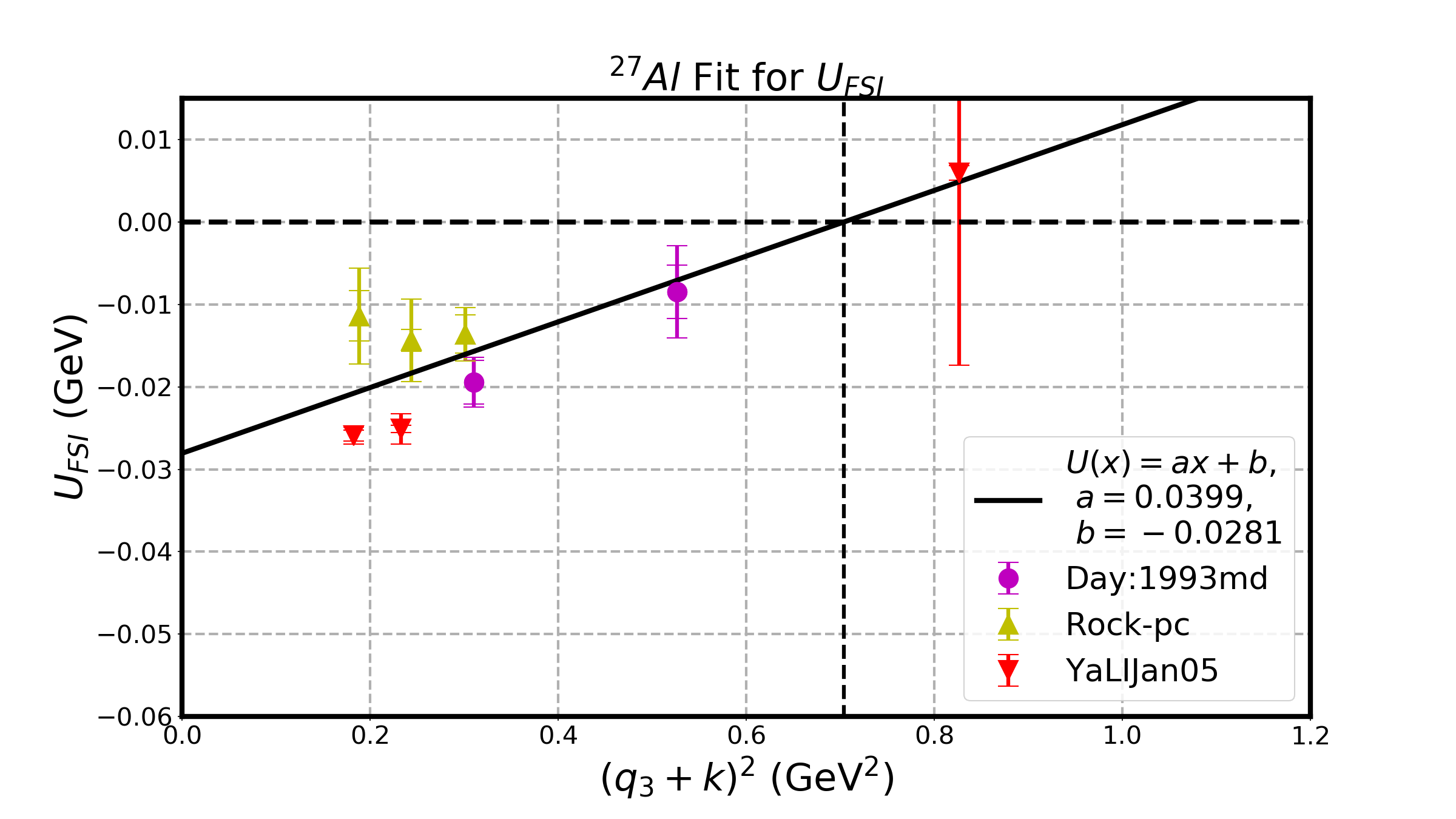}

\caption{
%\footnotesize\addtolength{\baselineskip}{-1\baselineskip} 
Extracted values of $U_{FSI}$ versus $(\vec q_3+\vec k)^2$ for 8 Aluminum ($_{13}^{27}$Al) spectra.}
   \label{Al27vskq32}
\end{figure*}

\begin{figure*}
\centering
       \includegraphics[width=17.cm,height=7.cm]{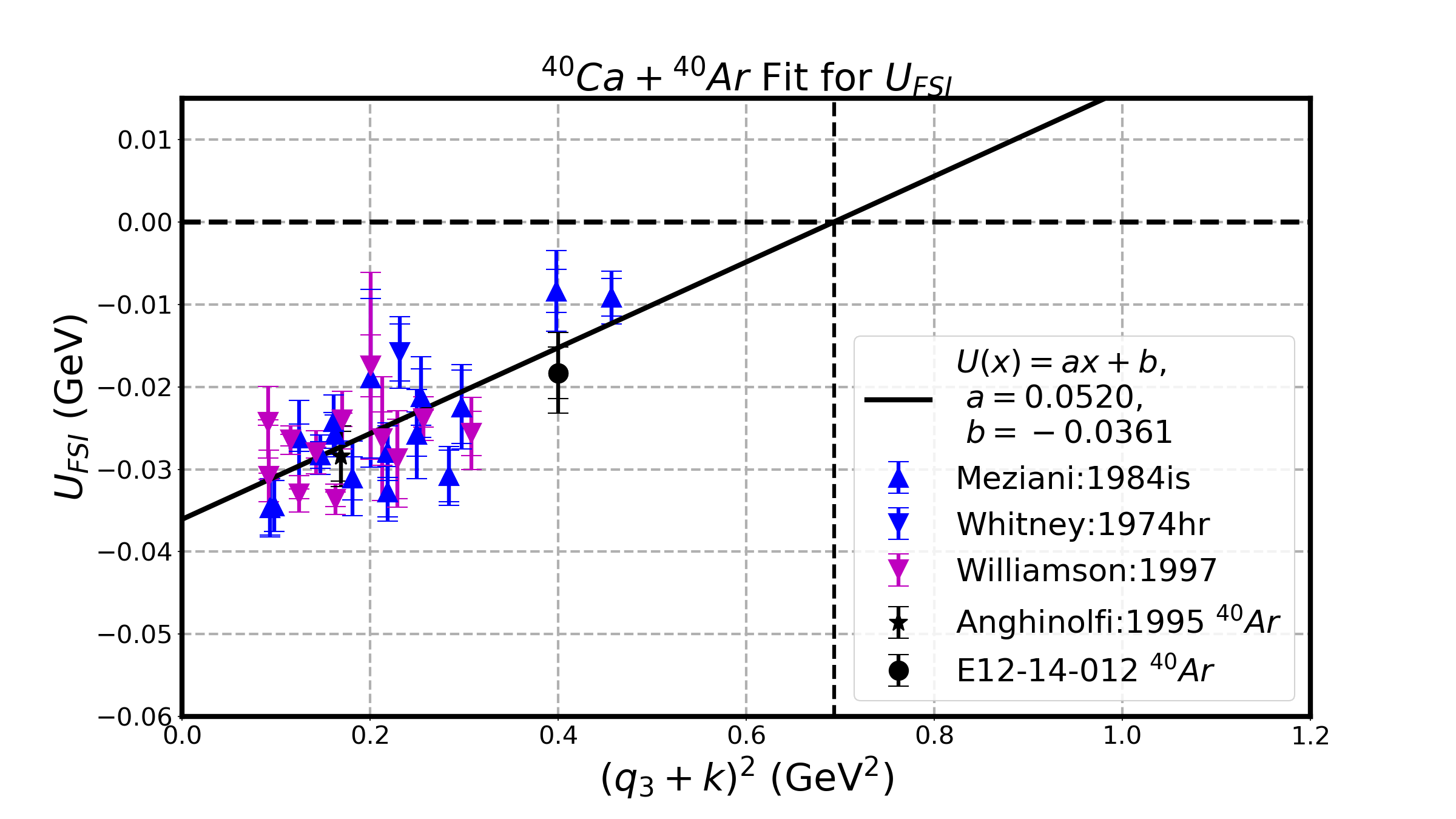}
\caption{
%\footnotesize\addtolength{\baselineskip}{-1\baselineskip} 
Extracted values of $U_{FSI}$ versus $(\vec q_3+\vec k)^2$ for 29 Calcium ($_{20}^{40}$Ca) and 2 Argon ( ($_{18}^{40}$Ar) spectra}
   \label{Ca40vskq32}
\end{figure*}

\begin{figure*} 
\centering
       \includegraphics[width=17.cm,height=7.cm]{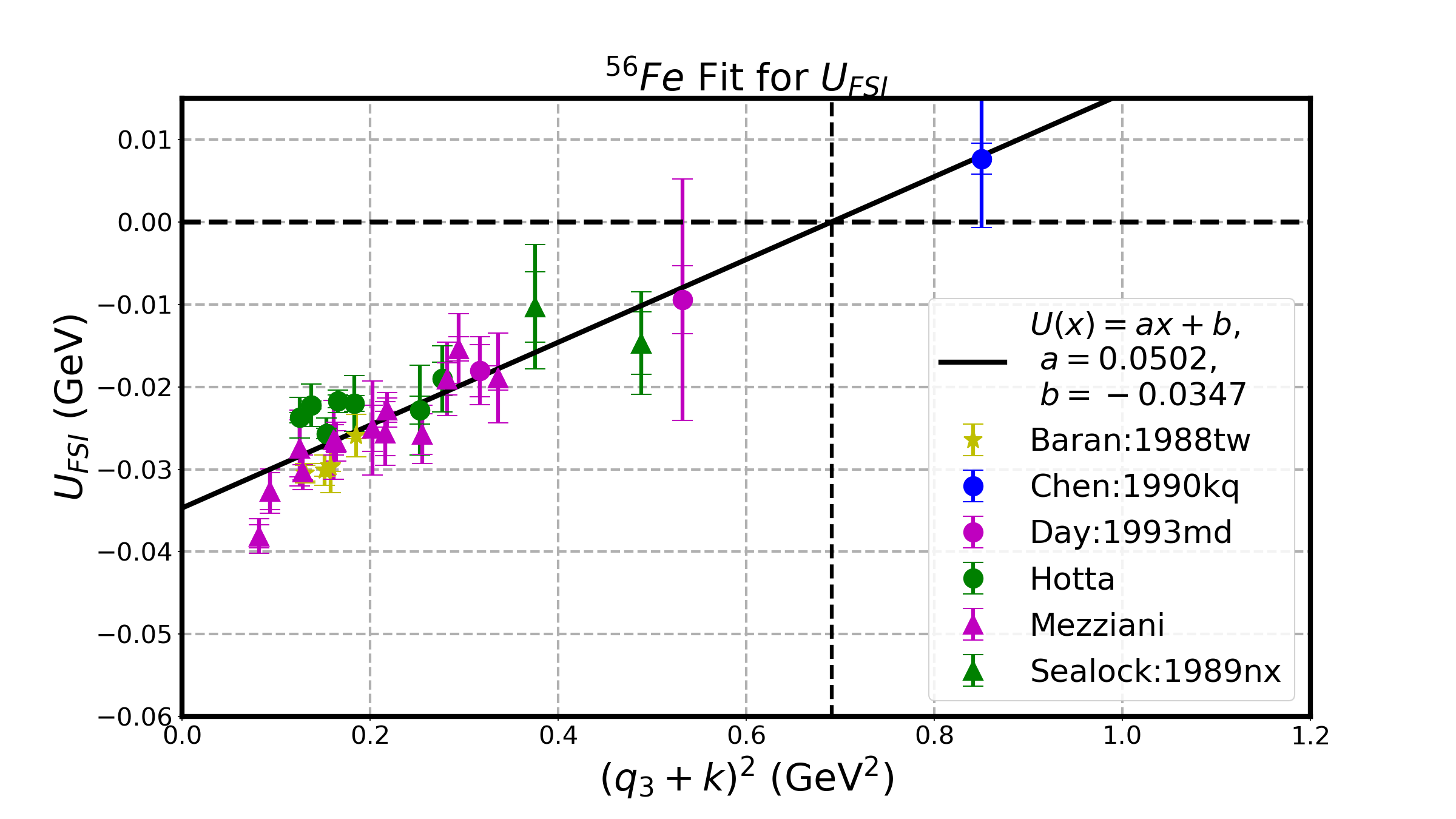}
\caption{ Extracted values of $U_{FSI}$ versus $(\vec q_3+\vec k)^2$ for 30  Iron ($_{26}^{56}$Fe) spectra.}
 \label{Fe56vskq32}
\end{figure*}

\begin{figure*}
    %[ht]
\centering
       \includegraphics[width=17.cm,height=7.cm]{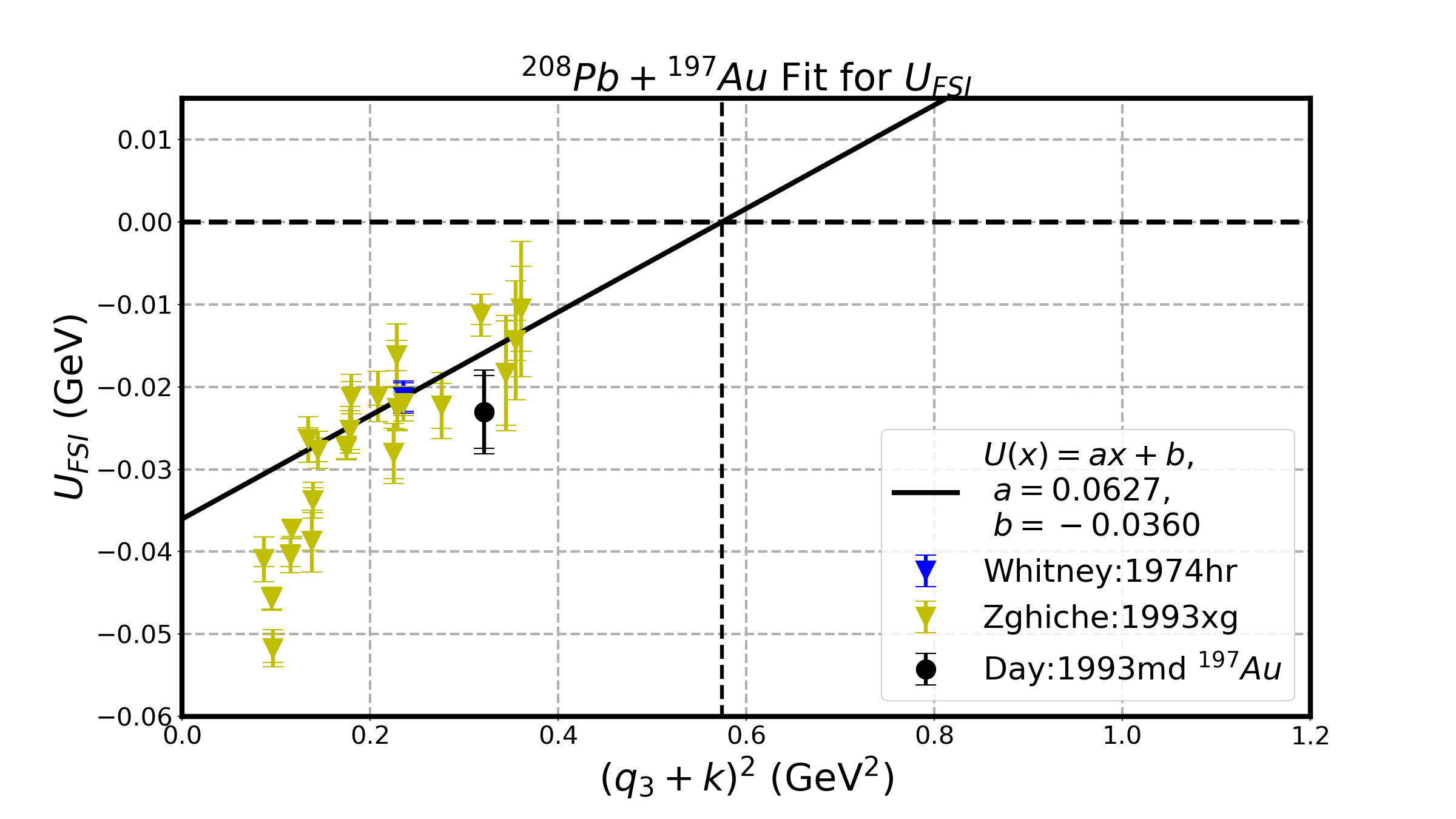}
\caption{
Extracted values of $U_{FSI}$ versus $(\vec q_3+\vec k)^2$ 
 for 22 Lead ($_{82}^{208}$Pb) and one Gold ($_{79}^{197}$Au) spectra.}
   \label{Pb208vskq32}
\end{figure*}

 The extracted  values of  $U_{FSI}$ versus $(\vec q_3+\vec k)^2$ for Lithium, Carbon+Oxygen, Aluminum, Calcium +Argon, iron, and Lead+Gold are shown in Figures \ref{Li6vskq32},  \ref{C12vskq32},   \ref{Al27vskq32},  \ref{Ca40vskq32},  \ref{Fe56vskq32},  and  \ref{Pb208vskq32}, respectively.
  Here $(\vec q_3+\vec k)^2$ is evaluated at the peak of the QE distribution.
 We fit the extracted values of $U_{FSI}( (\vec q_3+\vec k)^2)$ versus $(\vec q_3+\vec k)^2$ 
 for $(\vec q_3+\vec k)^2 >0.1~GeV^2$ to a linear function. The intercepts  at
$(\vec q_3+\vec k)^2=0$ and the slopes of $U_{FSI}((\vec q_3+\vec k)^2)$ are given in Table \ref{U-FSI} for various nuclei.

 For the Relativistic Fermi Gas  ($\textsc{rfg}$) the probability distribution $P_{rfg}(k_z)$ and the average $\langle \vec k^2(k_z) \rangle_{rfg}$ 
    are given in Appendix \ref{Arfg}.  
  We compare the e-A QE  cross sections versus $\nu$ to  the $\textsc{rfg}$ model for QE scattering.
 We account for the nucleon $Q^2$ dependent form factors and for Pauli suppression (discussed in Appendix B.2)  at low $\vec q_3^2$.  
 
\begin{figure*}
    %[ht]
\centering
\includegraphics[width=5.5cm,height=4.5cm]{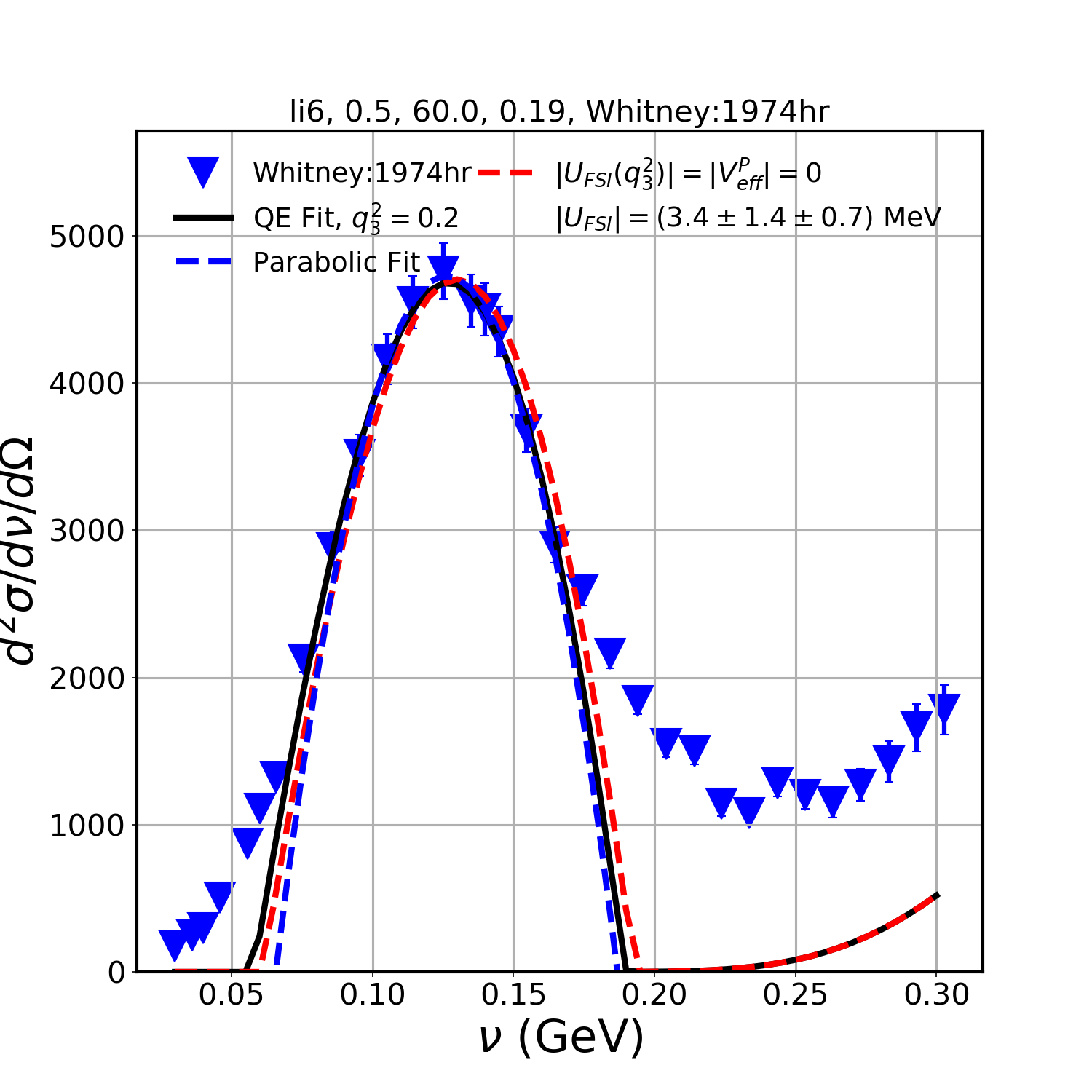}
\includegraphics[width=5.5cm,height=4.5cm]{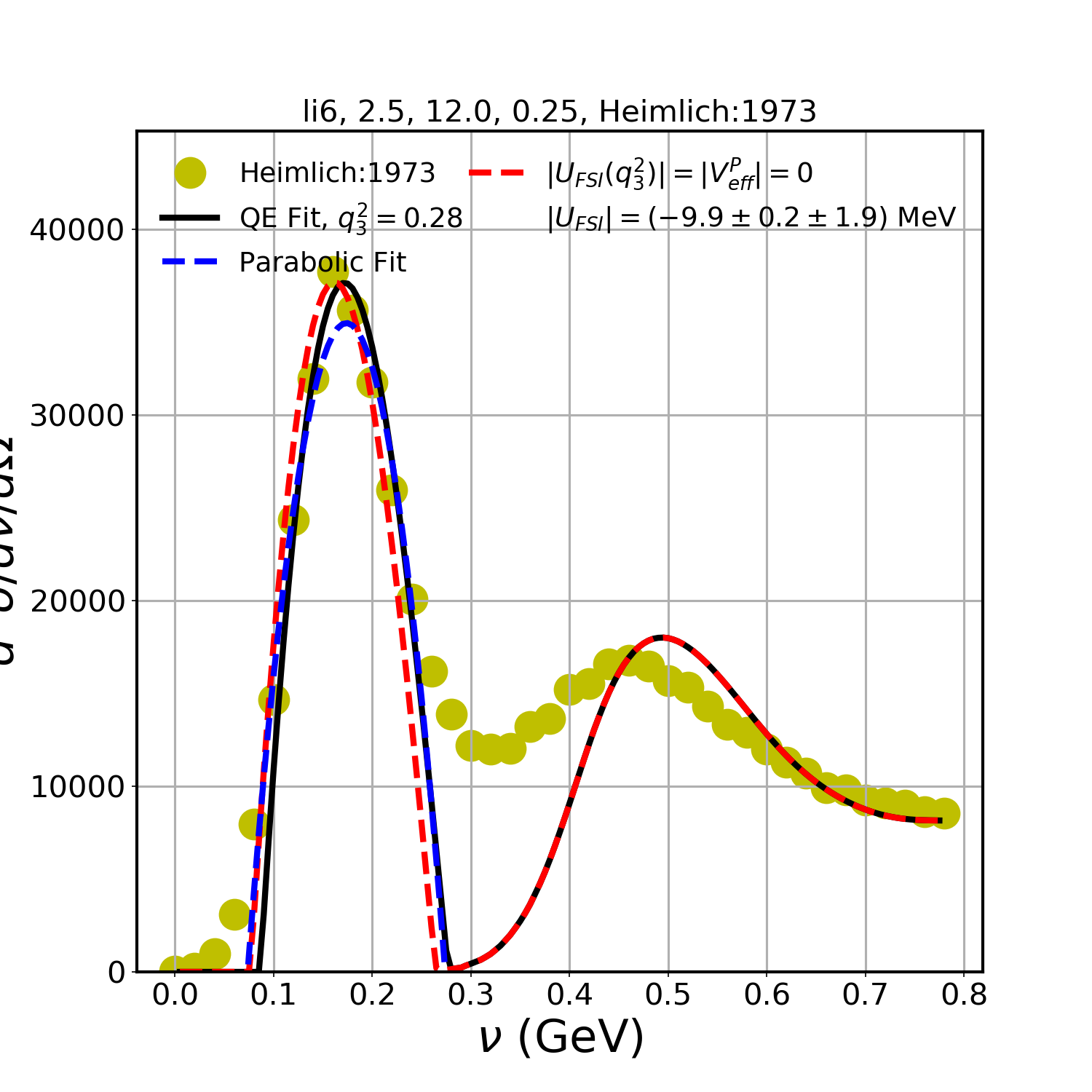}
\includegraphics[width=5.5cm,height=4.5cm]{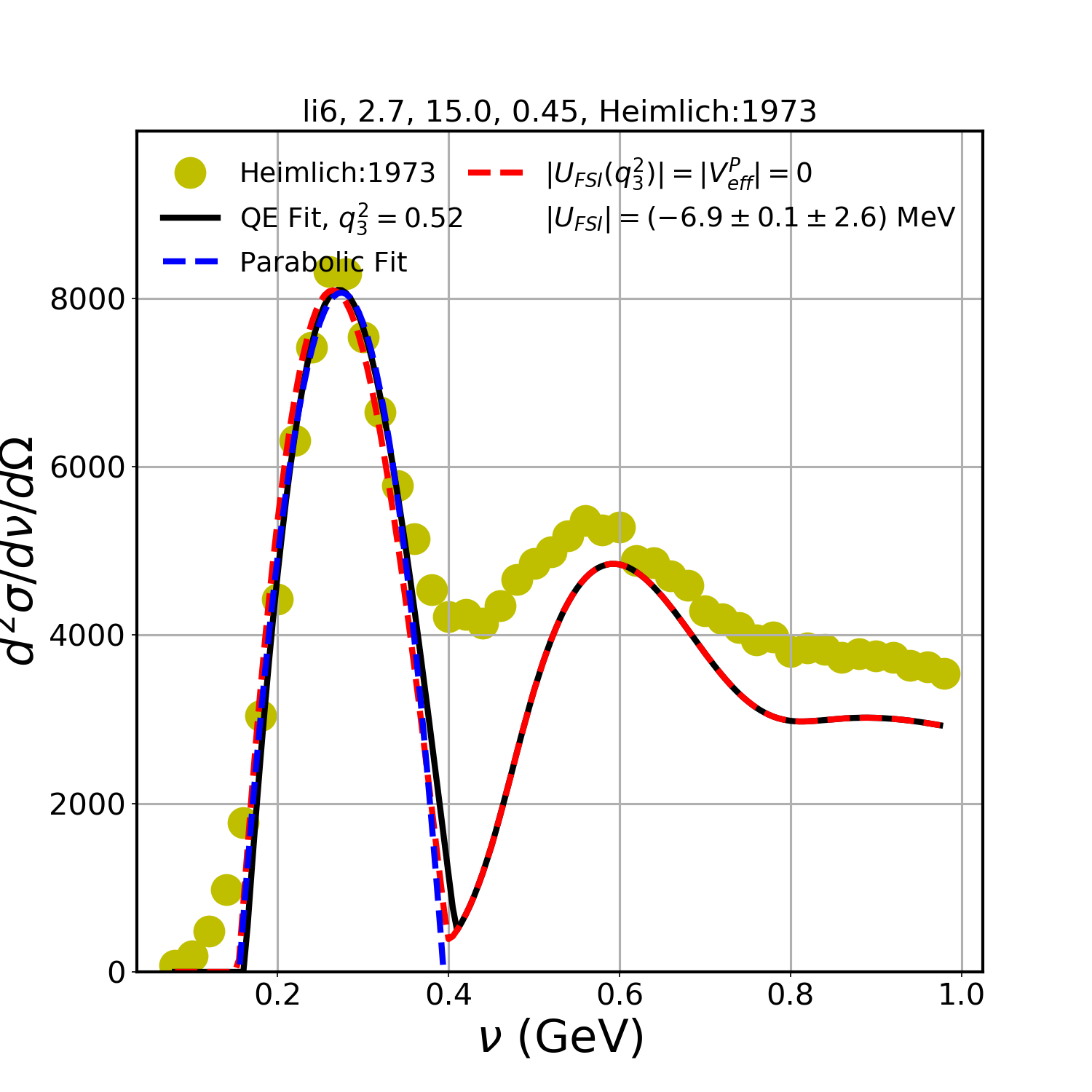}
\caption{
%\footnotesize\addtolength{\baselineskip}{-1\baselineskip} 
Examples of fits for  three out of four $_{3}^{6}$Li  ($k_F^P$ = 0.169 GeV) QE differential cross sections. The solid blue curve is the RFG fit with the best value of $U_{FSI}$. The black dashed curve is the simple parabolic fit used to estimate the systematic error.  The red dashed curve is the RFG model  with  $U_{FSI}=V_{eff}=0$.  Above each panel we show:  The element,  $E_0$ in GeV,  $\theta$ in degrees,  $Q^2$ in GeV$^2$, First Author, and year of publication.}
	\label{Li6_fits}
\end{figure*}

\begin{figure*}
    %[ht]
\centering
\includegraphics[width=5.5cm,height=4.5cm]{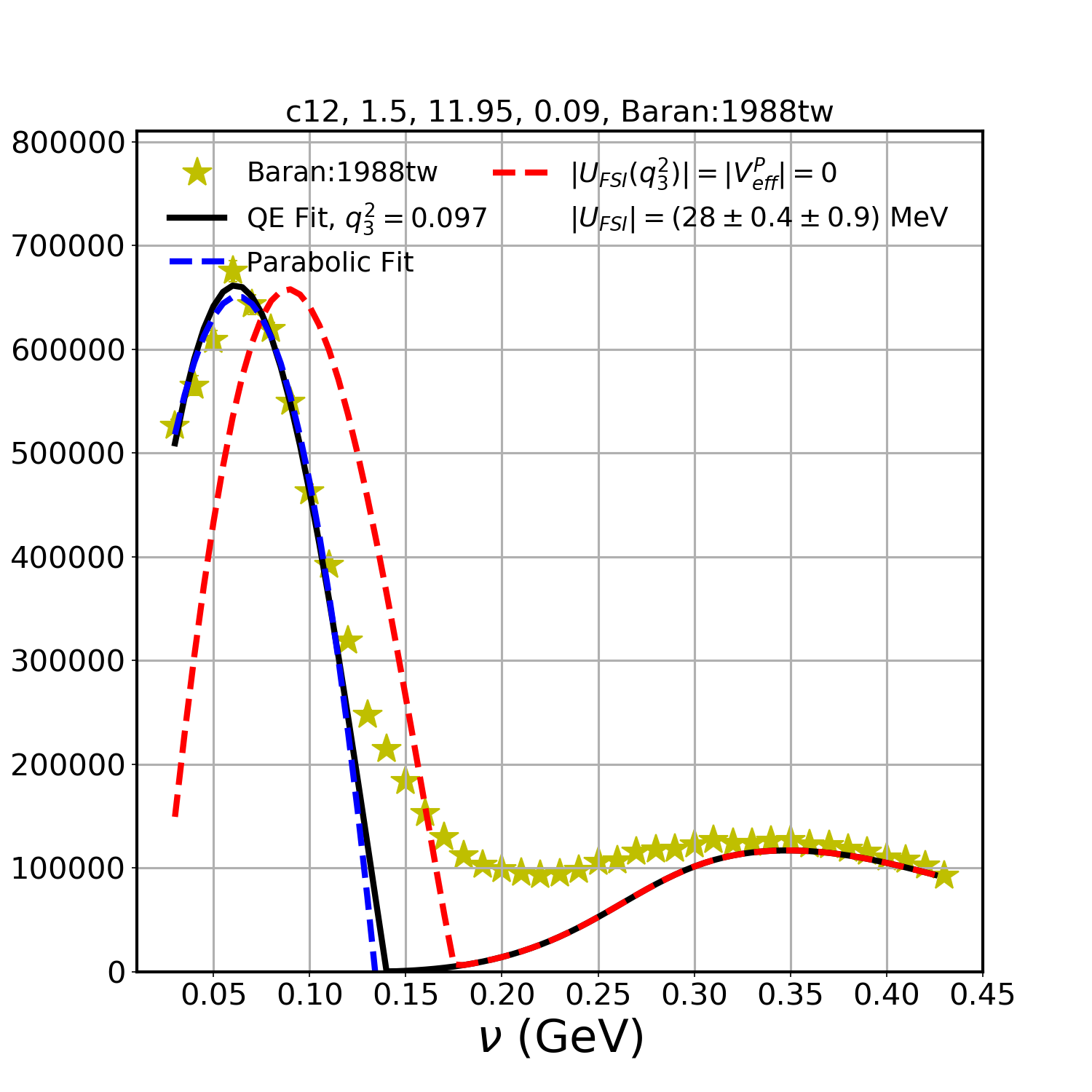}
\includegraphics[width=5.5cm,height=4.5cm]{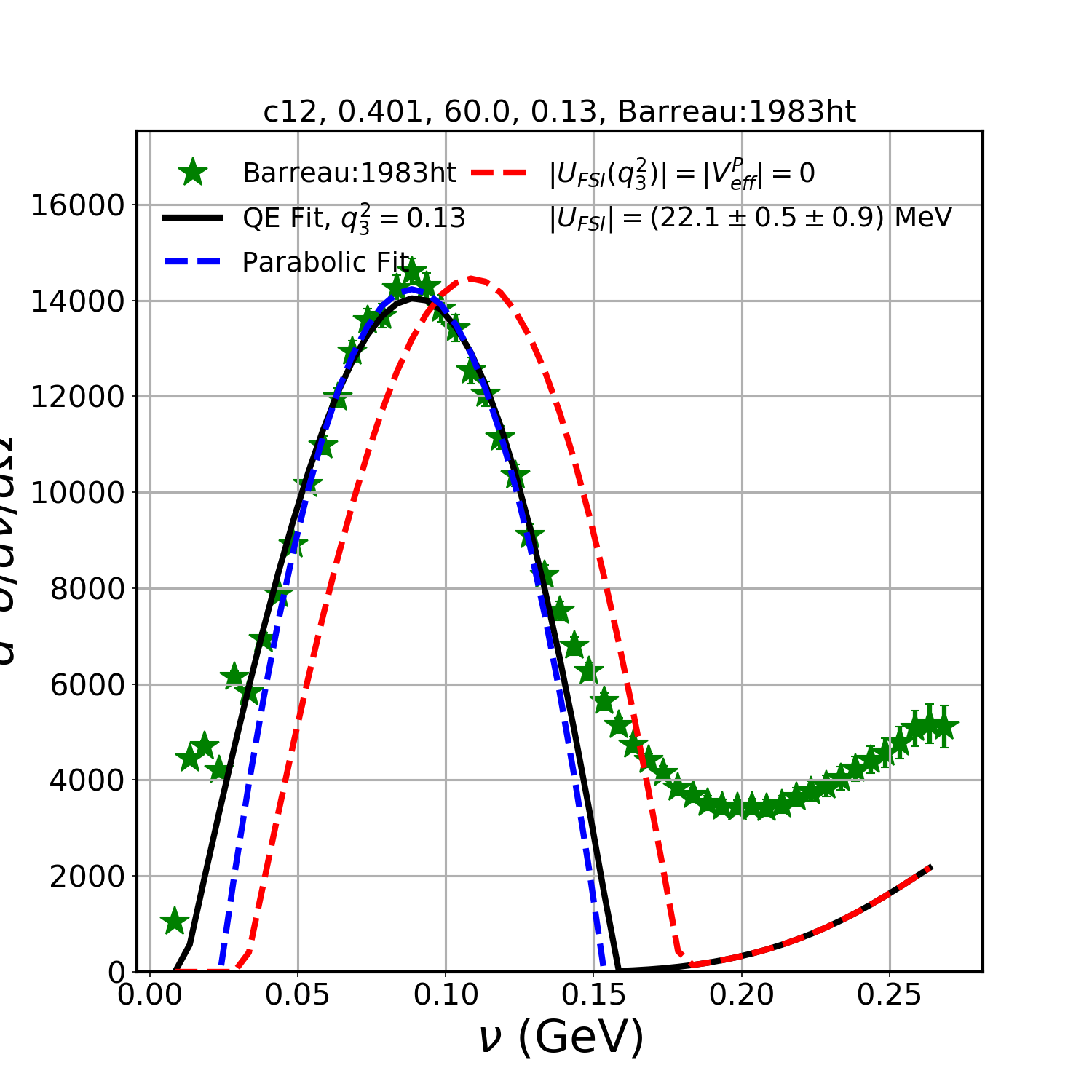}
\includegraphics[width=5.5cm,height=4.5cm]{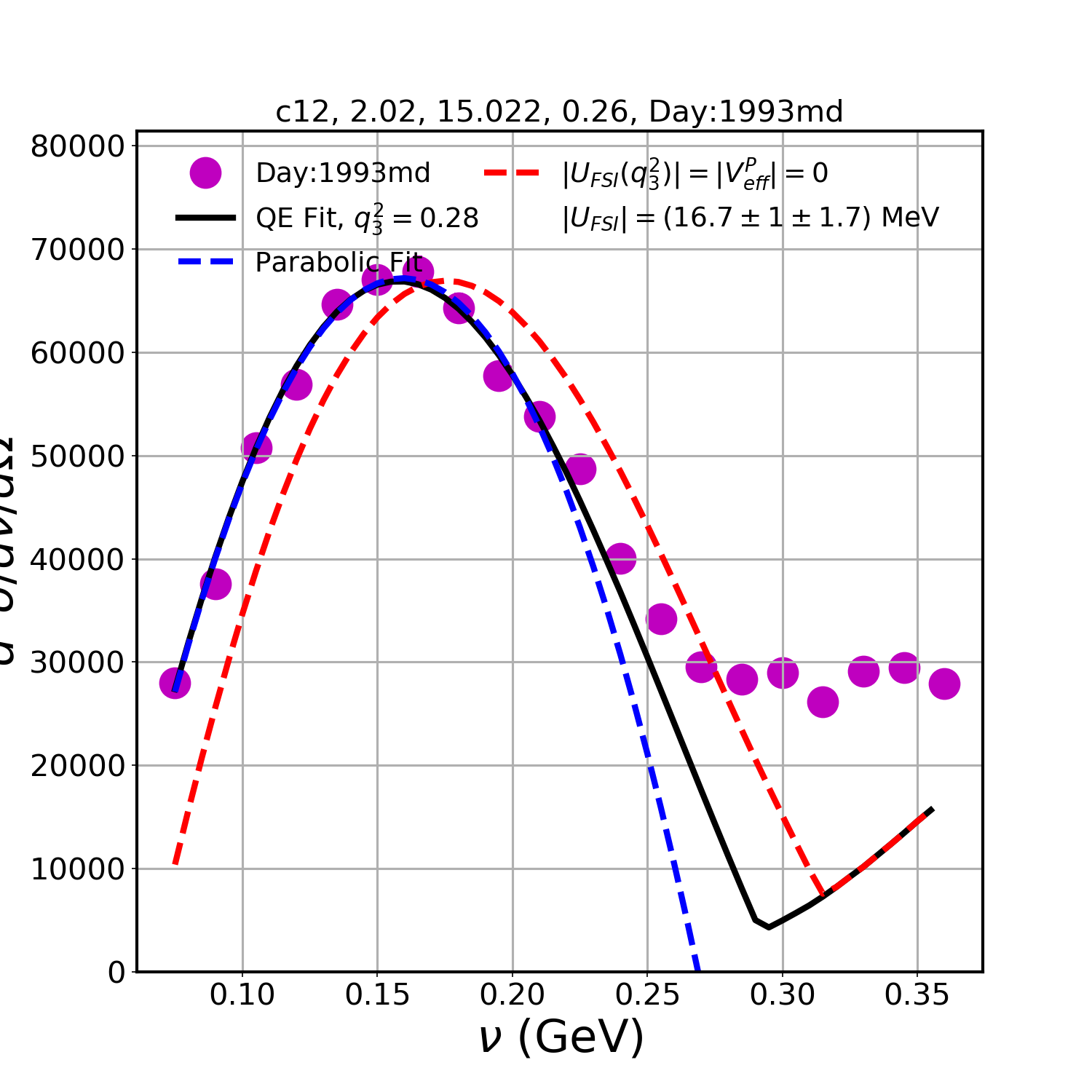}
\caption{
%\footnotesize\addtolength{\baselineskip}{-1\baselineskip} 
Same as Fig. \ref{Li6_fits} for three out of  33  $_{6}^{12}$C ($k_F^P$ = 0.221 GeV) QE differential cross sections.}
\label{C12_fits}
\end{figure*}

\begin{figure*}
    %[ht]
\centering
\includegraphics[width=5.5cm,height=4.5cm]{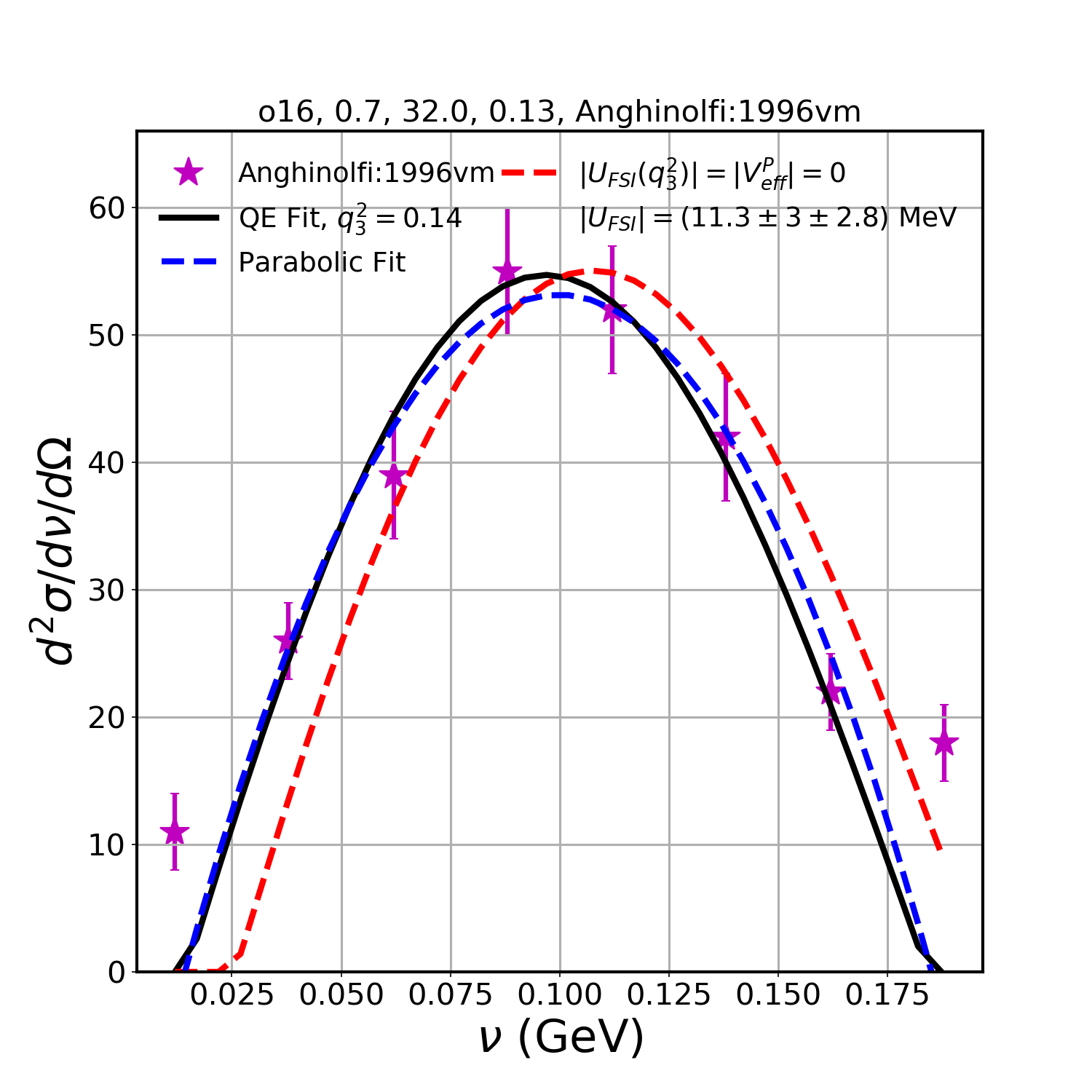}
\includegraphics[width=5.5cm,height=4.5cm]{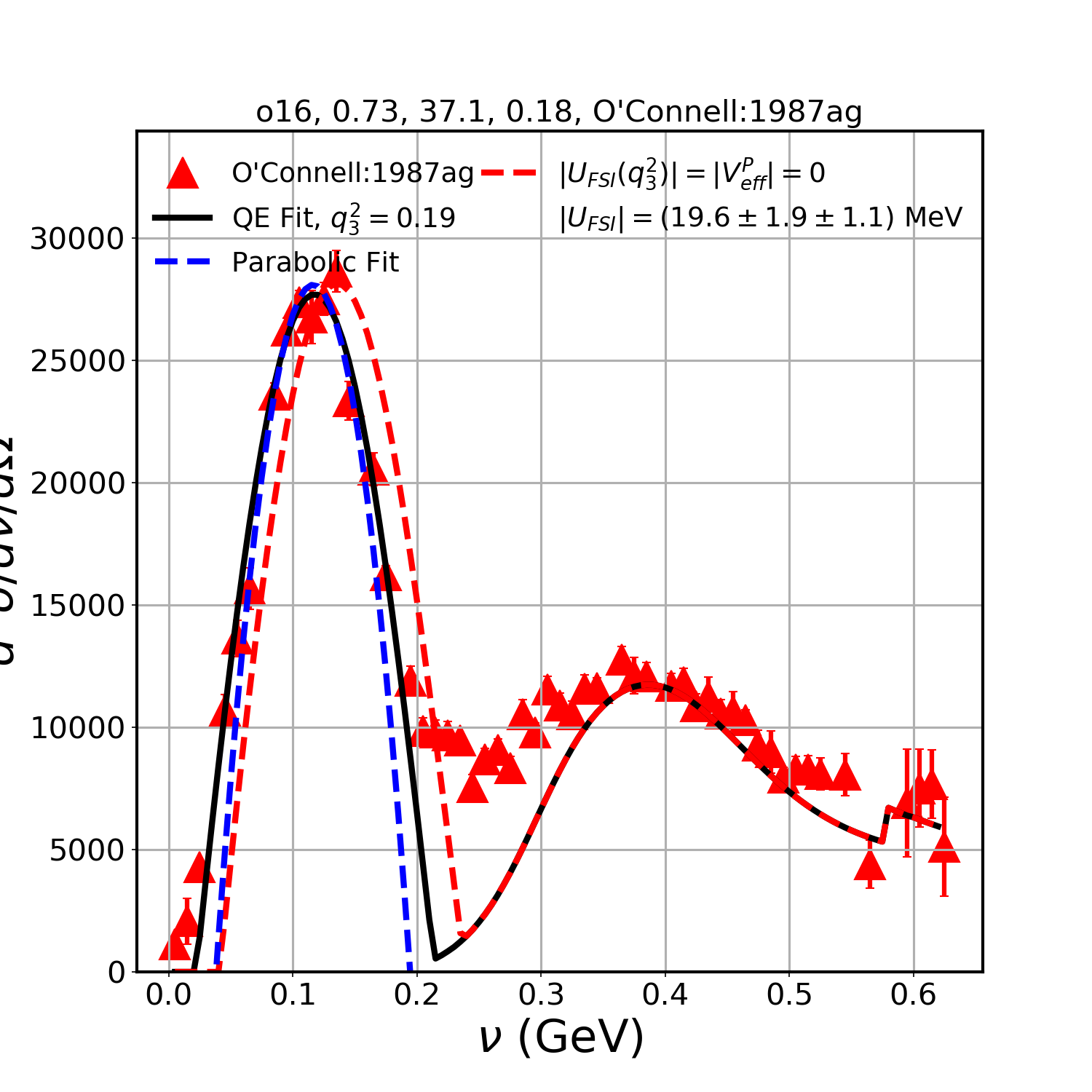}
\includegraphics[width=5.5cm,height=4.5cm]{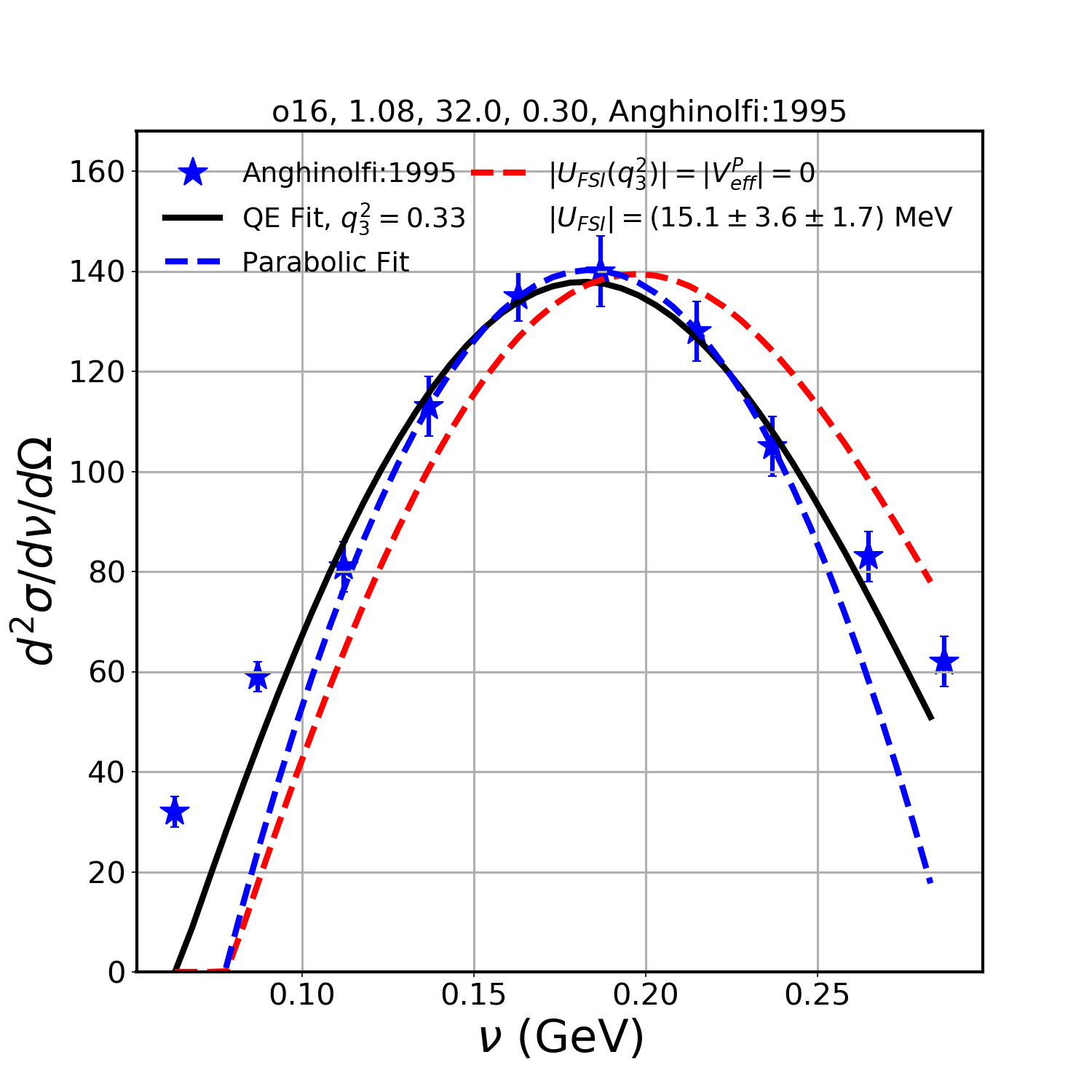}
\caption{
%\footnotesize\addtolength{\baselineskip}{-1\baselineskip} 
Same as Fig. \ref {Li6_fits} for three out of 8  $_{8}^{16}$O ($k_F^P$ = 0.225 GeV) QE differential cross sections.
}
\label{O16_fits}
\end{figure*}

\begin{figure*}
    %[ht]
\centering
\includegraphics[width=5.5cm,height=4.5cm]{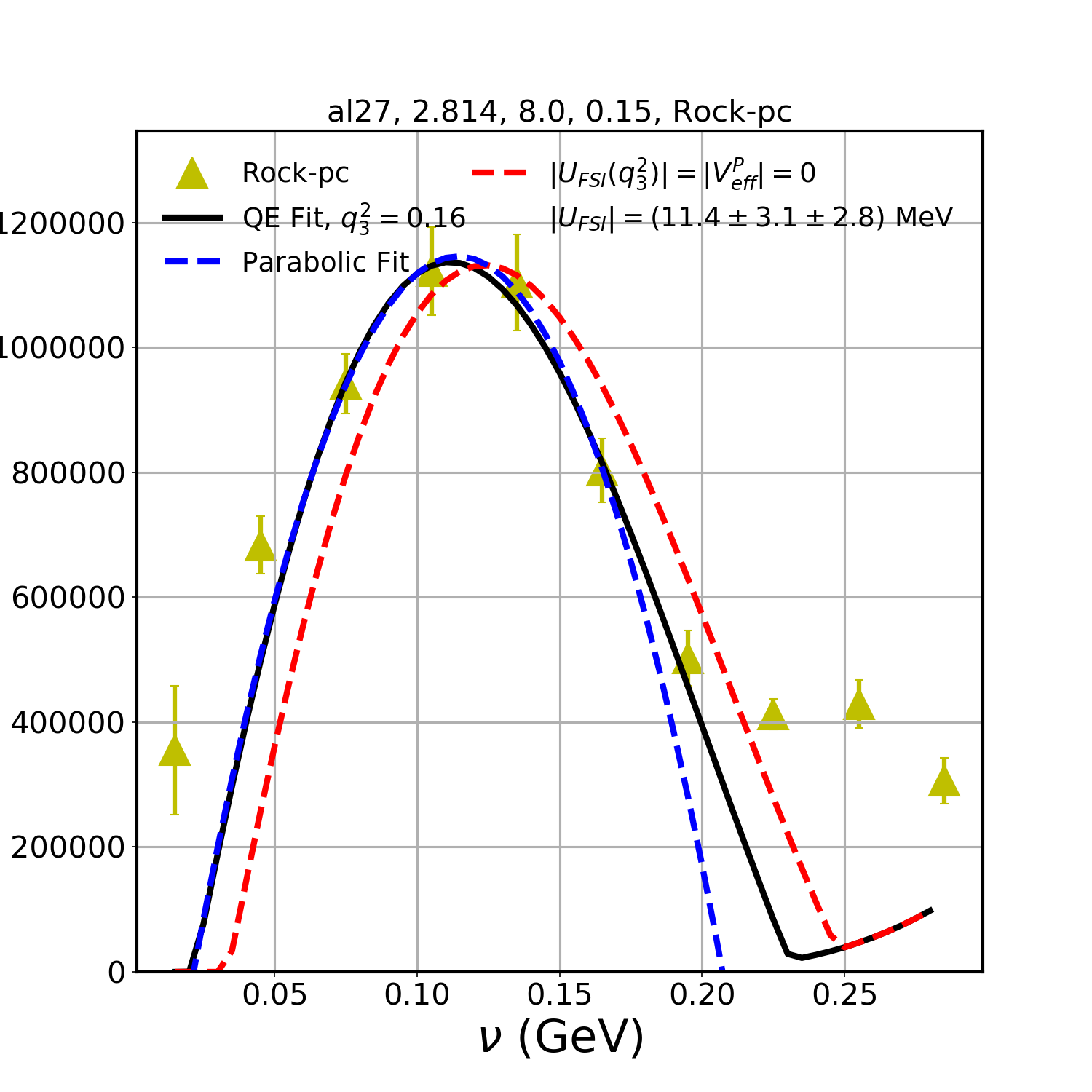}
\includegraphics[width=5.5cm,height=4.4cm]{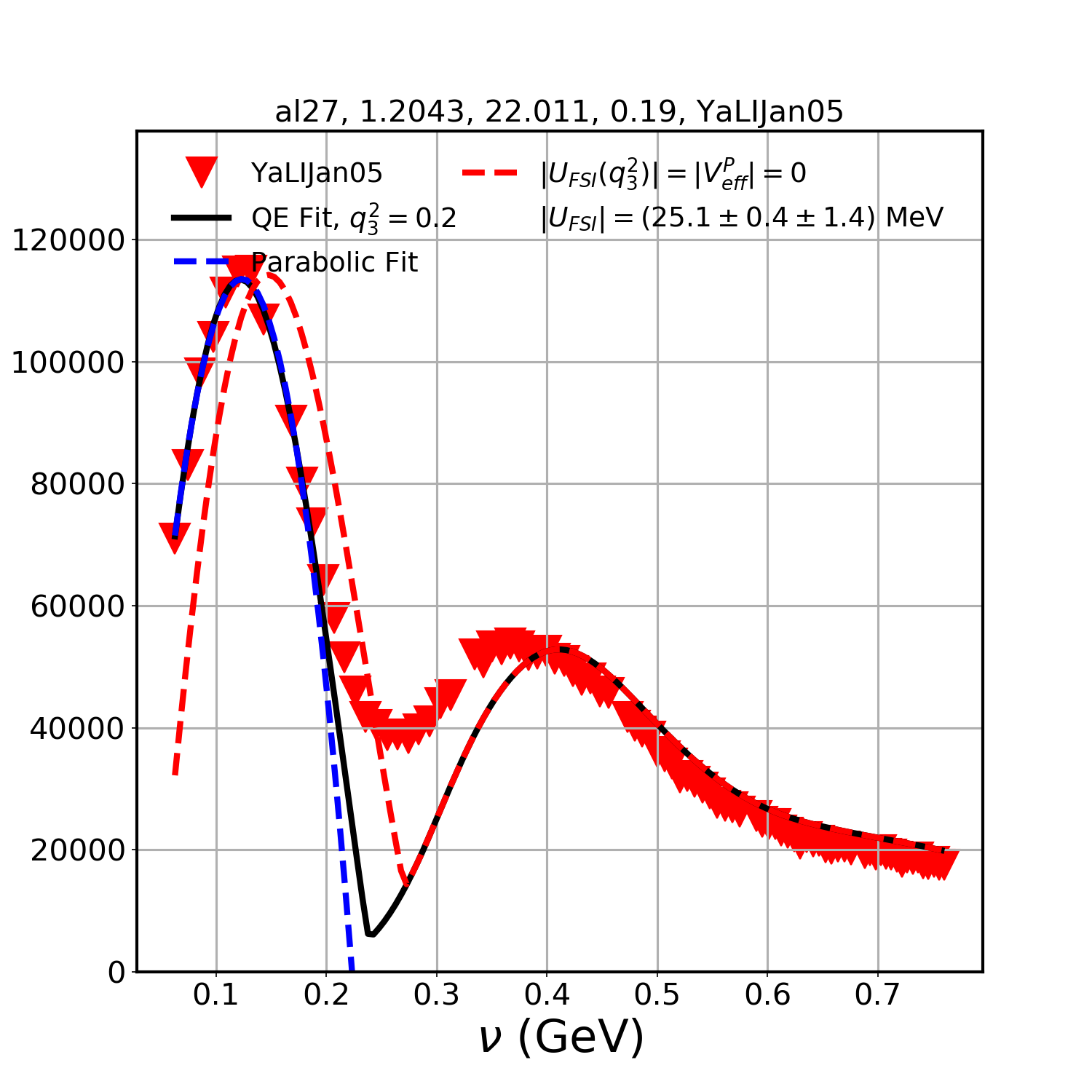}
\includegraphics[width=5.5cm,height=4.4cm]{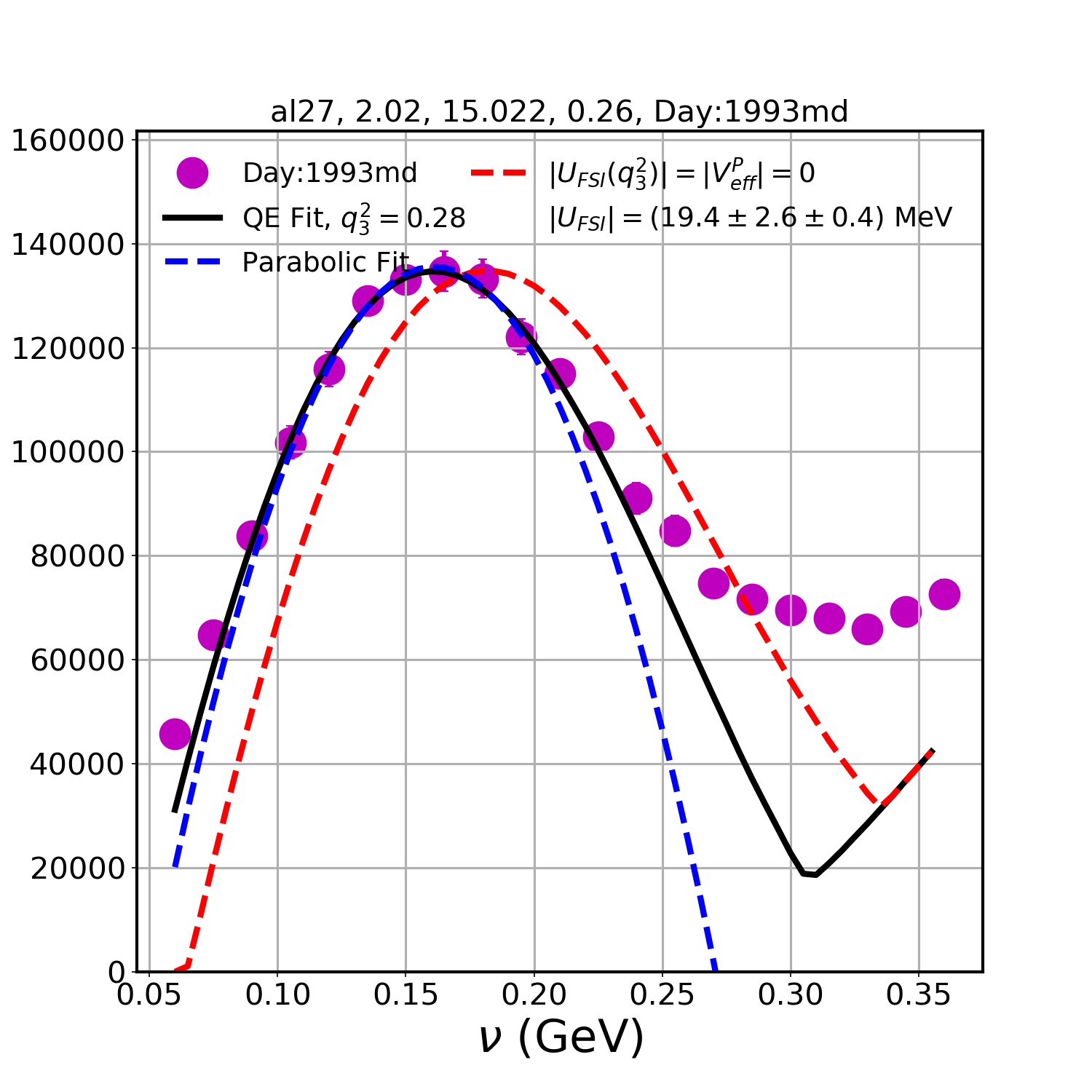}
\caption{
%\footnotesize\addtolength{\baselineskip}{-1\baselineskip} 
Same as Fig. \ref {Li6_fits} for three out of 8 $_{13}^{27}$Al ($k_F^P$ = 0.238 GeV)  QE differential cross sections.
}
\label{Al27_fits}
\end{figure*}

\begin{figure*}
    %[ht]
\centering
\includegraphics[width=5.5cm,height=4.5cm]{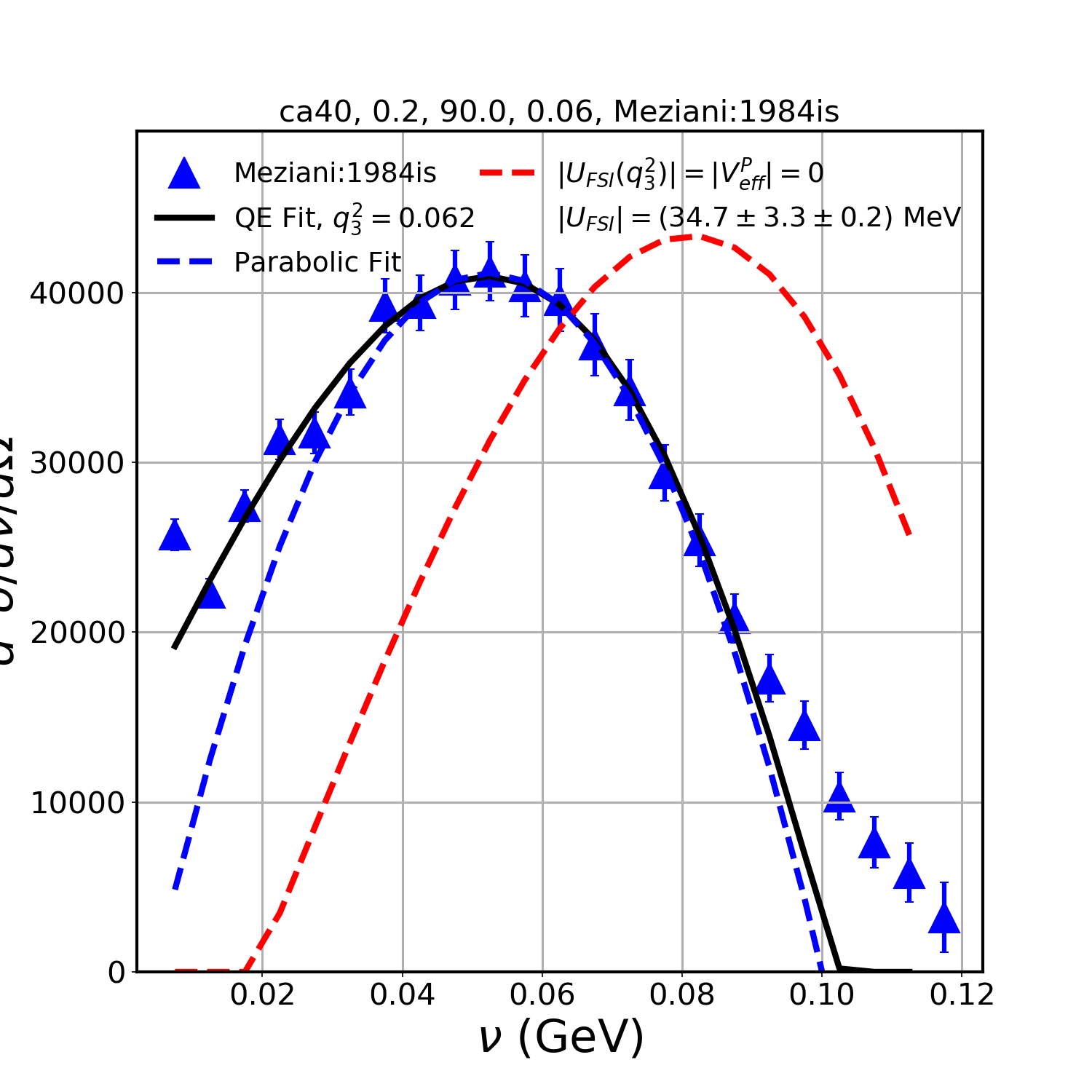}
\includegraphics[width=5.5cm,height=4.5cm]{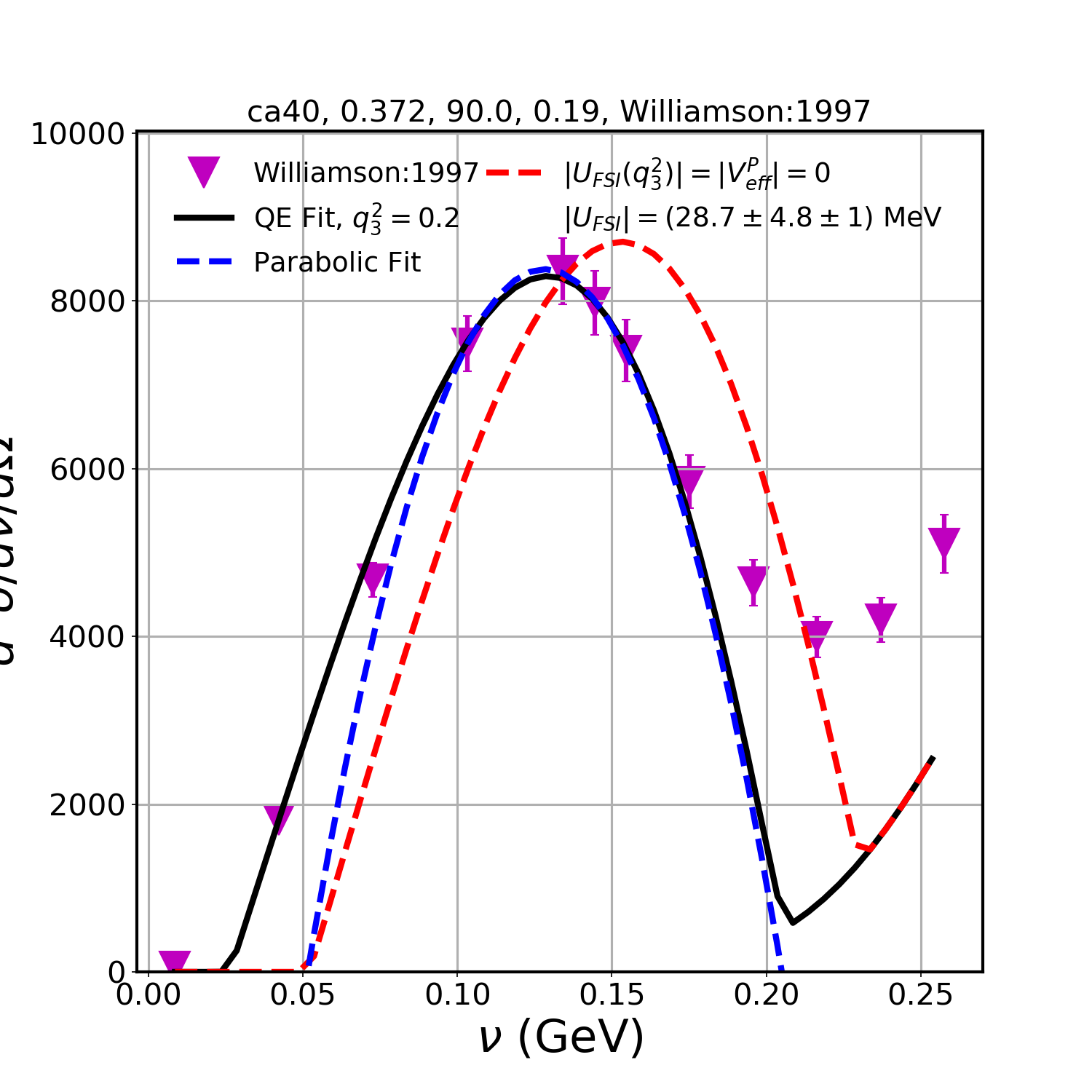}
\includegraphics[width=5.5cm,height=4.5cm]{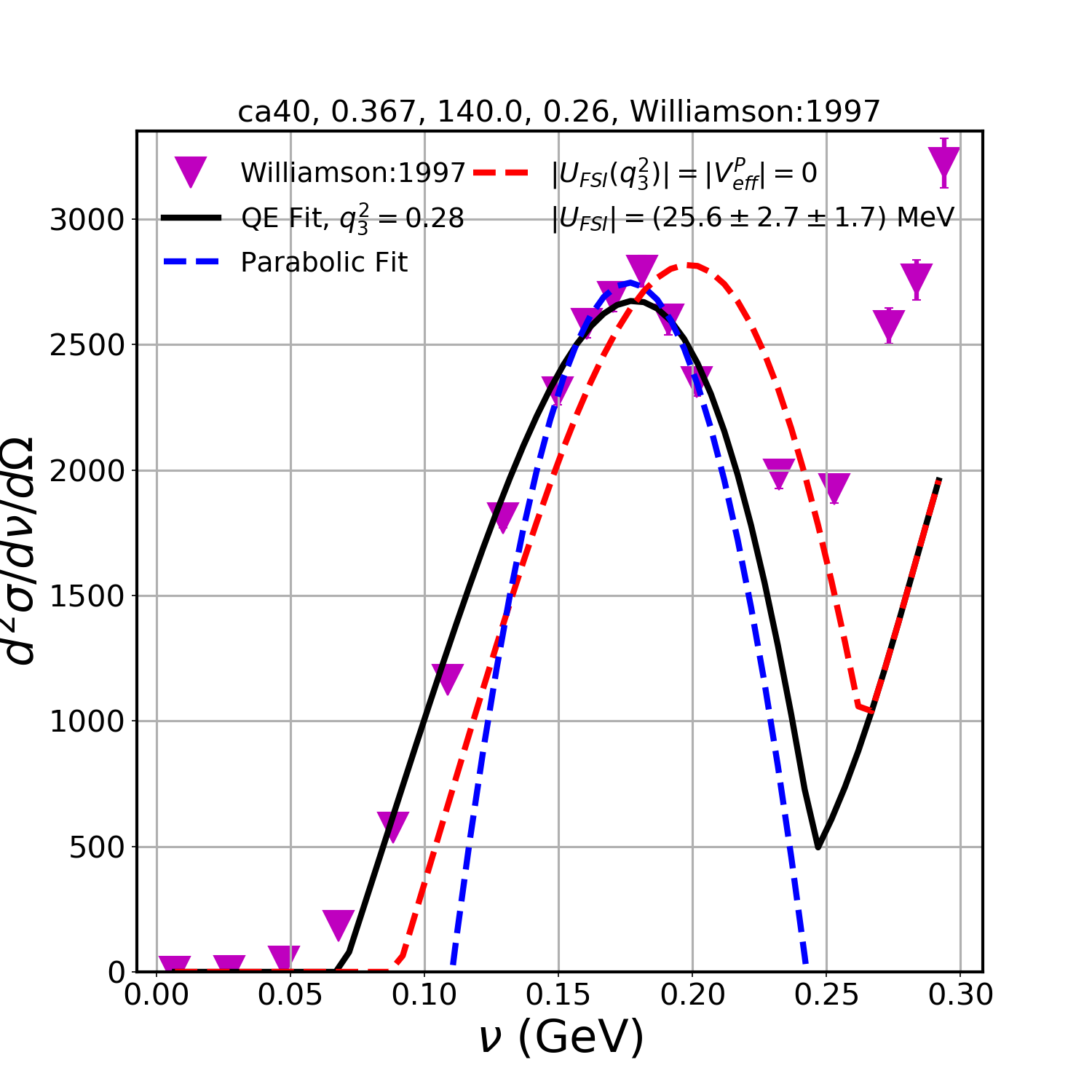}
\caption{
%\footnotesize\addtolength{\baselineskip}{-1\baselineskip} 
Same as Fig. \ref {Li6_fits} for three out of  29  $_{20}^{40}$Ca  ($k_F^P$ = 0.251 GeV)  QE differential cross sections.
}
\label{Ca40_fits}
\end{figure*}

\begin{figure*}
    %[ht]
\centering
\includegraphics[width=5.5cm,height=4.5cm]{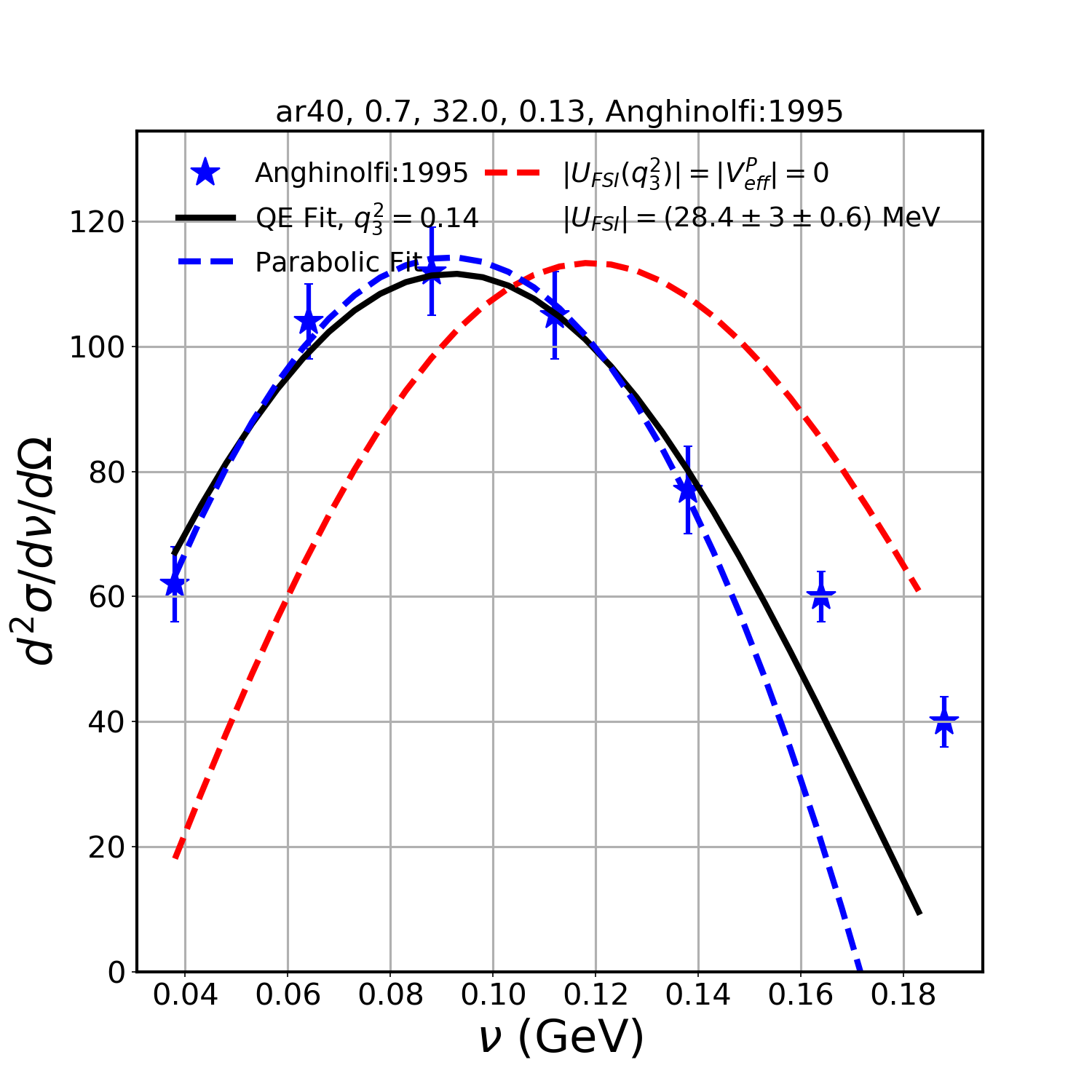}
\includegraphics[width=5.5cm,height=4.5cm]{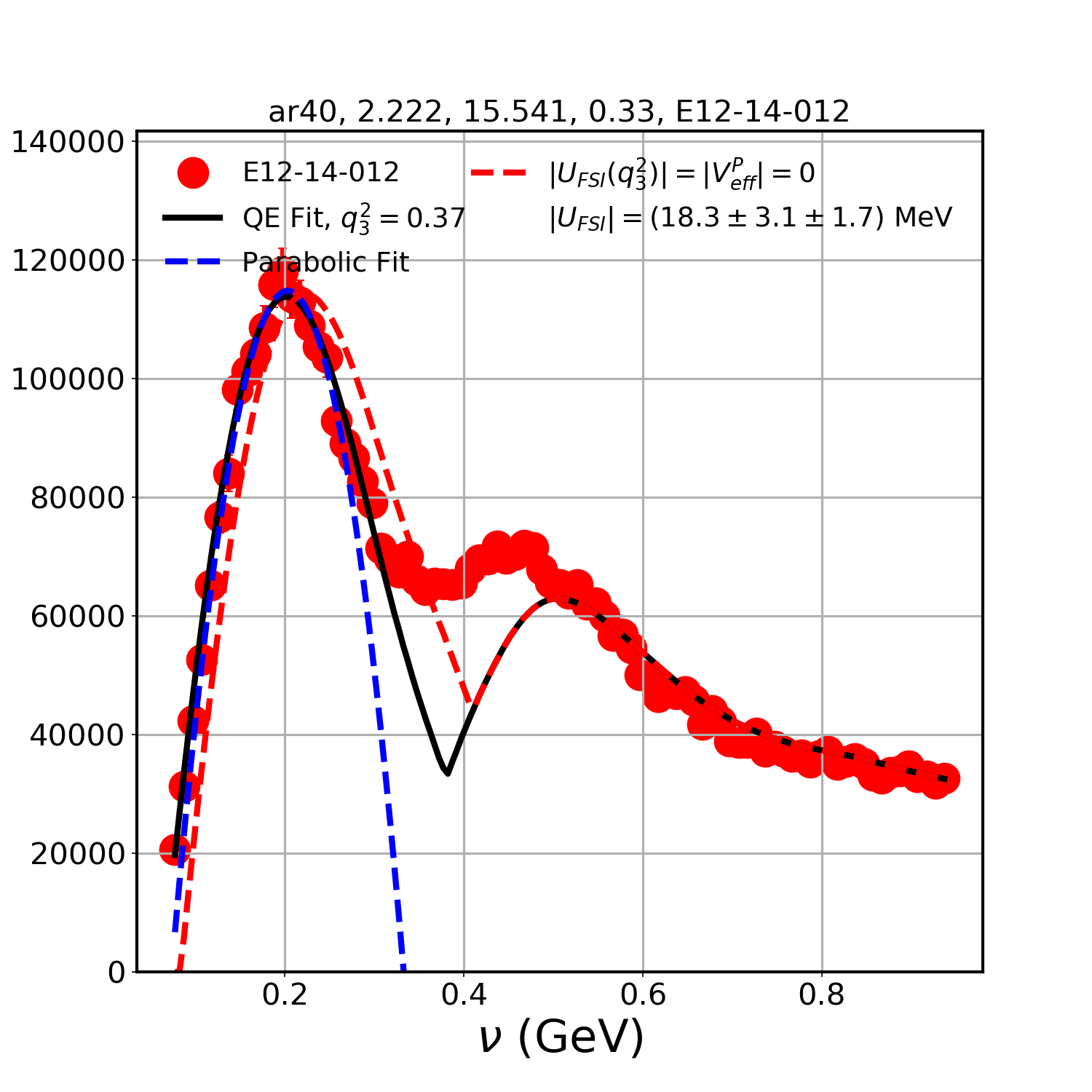}
\caption{
%\footnotesize\addtolength{\baselineskip}{-1\baselineskip} 
Same as Fig. \ref {Li6_fits} for two $_{18}^{40}$Ar ($k_F^P$ = 0.251 GeV)  QE differential cross sections.
}
\label{Ar40_fits}
\end{figure*}

\begin{figure*}
    %[ht]
\centering
\includegraphics[width=5.5cm,height=4.5cm]{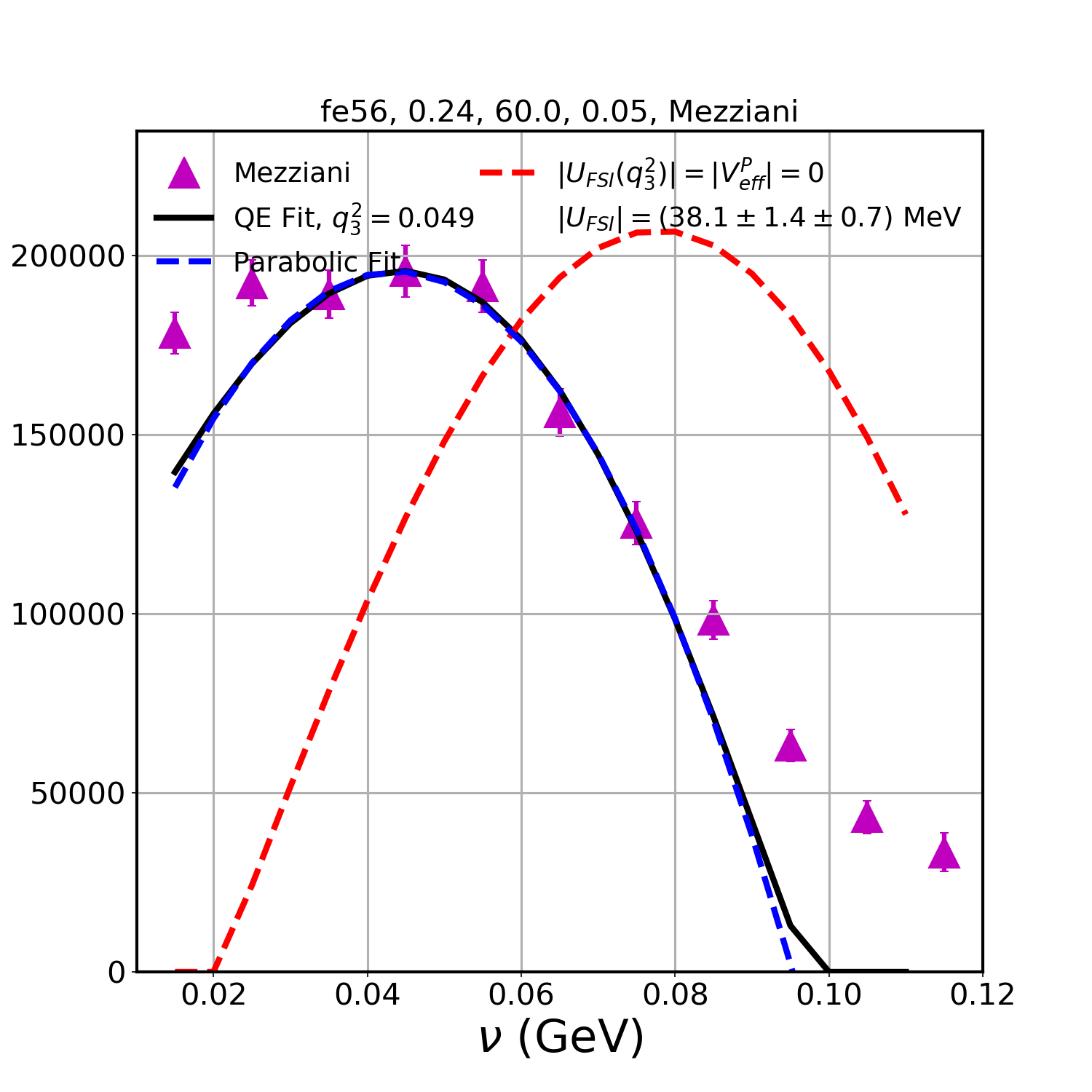}
\includegraphics[width=5.5cm,height=4.5cm]{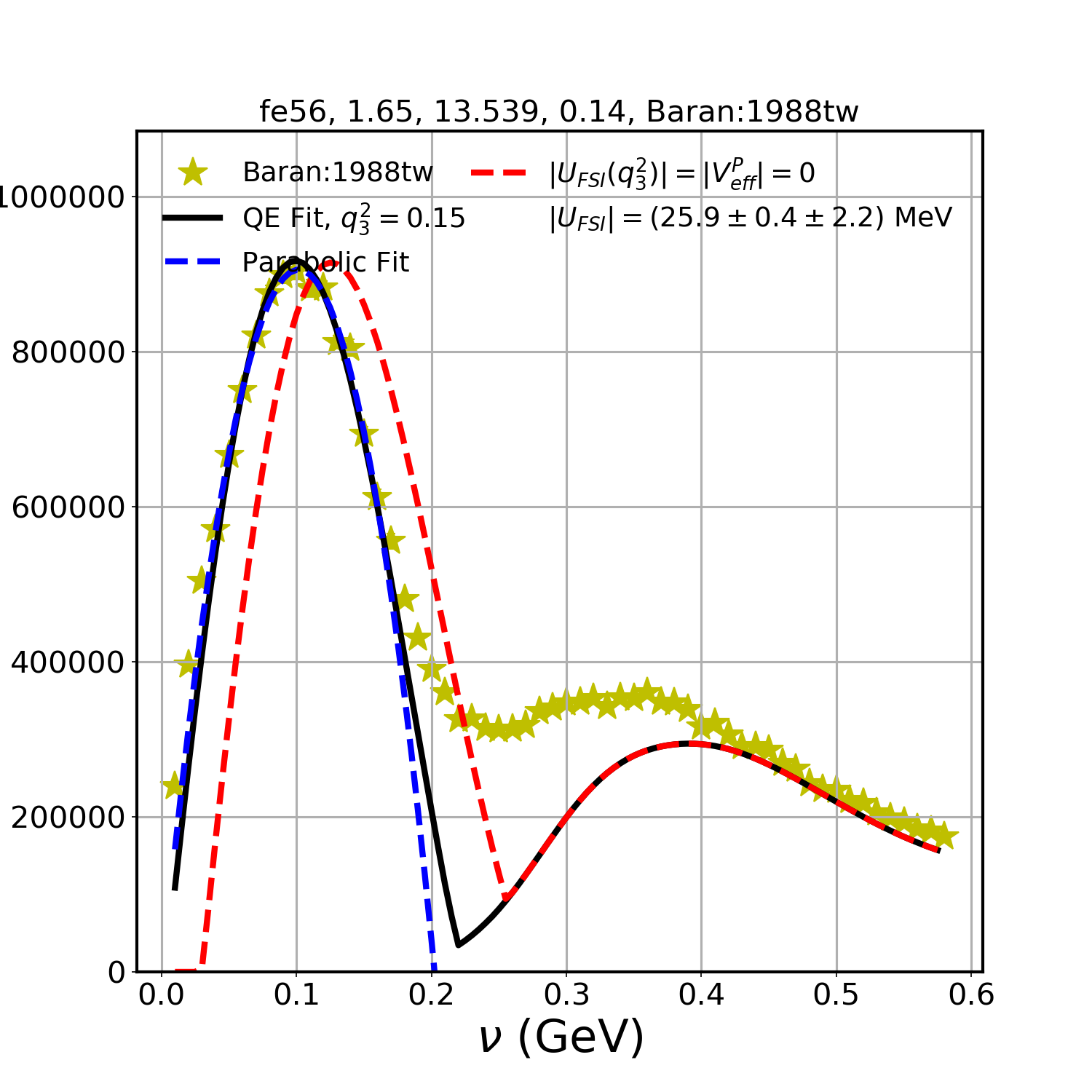}
\includegraphics[width=5.5cm,height=4.5cm]{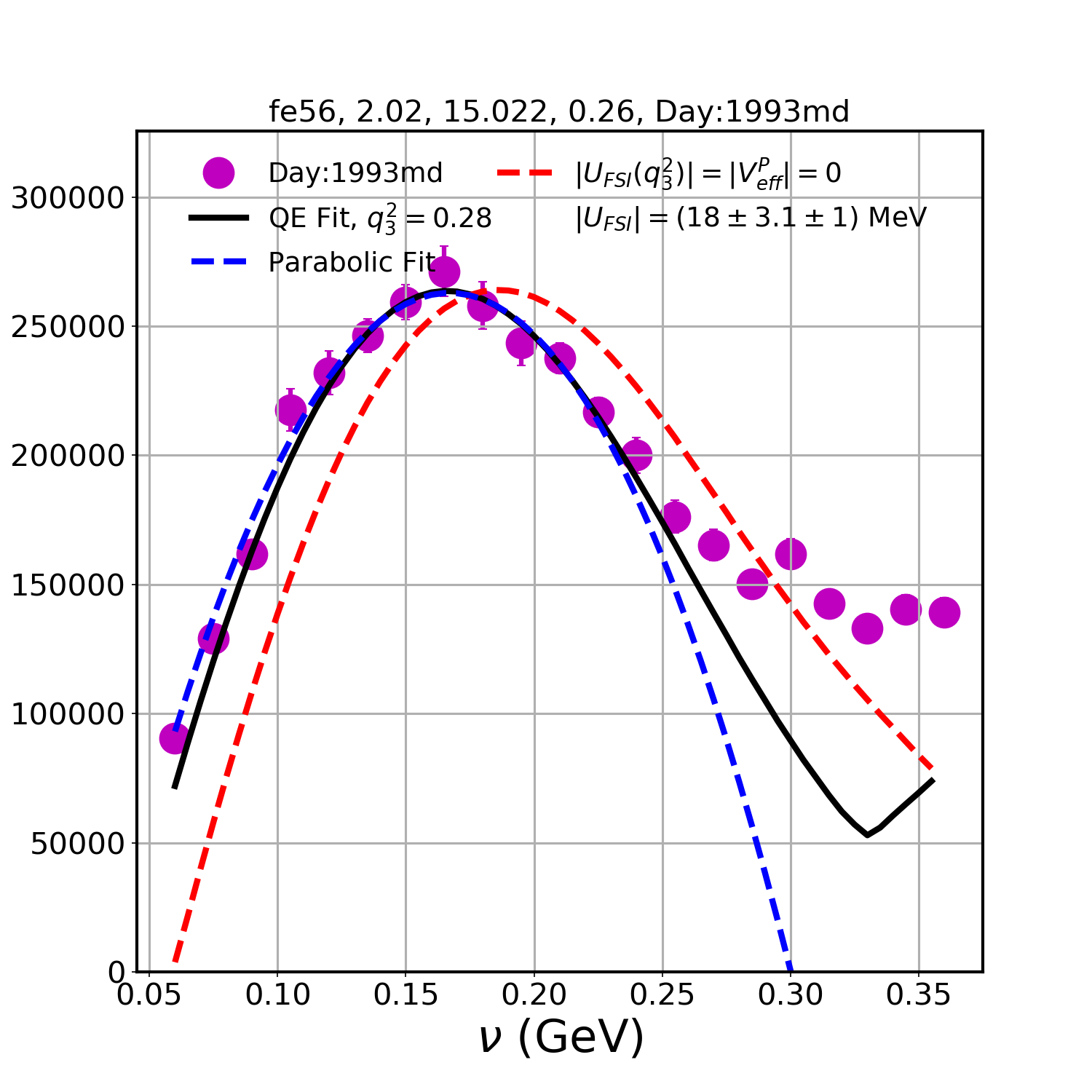}
\caption{
%\footnotesize\addtolength{\baselineskip}{-1\baselineskip} 
Same as Fig. \ref {Li6_fits} for three of 30  $_{20}^{56}$Fe ($k_F^P$ = 0.254 GeV) QE differential cross sections.
}
\label{Fe56_fits}
\end{figure*}

\begin{figure*}
    %[ht]
\centering
\includegraphics[width=5.5cm,height=4.5cm]{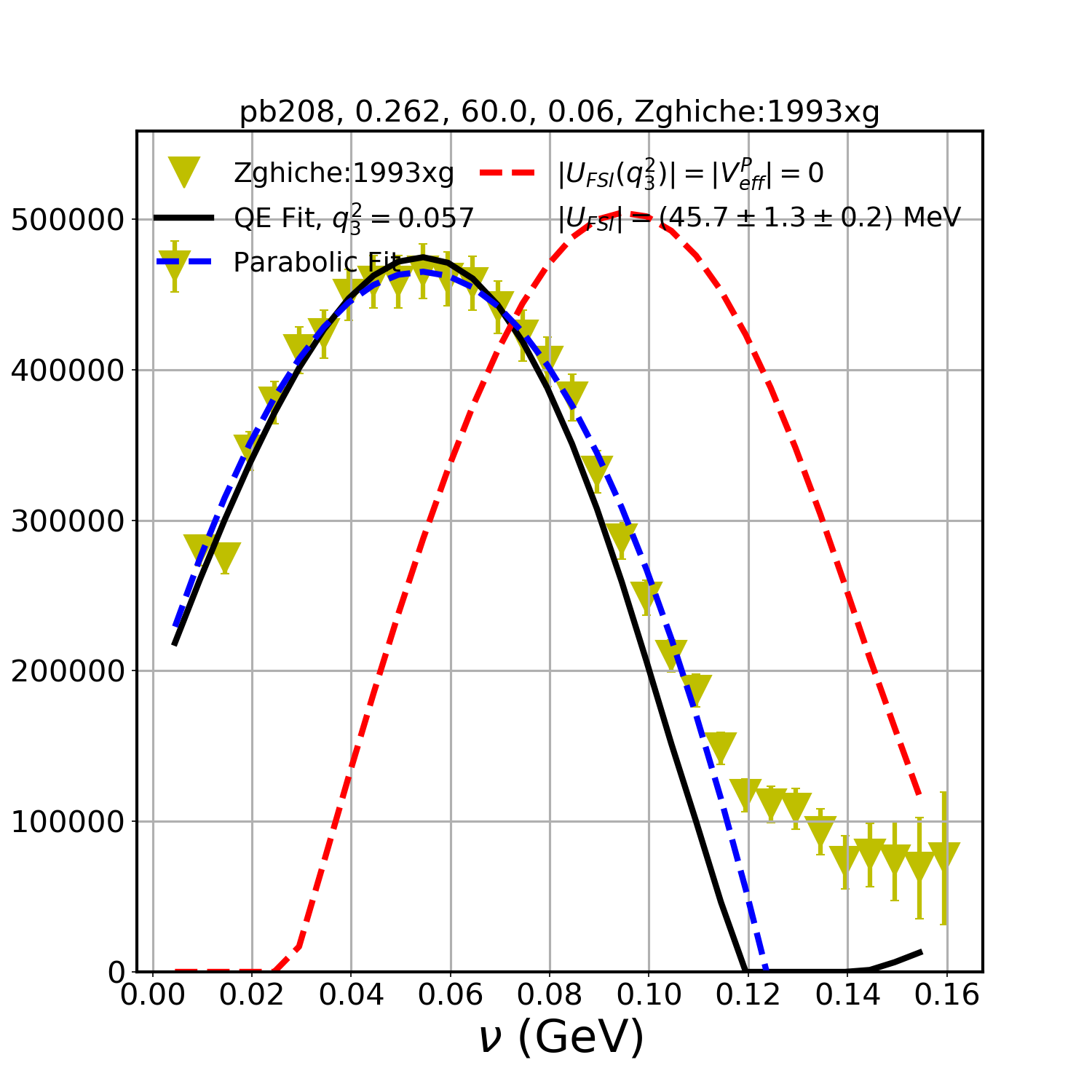}
\includegraphics[width=5.5cm,height=4.5cm]{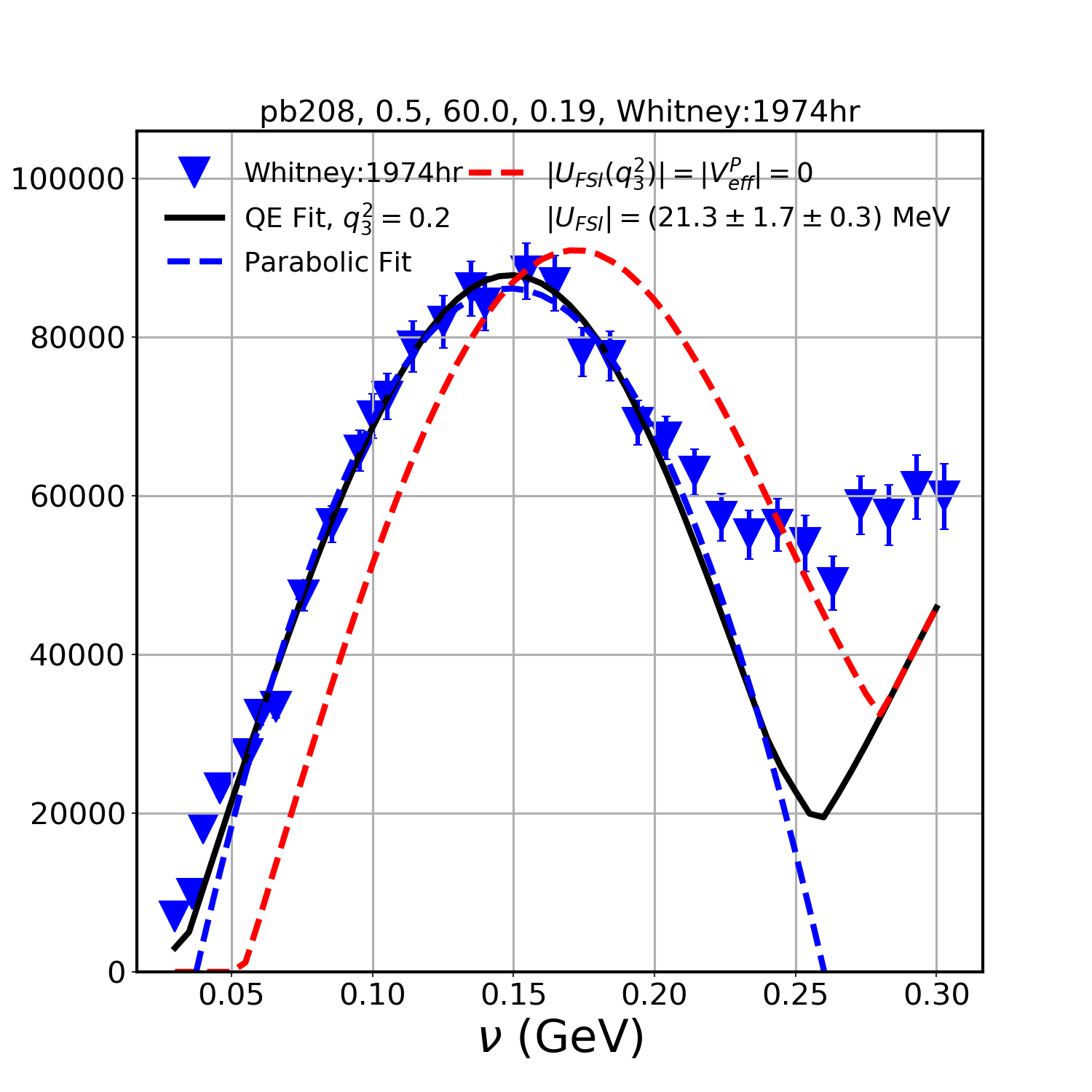}
\includegraphics[width=5.5cm,height=4.5cm]{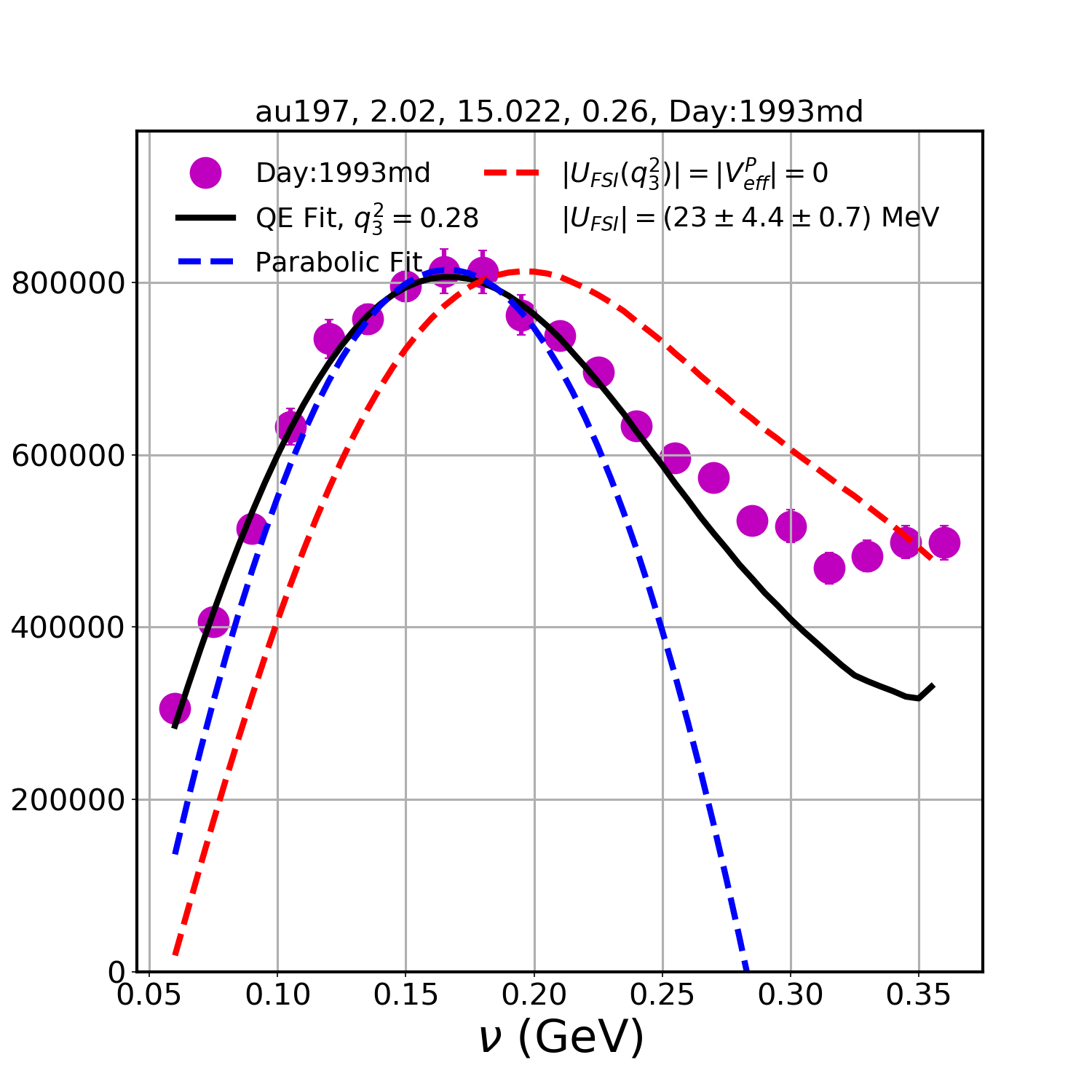}
\caption{
%\footnotesize\addtolength{\baselineskip}{-1\baselineskip} 
Same as Fig. \ref {Li6_fits} for two of  22  $_{82}^{208}$Pb ($k_F^P$ = 0.275 GeV)  and one $_{79}^{197}$Au ($k_F^P$ = 0.275 GeV) QE differential cross sections.
}
\label{Pb208_fits}
\end{figure*}

     We only fit to the data in the top 1/3 of the QE distribution to extract  the best value of  $U_{FSI}$ for $(\vec q_3+\vec k)^2$ at the peak.
  In the fit we let the normalization of the QE peak float to agree with data. 
  For the estimate of the systematic error we also fit the QE differential cross section versus $\nu$ near the peak region   to a simple parabola and extract the value of $\nu_{peak}^{parabola}$.  We use the difference between $\nu_{peak}^{parabola}$ and $\nu_{peak}^{rfg}$ as a systematic error in our
 extraction of $U_{FSI}((\vec q_3+\vec k)^2)$. 
 %
 %     subsection  6.3 
  %
  \subsection{The $\Delta (1232)$ resonance shown in Figures  \ref{Li6_fits}-\ref{Pb208_fits}} 
 A simple calculation of the cross section for the production of $\Delta (1232)$ resonance is shown in Figures  \ref{Li6_fits}-\ref{Pb208_fits}.
 The calculation uses Jlab fits to the structure functions in the  resonance region for protons and neutrons. 
 These structure functions were extracted from hydrogen and deuterium data.  
 
 The proton and neutron structure functions in the resonance region were used
 as input to a simple Fermi Gas smearing model. In the calculation,  $U_{FSI}$  for the  $\Delta (1232)$ resonance
 is assumed to be the same as $U_{FSI}$ for QE scattering.

   The curves shown  Figures  \ref{Li6_fits}-\ref{Pb208_fits}  
   do not include the contributions of  2p2h final states from meson exchange currents (MEC)  and isobar excitation.
 These 2p2h contributions yield additional cross section
 in the region between the QE peak and the $\Delta (1232)$ resonance.  The 2p2h contributions are primarily transverse and
 therefore are more significant  for electron scattering at larger angles  than at small angles (as observed in the figures).
The investigations of MEC  (which is model dependent) and  the values of $U_{FSI}$ for a  $\Delta (1232)$ resonance  
in the final state are the subject of a future investigation.

%     %        Figure 8
%\vspace{-001i}
    %    %        Figure 9
%
%\vspace{-001i}
%      %        Figure 10
%
 %
 %    %        Figure 11
%
 %     %        Figure 12
%        %        Figure 13
%      %        Figure 14
%    %        Figure 15
 %    %        Figure 16
%    %        Figure 17
  %   %        Figure 18
%    %        Figure 19
%        Figure 20
%      Figure 21
%
 %         Figure  22
 \begin{figure*}
 %[ht]
\begin{center}
\includegraphics[width=3.5in,height=2.5in]{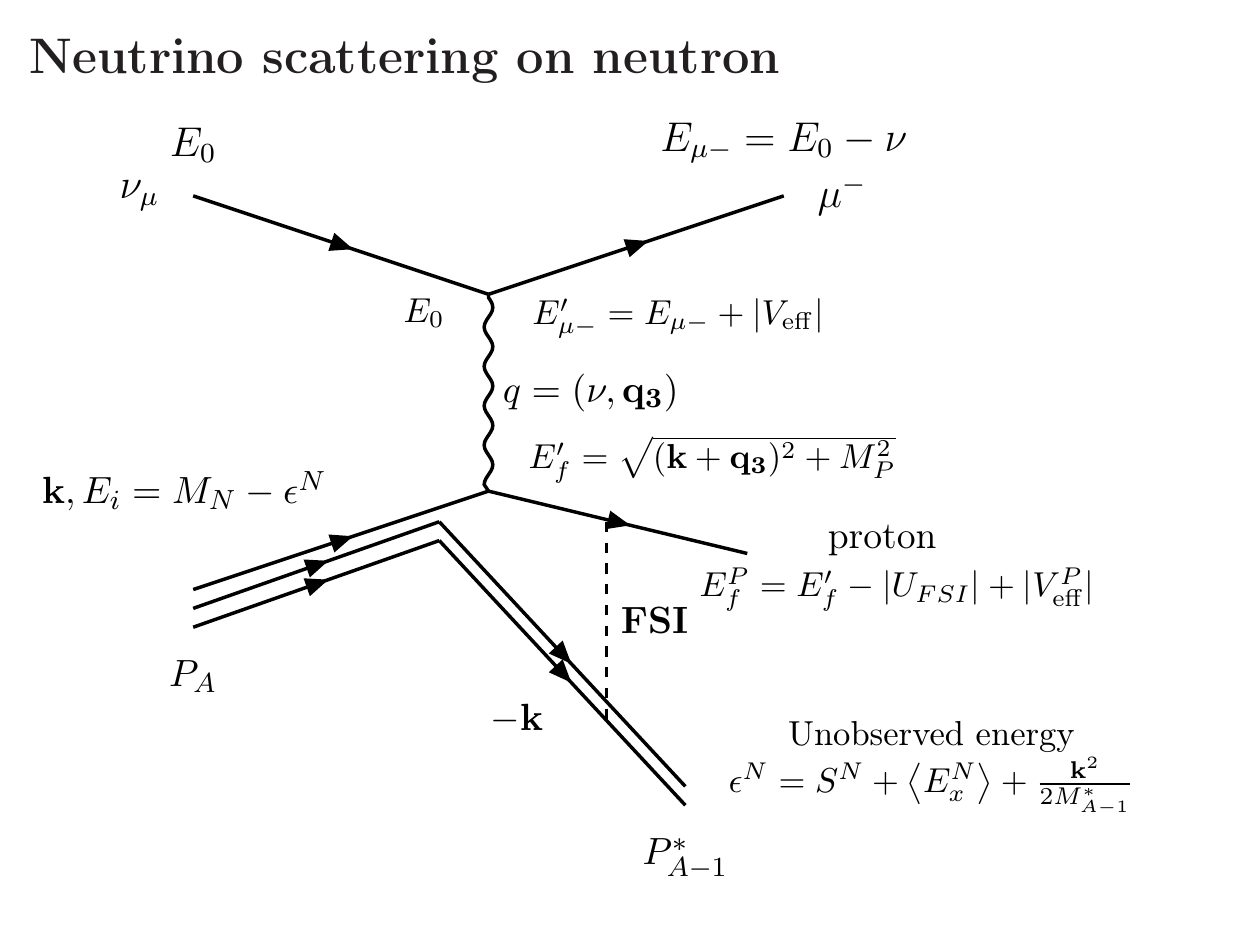}
\includegraphics[width=3.5in,height=2.5in]{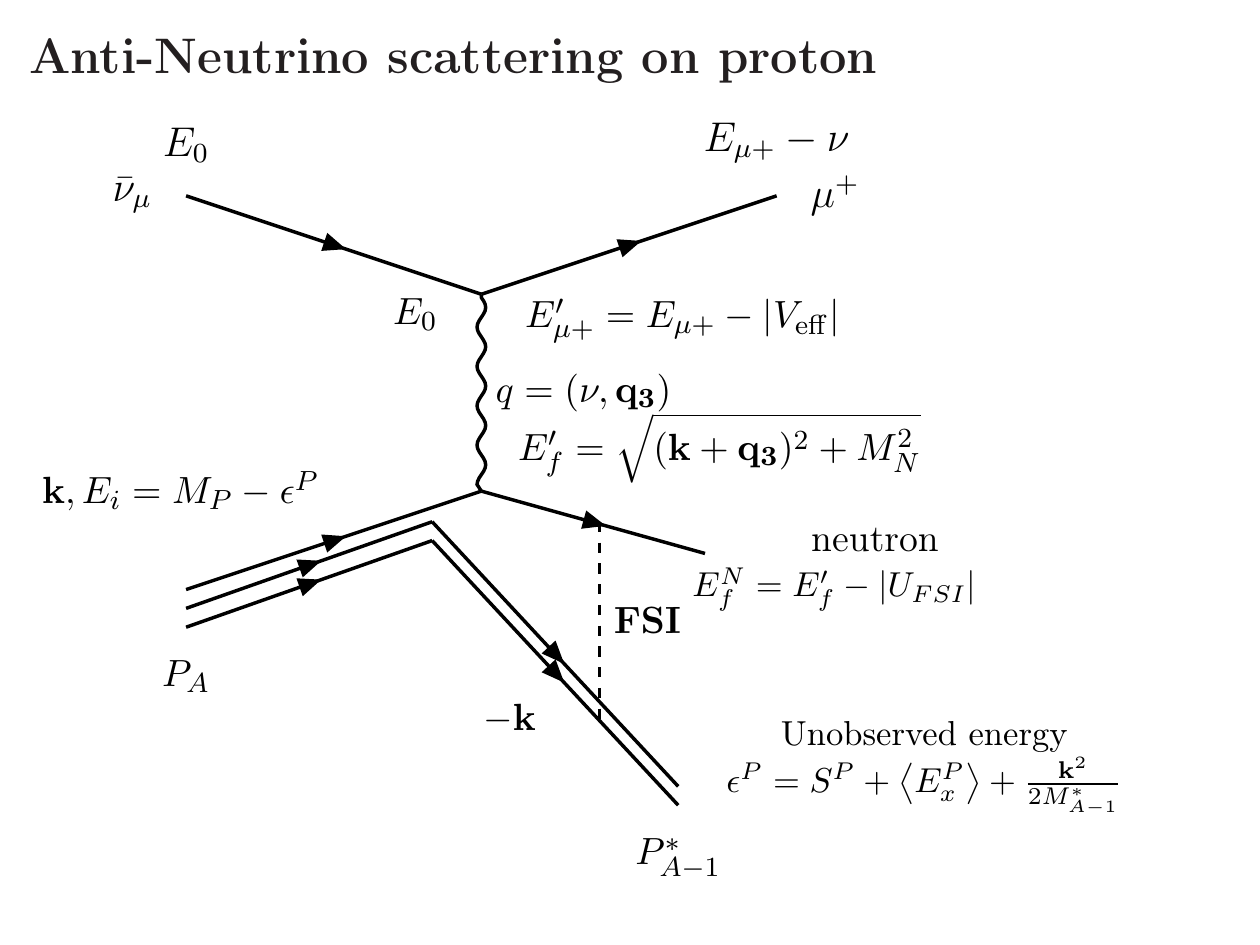}
\caption{ 
1p1h process: Neutrino (left) and antineutrino (right) QE scattering from an off-shell bound nucleon of momentum $\vec {p_i}$=$\vec k$ in 
a nucleus of mass A.   The off-shell energy of the interacting nucleon is 
$E_i^{N,P}  =  M_{N,P} -\epsilon^{N,P} $, where 
$\epsilon^{N,P} = S^{N,P} +\langle E_x^{N,P}  \rangle+\frac{\vec k^2}{2M^*_{A-1}}$. 
  We model the effect of FSI (strong and EM interactions) by setting the energy of the final state proton  $E_f^P=\sqrt{(\vec {k}+\vec {q_3})^2+M_P^2} -|U_{FSI}|$+$|V_{eff}^P|$ for neutrino QE scattering on bound neutrons, and the energy of the final state neutron $E_f^N=\sqrt{(\vec {k}+\vec {q_3})^2+M_P^2} -|U_{FSI}|$ for antineutrinos scattering on bound protons. Here  $U_{FSI}= U_{FSI}( (\vec q_3+\vec k)^2)$.
  For neutrino QE scattering  on bound neutrons   $|V_{eff}^P|=|V_{eff}|$,  and the effective  ${\mu-}$ energy at the vertex is
 $E^\prime_{\mu-}=E_{\mu-}+V_{eff}$.   For antineutrino QE scattering  on bound protons   $|V_{eff}^N|=0$ for the final state neutron  and the effective  ${\bar \mu}$ energy at the vertex is  $E^\prime_{\bar\mu}=E_{\bar\mu}-V_{eff}$.
  }
\label{neutrino-1p1h}
\end{center}
\end{figure*}
%%%

%Element 	slope 	intercept( q3+k )^2 	NPMODE 	slope 	intercept( q3+k )^2
%Li6 	0.0278 	-0.0045 	6        			0.0281 	-0.0043
%C12/O16 	0.0398 	-0.0292 	12 			0.0409 	-0.0291
%Au27 	0.0386 	-0.0285 	27 				0.0399 	-0.0281
%Ca40/Ar40 	0.048 	-0.0369 	40 			0.0520 	-0.0361
%Fe56 	0.0487 	-0.0362 	56 				0.0502 	-0.0347
%Pb208/Au197 	0.0443 	-0.0383 	208 		0.0627 	-0.0360

 %       Table 9
 \begin{table}[]
 %[ht]					
 \begin{center}					
\begin{tabular}{|c|ccc|}					
\hline					
$\bf_Z^{A}Nucl$	&$|V_{eff}|$ &$U_{FSI}$ &$U_{FSI}$\\ 
	& &	intercept 	& slope vs \\
	&GeV&	$(\vec q_3+\vec k)^2$=0.	& $(\vec q_3+\vec k)^2$   \\ 
\hline \hline
	 &  &  &    \\ 
$\bf_3^{6}Li$ &0.0014                       &-0.0043& 0.0281\\  
 &  &  &    \\
$\bf_6^{12}C/ \bf_8^{16}O$ &0.0031	&-0.0291& 0.0409	\\ 
 &  &  &    \\
$\bf_{13}^{27}Al$ &0.0051	         &-0.0281& 0.0399	\\ 
 &  &  &    \\
$\bf_{20}^{40}Ca/\bf_{18}^{40}Ar$ &0.0074/0.0063	 &-0.0361& 0.0520	\\ 
 &  &  &    \\
$\bf_{26}^{56}Fe$ &0.0089	          &-0.0347& 0.0502	\\  
 &  &  &   \\
$\bf_{82}^{208}Pb/ \bf_{79}^{197}Au$ &0.0189/0.0185	  &-0.0360& 0.0627	\\ 
 &  &  &     \\
\hline  
\end{tabular}
\caption{ 
The intercepts (GeV) at $\vec {q_3^2}$=0 and slopes (GeV/GeV$^2$) of  fits to  $U_{FSI}$ versus $(\vec q_3+\vec k)^2$. 
  The overall  systematic errors on  $U_{FSI}$ are estimated at $\pm$0.005 GeV.}	
\label{U-FSI} 					
\end{center}					
\end{table}
%                            section 7
 \section{Implementation for neutrino experiments}
  For QE scattering of $neutrinos$ ($antineutrinos$) on bound neutrons (protons) the final state nucleon is a proton (neutron). The following equations should be used in neutrino/antineutrino MC generators:
  
  For neutrino QE  scattering on bound neutrons:
  \begin{equation}
   \label{nu-equation}
    \nu +(M_N-\epsilon^N)= \sqrt{{\bf ( k + \vec q_3)^2} + M_P^2} -|U_{FSI}| +|V_{eff}^P|
  \end{equation}
  where $|V_{eff}^P|=|V_{eff}|$.
  
  For antineutrino QE scattering on bound protons:
  \begin{equation}
  \label{nubar-correct}
    \nu +(M_P-\epsilon^P)= \sqrt{{\bf ( k + \vec q_3)^2} + M_N^2} -|U_{FSI}|
  \end{equation}
  Where
  \begin{equation}
   \epsilon^{N,P}= S^{N,P}+\langle E_x^{N,P}\rangle + \frac { \langle k^2 \rangle}{2M_{A-1}^*}
  \end{equation}
  Rearranging, we have
  \begin{equation}
  \nu +(M_{N,P}-x^{\nu,\bar\nu}) =   \sqrt{{\vec (k + \vec q_3)^2} + M_{P,N}^2}
  \end{equation}
  where for neutrinos and antineutrinos we have:
  \begin{eqnarray}
    x^{\nu}((\vec q_3+\vec k)^2)&=& \epsilon^N- |U_{FSI}|+|V_{eff}^P| \\
    x^{\bar\nu}((\vec q_3+\vec k)^2)&=& \epsilon^P- |U_{FSI}| 
    \label{xnu-nubar}
  \end{eqnarray}
and
  \begin{eqnarray}
    &Q^2&=-m_{\mu,\bar\mu}^2+2E_{\nu,\bar\nu} (E_{\mu,\bar\mu}^{\prime}-\sqrt{ (E_{\mu,\bar\mu}^{\prime}) ^2-m_{\mu,\bar\mu}^2}\cos\theta_{\mu,\bar\mu})\nonumber\\
   &E_{\mu,\bar\mu} &=E_{\nu,\bar\nu} - \nu~; ~~~~~~~  \vec q_3^2=Q^2+\nu^2 \nonumber\\
     &E^\prime_{\mu} &= E_{\mu} + |V_{eff}|~;~~~~~~~E^\prime_{\bar \mu} = E_{\bar \mu} - |V_{eff}| \nonumber\\
 &E_f^{P,N} &= \nu - \epsilon^{N,P} 
  \end{eqnarray}
For both neutrinos and antineutrinos  $\epsilon^{P,N}$ is the unobserved removal energy. 

In neutrino experiments in which both the final state lepton and final state proton (or neutron) are 
measured (e.g. \textsc{nova}, \minerva, \textsc{dune}) the neutrino energy can be calculated as follows:
  \begin{eqnarray}
E_{\nu,\bar \nu}&=&  E_{\mu,\bar\mu} +  E_f^{P,N}  + \epsilon ^{N,P} \nonumber \\
\epsilon ^{N,P} &= &S^{N,P}+\langle E_x^{N,P}\rangle + T_{A-1}^{N,P}\\
T_{A-1}^{N,P}& = & \frac{\langle (k^{N,P})^2\rangle}{2 M_{A-1}} =\frac{3}{5} \frac {(k_F^{N,P})^2}{2M_{A-1}} \nonumber
  \end{eqnarray}
%
%
%                   section 8.
 \section {Corrections to   $\textsc{GENIE}$ version 2}
  The generation of events in  $\textsc{genie}~2$ as currently done is equivalent to using equations \ref{nu-equation} and \ref{nubar-correct}, but with 
  $V_{eff}=0$,  $U_{FSI}=0$, and  $\langle E_x^{P,N}\rangle$=0.  In addition, an amount  $\Delta^\text{nucleon}_\textsc{genie}$ (25 MeV for $^{12}_6C$) is subtracted from  the  energy of the final state nucleon (or quark for inelastic events) to account for ``binding energy'' in $\textsc{genie}~2$.  For neutrino QE scattering on bound neutrons events are  generated in $\textsc{genie}~2$  using the following equations:
  \begin{eqnarray}
  \nu_\textsc{genie}^{\nu,\bar \nu} +(M_{N,P}- x_\textsc{genie}^{\nu,\bar \nu}) &=&   \sqrt{{\bf ( k + \vec q_3)^2} + M_{P,N}^2} \nonumber \\
  \epsilon_\textsc{genie}^{\nu,\bar \nu} = x_\textsc{genie}^{\nu,\bar \nu};~~~~~~~x_\textsc{genie}^{\nu,\bar \nu} &=& S^{N,P}+ \frac { \langle k^2 \rangle}{2M_{A-1}}  
%  \epsilon_\textsc{genie}^{\nu,\bar \nu} &=& x_\textsc{genie}^{\nu,\bar \nu} \nonumber
 \end{eqnarray}
\noindent  Where $  \epsilon_\textsc{genie}^{\nu,\bar \nu}$ is the removal energy for neutrino (antineutrino) assumed in  $\textsc{genie}$ . Therefore the difference between the correct muon energy and the
  muon energy generated by   $\textsc{genie}$ is approximately  equal to $\Delta x^{\nu}_\textsc{genie}= x^{\nu}-x_\textsc{genie}$.
  \begin{eqnarray}
    \label{genie-corr}
  \Delta x^{\nu}_\textsc{genie}& =& \langle E_x^N \rangle -|U_{FSI}| +|V_ {eff}^P|  \\
  \Delta x^{\bar \nu}_\textsc{genie}& =& \langle E_x^P \rangle -|U_{FSI}|  \nonumber\\
  \Delta \epsilon_\textsc{genie}^{\nu,\bar \nu} &=&  \langle E_x^{N,P}  \rangle \nonumber \\
 \Delta \nu_\textsc{genie} ^{\nu,\bar \nu}&=& \nu^{\nu,\bar \nu}-\nu_\textsc{genie}^{\nu,\bar \nu}\nonumber\\ &\approx&    \Delta x^{\nu,\bar \nu}_\textsc{genie} \nonumber\\
 \Delta (E_\mu)_\textsc{genie} &= & E_\mu-(E_\mu)_\textsc{genie}\nonumber\\
			       &=&  -  \Delta \nu_\textsc{genie}^{\nu,\bar \nu} \nonumber \\
 \Delta(E_f^{P,N})_\textsc{genie} & = & E_f^{P,N}- (E_f^{P,N})_\textsc{genie}\nonumber \\
 &= &  \Delta \nu_\textsc{genie}^{\nu,\bar \nu} -\langle E_x^{N,P} \rangle+\Delta^\text{nucleon}_\textsc{genie}
 \nonumber
 \end{eqnarray}
  Where  $\Delta^\text{nucleon}_\textsc{genie}$= 25 MeV is used in $\textsc{genie}$.  
  
As given in the Tables  \ref{TMonizNew} and   \ref{TMoniz2}, 
 $\langle E_x^{P,N} \rangle$ for Carbon is equal to 10.1 MeV and 10.0 MeV for protons and neutrons, respectively and $|V_{eff}|$=3.1 MeV.
  
 For Oxygen $\langle E_x^{P,N} \rangle$ is equal to 10.9 MeV and 10.2 MeV for protons and neutrons, respectively, and   $|V_{eff}|$=3.4 MeV.
 
 For Argon  $\langle E_x^{P,N} \rangle$ is equal to 17.8 MeV and 21.8 MeV for protons and neutrons, respectively, and   $|V_{eff}|$=6.3 MeV.

Tables  \ref{errorC12} and  \ref{errorO16}   show the differences between the correctly simulated  muon,  final state nucleon and removal (unobserved)  energies and those generated by   $\textsc{genie}$ 2 for QE events in carbon, and oxygen respectively. These differences are shown for the
case of  $\vec q_3^2$ = 0.2 GeV$^2$ 	 
 ($|U_{FSI}|$ = 20 MeV)  and for  $\vec q_3^2$=  0.8 GeV$^2$  ($|U_{FSI}|$ = 0). 
\begin{table}[]
\label{errorC12} 
 %[ht]					
 \begin{center}					
\begin{tabular}{|cc|ccc|}					
\hline					
Carbon	&$\vec q_3^2$ & $\Delta E_{\mu+,\mu-}$ &$\Delta E_f^{P,N}$ & $\Delta \epsilon^{P,N}$ \\
MC&GeV$^2$&$MeV$ &$MeV$& MeV \\  
\hline \hline
$U_{FSI}$= 20.0 MeV &0.2	&  & & \\ 
$\textsc{genie}~\nu$ &	&+6.9  &+8.1& +10.0 \\ 
$\textsc{genie}~\bar\nu$ &	& +9.9 & +5.0& +10.1\\  
\hline
$U_{FSI}$= 0.0 MeV &0.8	&  & &\\ 
$\textsc{genie}~\nu$ &&       -13.1	& +28.1& +10.0 \\ 
$\textsc{genie}~\bar\nu$ &	&-10.1  & +25.0 & +10.1 \\ 
%\hline \hline
%$U_{FSI}$= 20.0 MeV &0.2	&  & &  \\ 
%$\textsc{neut}~\nu$ &	& -1.7 &-16.9& +18.6 \\  
%$\textsc{neut}~\bar\nu$ &	& +4.0 &-20.0 & +16.0 \\
%\hline
%$U_{FSI}$= 0.0 MeV &0.8	&  && \\ 
%$\textsc{neut}~\nu$ &	&  -21.7&+3.1&+18.6 \\  
%$\textsc{neut}~\bar\nu$ &	&  -16.0 &0.0& +16.0 \\
\hline  \hline
\end{tabular}
\caption{ Estimates of the  difference between the correctly simulated  muon, final state nucleon and removal (unobserved)  energies and those generated by   $\textsc{genie}~2$ for QE events in  carbon.   
% Here,  $E_x^{P,N}$=10.1 and 10.0 MeV,  $V_{eff}=3.1$ MeV,  $\Delta^\text{nucleon}_\textsc{genie}$=25 MeV,
% ($\epsilon_{SM}^{\prime {P,N}})_\textsc{neut}$=27 MeV,  $\epsilon^{\prime P.N}_{SM}$= 43.0 and 45.6 MeV, respectively, and  $\epsilon^{P,N}$= 27.5 and 30.1 MeV, respectively.
 }						
\end{center}					
\end{table}
\begin{table}[]
 %[ht]	
 \label{errorO16} 				
 \begin{center}					
\begin{tabular}{|cc|ccc|}					
\hline					
Oxygen	&$\vec q_3^2$ & $\Delta E_{\mu+,\mu-}$ &$\Delta E_f^{P,N}$ & $\Delta \epsilon^{P,N}$ \\
MC&GeV$^2$&$MeV$ &$MeV$& MeV \\  
\hline \hline
$U_{FSI}$= 20.0 MeV &0.2	&  & & \\ 
$\textsc{genie}~\nu$ &	&+6.4  &+8.4& +10.2 \\ 
$\textsc{genie}~\bar\nu$ &	& +9.1 & +5.0 & +10.9 \\  
\hline
$U_{FSI}$= 0.0 MeV &0.8	&  && \\ 
$\textsc{genie}~\nu$ &&       -13.6	& +28.4& +10.2 \\ 
$\textsc{genie}~\bar\nu$ &	&-10.9  & +25.0&  +10.9 \\ 
%\hline \hline
%$U_{FSI}$= 20.0 MeV &0.2	&  & &\\ 
%$\textsc{neut}~\nu$ &	& +0.9 &-16.9 &+16.0 \\  
%$\textsc{neut}~\bar\nu$ &	& +6.9 &-20.0&  +13.1 \\
%\hline
%$U_{FSI}$= 0.0 MeV &0.8	&  &&  \\ 
%$\textsc{neut}~\nu$ &	&  -19.1&+3.1& +16.0 \\  
%$\textsc{neut}~\bar\nu$ &	&  -13.1& 0.0&+13.1 \\
\hline  \hline
\end{tabular}
\caption{ Difference between the correctly simulated  muon,  nucleon and removal (unobserved)  energies and those generated by   $\textsc{genie}~2$ 
 for QE events in Oxygen. 
% Here,  $E_x^{P,N}$=10.9 and 10.2 MeV,  $V_{eff}=3.4$ MeV,  $\Delta^\text{nucleon}_\textsc{genie}$=25 MeV, 
%$(\epsilon_{SM}^{\prime P,N})_\textsc{genie}$=27 MeV, and    $\epsilon^{\prime P,N}_{SM}$= 40.1 and 43.0 MeV, respectively, and  $\epsilon^{P,N}$= 24.1 and 27.0  MeV, respectively.
}							
\end{center}					
\end{table}
 %            %subsection 9
  \section {Conclusion}
 We investigate the  binding energy  parameters that should be used in modeling  electron and neutrino scattering from nucleons bound in a nucleus within the framework of the impulse approximation. We discuss the relation between binding energy, missing energy,  removal energy ($\epsilon$),   spectral functions and  shell model energy levels and extract updated removal energy parameters from  ee$^{\prime}$p spectral function data. We address the difference in parameters for scattering from bound protons and neutrons. We also use inclusive e-A data to extract an  empirical parameter  $U_{FSI}((\vec q_3+\vec k)^2)$  to account for the interaction of final state nucleons (FSI) with the optical potential of the nucleus. Similarly we use  $V_{eff}$ to account for the  Coulomb potential  of the nucleus. 
 
With  three parameters  $\epsilon$,  $U_{FSI}((\vec q_3+\vec k)^2)$and $V_{eff}$ we can describe the energy of final state electrons for all available electron QE scattering data. The use of the updated parameters in neutrino Monte Carlo generators reduces the systematic uncertainty in the combined removal energy (with FSI corrections) from $\pm$ 20 MeV to $\pm$ 5 MeV.
%
% Appendix A
\appendix
\section {Appendix: Coulomb corrections}
\label{Coulomb}
For targets with atomic number Z greater than one we should take into
  account the effect of the  electric field of the nucleus on
  the incident and scattered electrons (and also on the final state proton
  in QE events).  These corrections are
  called Coulomb corrections. For atomic weight A and
  atomic number Z the protons create an electrostatic potential V(r). 
   In the effective momentum approximation (EMA), 
  the effective potential  for an incident electron is  $V_{eff}$, which can
  be calculated as follows:
  \begin{eqnarray}
V (r) &=& \frac{3\alpha (Z) }{2R}+\frac{r \alpha (Z) }{2R^2} \nonumber \\
  R&=&1.1 A^{1/3}+0.775 A^{-1/3} \nonumber  \\
  \label{coul}
  V_{eff}&=&-0.8 V (r=0) = -0.8 \frac{3\alpha (Z) }{2R}. 
\end{eqnarray}
The values for $|V_{eff}|$ calculated from equation \ref{coul}  agree (within errors)  with values extracted from a comparison of 
the  {\it peak} positions and cross sections of positron and electron QE scattering\cite{gueye}.
For our estimates of $|V_{eff}|$  shown in Table \ref{TMonizNew} we use the experimental values for the  nuclei
    that were measured in ref.\cite{gueye}. We use equation \ref{coul} to interpolate to other nuclei.
   %shown in Table \ref{TMoniz}.
  
For electrons scattering on bound nucleons  the  effective incident energy is $E_{eff}=E_0+|V_{eff}|$,
  and the effective scattered energy is $E^\prime_{eff}= E^\prime+|V_{eff}|$. This implies that the effective square of the momentum transfer is increased.  
  For positrons scattering on bound nucleons  the  effective incident energy is $E_{eff}=E_0-|V_{eff}|$,
  and the effective scattered energy is $E'_{eff}= E^\prime-|V_{eff}|$. This implies that the effective square of the momentum transfer is decreased.  
  
 For electron QE scattering on bound protons $|V_{eff}^P|=\frac{Z-1}{Z}|V_{eff}|$.   For neutrino QE scattering on bound neutrons
 $|V_{eff}^P|= |V_{eff}|$.   For neutrino QE scattering on bound protons
 $|V_{eff}^N|= 0$.

For completeness, though not relevant in this analysis,   there is also a focusing factor $F_{foc}=(\frac{E_0+|V_{eff}|}{E_0})^2$ that enhances the cross section for electrons and reduces the cross section for positrons.   The focussing factor cancels the $1/E_0^2$ factor in the Mott cross section.  Therefore, the  Coulomb correction  should only be applied to the structure functions $W_1$ and $W_2$. 
%           Appendix B
\section{Appendix: Relativistic Fermi Gas (RFG)}
\label{Arfg}
For the Fermi gas model the   momentum distribution is zero for   $k>k_F$, and  for $k<k_F$ it is given by
\begin{eqnarray}
|\phi(k)|^2&=& \frac {1}{N_{rfg}} % \nonumber \\, 
~~~~~~~~N_{rfg}= \frac {4}{3} {\pi }  K_{F}^{3}  \nonumber \\
P_{rfg}(k)dk &=&  |\phi(k)|^2 4\pi \vec k^2 dk   =  \frac {1}{N_{rfg}} {4\pi \vec k^2 dk, } 
\label{eqFermi}
\end{eqnarray}
\noindent and $\langle \vec k^2\rangle= (3/5)k_F^2$. 
%       B.1
\subsection{Distributions and parameters of RFG versus $k_z$}
Here we do the calculation in  cylindrical coordinates 
%\begin{eqnarray} 
%k_z&=&k \cos\theta \nonumber \\
$$(2\pi \vec k^2~d\cos\theta ~dk =  \pi dk_T^2~ dk_z)$$
$$k=\sqrt{k_T^2+k_z^2}.$$
%Here   $P(k_z)=0$  for $k_z>k_F$ ($\xi>2$)  and  $k_z<-k_F$ ($\xi<-2$)
 the probability distribution of the Z component
of the momentum $k_z$  $P(k_z)~dk_z$,  and average square of the  transverse momentum 
 $\langle k_T^2 (k_Z)\rangle$ as a function of $k_z$ are given below.   
\begin{eqnarray}
P(k_z)_{rfg} dk_z &=&\frac{3(1-k_z^2 /k_F^2)}{4k_F} dk_z  \\
 \langle k_T^2(k_z)_{rfg}\rangle&=& \frac{1}{2}{k_F^2}(1-k_z^2 /k_F^2) \\
 \label{kq32peak}
 \langle \vec k^2(k_z)_{rfg}\rangle&=&\langle k_T^2(k_z)_{rfg}\rangle+k_z^2
\label{Fermi}
\end{eqnarray}
%
%\subsection{Distributions and parameters of RFG versus $\xi$}
%
% The same distribution in terms of the variable $\xi$ where $\xi=2(k_z/k_F$), and 
% $\xi$ ranges from 2 to -2 are given below.
% \begin{eqnarray}
%P(\xi)_{rfg}d\xi &= &\frac{3}{8} (1-\frac{\xi^2} {4})d\xi    \\
%\langle k_T^2(\xi))_{rfg}\rangle&=& \frac{1}{2}{k_F^2}(1-\xi^2/4)\\
% \langle \vec k^2(\xi)_{rfg}\rangle &=&\langle k_T^2(\xi)\rangle+\frac{k_F^2}{4}\xi^2
%\label{Fermixi}
%\end{eqnarray}
%
 %      B.2
\subsection{Pauli Blocking}
We multiply the QE differetial cross sections by a  Pauli blocking factor $K_{Pauli}^{nuclei}(\vec q_3^2)$
%done by a  $\vec q_3^2$-dependent multiplicative Pauli suppression factor ($K_{Pauli}^{nuclei}$).
which   reduces the predicted cross sections at low $\vec q_3^2$. The Pauli suppression
factor shown below is from  Eq. B54 of reference \cite{Tsai}.
% Y. S. Tsai, Rev. Mod. Phys. 46, 815 (1974). 123
\begin{equation}
%K_{Pauli}^{nuclei} =(3/4) (|\vec q|/ k_F)(1 - (|\vec q|/ k_F)^2)/12)\\
K_{Pauli}^{nuclei} =\frac{3}{4} \frac{|\vec q_3|}{k_F}(1 - \frac{1}{12} (\frac {|\vec q_3|}{k_F})^2)
\label{nucpauli}
\end{equation}
For $ |\vec q_3| < 2k_F$, otherwise no Pauli suppression correction is made.  Here  $ |\vec q_3| =\sqrt{Q^2+\nu^2}$ is the absolute magnitude of the 3-momentum transfer to the target nucleus,
% and  $Q^2=-q^2$ is the square of the four-momentum transfer to the target  nucleus. 
% For scattering from a stationary nucleon  $Q^2= 2M\nu$ and $|\vec q|=\sqrt{Q^2(1+Q^2/4M^2)}$, where M is the mass of the nucleon. 
%The vector  $k$ is the initial momentum of the neutron.
%
 %           Appendix C
\section{Reconstruction of $E_\nu^{QE{\text -}\mu}$,  $Q^2_{QE{\text -}\mu}$ and $Q^2_{QE{\text -}P}$}
\label{neutrino}
In this section we update the expressions  for  the  {\it mean}  reconstructed neutrino energy $E_\nu^{QE{\text -}\mu}$ and  
 square of the four-momentum transfer 
$Q^2_{QE{\text -}\mu}$  extracted only from the kinematics of  final state muons in QE events.
 In addition we can also reconstruct the four momentum transfer $Q^2_{Q{\text -} (P,N) }$
from the kinematics of the final state recoil proton or neutron  in QE events. 

The expressions are updated to include: 
\begin{enumerate}
\item The contribution of final state interaction $|U_{FSI}|$.
\item  The contribution of Coulomb corrections $V_{eff}$.
\item  The contribution of the proton and neutron transverse momentum $\vec k_{T}$ at the location of the QE  {\it peak}.
\end{enumerate}
\noindent In the derivation of the expressions we use relativistic kinematics. 
The "primed" energies and momenta are at the vertex before FSI with the nuclear and Coulomb field.  
\begin{eqnarray}
\label{xnu-nubarA}
E_i^{N,P}&=&M_{N,P} -  \epsilon^{N,P}   \\
E_f^P &=&\sqrt {(\vec k+\vec q_3)^2+M_P^2)} -|U_{FSI}| +|V_{eff}^P|   \\
            &=& E_F^{P\prime} -|U_{FSI}| +|V_{eff}^P| \nonumber \\
E_f^N &=&\sqrt {(\vec k+\vec q_3)^2+M_N^2)} -|U_{FSI}|   \\
            &=& E_F^{N\prime} -|U_{FSI}| \nonumber  \\
E_{\mu{\text -}} &= & E_{\mu{\text -}}^{\prime} -|V_{eff}^P|,~~~~~  E_{\mu{\text +}}  =E_{\mu{\text +}}^{\prime} + |V_{eff}|       
\end{eqnarray}
%Hence,  the  {\it mean}   $  \langle {\epsilon_R^N} \rangle$  and  $  \langle {\epsilon_R^P} \rangle$.
% should be used for  the nuclear binding energy  parameters  for  QE scattering  on neutrons and protons, respectively. 
%
For neutrino scattering on bound neutrons $  |V_{eff}^P|= |V_{eff}|$.
 We define  $M_N$, $M_P$, $m_\mu$ as the neutron, proton and muon masses. 
At the peak location of the QE  distribution the  bound neutron momentum is perpendicular to  $\vec q$ (i.e. $k_z$=0).
In this case, the average of the square of transverse momenta
of the neutron  (proton)  for a Fermi gas momentum distribution (and also for a Gaussian distribution)  is 
$\langle\vec k_{T{\text -}N}^2\rangle=\frac{ (k_F^N) ^2}{2}$ for 
a bound neutron in the initial state and $\langle\vec k_{T{\text -}P}^2\rangle=\frac{ (k_F^P) ^2}{2}$)  for bound proton in the initial state.
%
%           subsection  C.1
\subsection{Using only the kinematics of the  $\mu^-$} 
For $neutrino$  QE  events we define $E_{\mu{\text -}}^{\prime} = T_{\mu{\text -}} + m_\mu +|V_{eff}|$ as the total  Coulomb corrected muon energy. 
We define  $ (M_P^{\prime}) ^2=M_P^2 +\langle\vec k_{T{\text -}N}^2\rangle$ to account for the fact that the final state $proton$
has  the same average transverse momentum as that of the initial state $neutron$ 
  $\langle\vec k_{T{\text -}N}^2\rangle$ with respect to the neutrino-muon
scattering plane.
From energy-momentum conservation we get:
\begin{eqnarray}
\label{e_p_cons}
E_\nu&=&p^\prime_{\mu{\text -}}\cos\theta_{\mu{\text -}} +  P^\prime_{P}\cos\theta_P \\
p^\prime_{\mu{\text -}} \sin\theta_\mu&=&P^\prime_P \sin\theta_P\nonumber\\
E_\nu+M_N -\epsilon^N &=& [\sqrt{ (P_P^\prime) ^2+ (M_P^{\prime}) ^2}] - |U_{FSI}| + |V_{eff}^P| \nonumber \\
                                   &+& E^\prime_{\mu{\text -}} - |V_{eff}| \nonumber\\
   E_\nu+ M_N^{\prime \prime} & =&    \sqrt{ (P_P^\prime) ^2+ (M_P^{\prime}) ^2}+ E^\prime_{\mu{\text -}} \nonumber \\
   M_N^{\prime \prime}& = &M_N -(\epsilon^N - |U_{FSI}|). \nonumber                        
\end{eqnarray}
Here. for neutrino scattering on bound neutrons $|V_{eff}^P|= |V_{eff}|$, 
$|U_{FSI}= |U_{FSI}(\vec q_3^2+\frac{1}{2}k_F^2)|$, 
 $p^\prime_{\mu{\text -}}= \sqrt{ (E^\prime_{\mu{\text -}})^2-m_\mu^2}$,
$E_{\mu{\text -}} =  E_{\mu{\text -}}^{\prime} -|V_{eff}|$, 
$P_P^\prime$ is the momentum
of the final state $proton$ (before FSI) in the $neutrino-muon$ plane, and $\theta_P$ is the angle of the
proton in the  $neutrino-muon$ scattering plane.  From equations \ref{e_p_cons}  we obtain the following expressions. 
\begin{eqnarray}
\label{QEequ}
E_\nu^{QE{\text -}\mu}&=& \frac{2 (M_N^{\prime \prime}) E_{\mu{\text -}}- ((M_N^{\prime \prime}) ^2+m_\mu^2- (M_P^{\prime}) ^2) }
{2\cdot[ (M_N^{\prime \prime}) -E_{\mu{\text -}} + \sqrt{ (E_{\mu{\text -}}) ^2-m_\mu^2}) \cos\theta_{\mu{\text -}}]} \nonumber\\
Q^2_{QE{\text {\text -}}\mu} 
&=&-m_\mu^2+2E_\nu^{QE} (E_{\mu{\text -}}^{\prime}-\sqrt{ (E_{\mu{\text -}}^{\prime}) ^2-m_\mu^2}\cos\theta_{\mu{\text -}}). \nonumber\\
\vec {q_3^2} &=& Q^2+(E_\nu^{QE{\text -}\mu}-E_{\mu{\text -}})^2
%M_N^{\prime \prime}& = &M_N -(\epsilon - |U_{FSI}(\vec q_3^2)| )
\nonumber
\end{eqnarray}
Note that because $ |U_{FSI}(\vec q_3^2+\frac{1}{2}k_F^2)|$ is $\vec q_3^2$ dependent, the above expressions should  be solved iteratively,
or an average value corresponding to the mean $\vec q_3^2$ should be used.
\subsection{Using only the kinematics of the  $\mu^+$} 
\label{mu_kinematics}
  For $antineutrino$ QE  events we define $E_{\mu+}^{\prime} = T_{\mu+} + m_\mu -|V_{eff}|$ as the total  Coulomb corrected muon energy. 
We define  $ (M_N^{\prime}) ^2=M_N^2 +\langle\vec k_{T{\text -}P}^2\rangle$ to account for the fact that the final state $neutron$ has  the same average transverse momentum as that of the initial state $proton$   $\langle\vec k_{T{\text -}P}^2\rangle$ with respect to the $antineutrino-muon$
scattering plane.
From energy-momentum conservation we get:
\begin{eqnarray}
\label{e_n_cons}
E_{\bar \nu}&=&p^\prime_{\mu{\text +}}\cos\theta_{\mu{\text +}} +  P^\prime_{N}\cos\theta_N \\
p^\prime_{\mu{\text +}} \sin\theta_\mu&=&P^\prime_N \sin\theta_N\nonumber\\
E_{\bar \nu}+M_P -\epsilon^P &=& [\sqrt{ (P_N^\prime) ^2+ (M_N^{\prime}) ^2}] - |U_{FSI}|  \nonumber \\
                                   &+& E^\prime_{\mu{\text +}} + |V_{eff}| \nonumber\\
   E_{\bar \nu}+ M_P^{\prime \prime} & =&    \sqrt{ (P_N^\prime) ^2+ (M_N^{\prime}) ^2}+ E^\prime_{\mu{\text +}} \nonumber \\
   M_P^{\prime \prime}& = &M_P -(\epsilon^P -  |U_{FSI}| + |V_{eff}|). \nonumber                               
\end{eqnarray}
Here, $|U_{FSI}= |U_{FSI}(\vec q_3^2+\frac{1}{2}k_F^2)|$, 
 $p_{\mu+}=\sqrt{ (E_{\mu+} ^2-m_\mu^2}$, $E_{\mu{\text +}}  =E_{\mu{\text +}}^{\prime} + |V_{eff}|$,      
 $P_N^\prime$ is the momentum
of the final state $neutron$  (before FSI) in the $antineutrino-muon$ scattering plane, and $\theta_N$ is the angle of the
neutron in the  neutrino-muon plane.  From equations \ref{e_n_cons}  we obtain the following expressions. 
\begin{eqnarray}
E_{\bar \nu}^{QE{\text -}\mu+}&=& \frac{2 (M_P^{\prime \prime}) E^\prime _{\mu+} - ((M_P^{\prime \prime}) ^2+m_\mu^2- (M_N^{\prime}) ^2) }
{2\cdot[ (M_P^{\prime \prime}) -E^\prime_{\mu+} + \sqrt{ (E^\prime_{\mu+})^2-m_\mu^2}) \cos\theta_{\mu+} ]} \nonumber\\
Q^2_{QE{\text -}\mu+} &=&-m_\mu^2+2E_{\bar \nu}^{QE} (E_{\mu+}^{\prime}-\sqrt{ (E_{\mu+}^{\prime}) ^2-m_\mu^2}\cos\theta_{\mu+}).\nonumber\\
\vec {q_3^2} &=& Q^2+(E_{\bar \nu}^{QE{\text -}\mu+}-E_{\mu{\text +}})^2 
\end{eqnarray}
%where  $M_P^{\prime \prime}= M_P -   \langle {\epsilon_R^N} \rangle $. 
Note that because $ |U_{FSI}(\vec q_3^2)+\frac{1}{2} k_F^2)|$ is $\vec q_3^2$ dependent, the above expressions should  be solved iteratively,
or an average value corresponding to the mean $\vec q_3^2$ should be used.
\subsubsection{Using only the kinematics of the final state nucleon}
For  $neutrino$  QE  events  the average reconstructed  $Q^2_{QE{\text -}P}$ 
 can  be extracted  from final state $proton$ variables only by using  following expression: 
\begin{eqnarray}
 M_P^2 &=&  (q+E_i^N)^2 = -Q^2+2(M_N-\epsilon^N)\nu+(M_N-\epsilon^N)^2 \nonumber \\
 Q^2_{QE{\text -}P} &= &(M_N-\epsilon^N)^2 - M_P^2 \nonumber \\
 &+&2(M_N-\epsilon^N) [M_P+T^P-(M_N-\epsilon^N)] 
%Q^2_{QE{\text -}P} &=& (M_N^{\prime \prime}) ^2- (M_P^{\prime}) ^2+2M_N^{\prime \prime}[M_P+T^P-M_N^{\prime \prime}] \nonumber\\
\end{eqnarray}
For  $antineutrino$  QE  events  the average reconstructed  $Q^2_{QE{\text -}N}$ 
 can  be extracted  from final state $neutron$ variables only by using  following expression: 
 \begin{eqnarray}
 M_N^2 &=&  (q+E_i^P)^2 = -Q^2+2(M_P-\epsilon^P)\nu+(M_P-\epsilon^N)^2 \nonumber \\
 Q^2_{QE{\text -}N} &= &(M_P-\epsilon^P)^2 - M_N^2 \nonumber \\
 &+&2(M_P-\epsilon^P) [M_N+T^N-(M_P-\epsilon^P)] 
 \end{eqnarray}
\subsection{Comparison to previous  analyses}
If we set $\vec k_T^2=0$, $U_{FSI}=0$, and $|V_{eff}|=0$, the  above equations  are reduced to the equations used in previous analyses
except  that  $  x^{\nu}$ and $  x^{\bar \nu}$  (equation \ref{xnu-nubar})  are used.


\begin{thebibliography}{}
% ref 1
\bibitem{genie}  C. Andreopoulos [$\textsc{genie}$], Acta Phys. Polon. B 40, 2461 (2009); ibid Nucl. Instr. Meth.A614, 87, 2010
%ref 2
\bibitem{neugen}H. Gallagher ($\textsc{neugen}$), Nucl. Phys. Proc. Suppl. 112 (2002)
%ref 3
\bibitem{neut}Y. Hayato ($\textsc{neut}$), Nucl. Phys. Proc. Suppl. 112, 171 (2002)
% ref 4
% \bibitem{nuance}  D. Casper ($\textsc{nuance}$), Nucl. Phys. Proc. Suppl. 112, 161 (2002).  (http://nuint.ps.uci.edu/nuance/).
\bibitem{nuwro}J. Sobczyk (NuWro), PoS $\textsc{nufact}$08, 141 (2008);  C. Juszczak, Acta Phys. Polon. B40 (2009)  2507 (http://borg.ift.uni.wroc.pl/nuwro/)
% ref 5 
\bibitem{gibuu}T. Leitner, O. Buss, L. Alvarez-Ruso, U. Mosel (GiBUU), Phys. Rev. C79, 034601 (2009) (arXiv:0812.0587)
% ref 6
    \bibitem{benhar} Omar Benhar, Patrick Huber, Camilo Mariani, Davide Meloni, Physics Reports 700, 1 (2017); 
    %Neutrino-nucleus unobserveds and the determination of oscillation parameters 
    %(nucl-th/1501.06448); 
    Omar Benhar, Donal day, Ingo Sick,Rev.Mod.Phys. 80, 189 (2008);
    %   (nucl-ex/0603029). 
     O. Benhar and S. Fantoni and G. Lykasov, Eur Phys. J.  A 7 (2000), 3, 415;  O. Benhar et al., Phys. Rev. C55. 244 (1997)
% ref 7
  \bibitem{spectral-theory2} A. M. Ankowski and J. T. Sobczyk, Phys. Rev. C 74, 054316 (2006)
  % ref 8
  \bibitem{Sakuda}   	S. X. Nakamura et. al, Reports on Progress in Physics 80, 056301 (2017) 
  %ref 9
 \bibitem{optical} A M. Ankowski, Omar Benhar, and Makoto Sakuda,  Phys. Rev. D 91, 033005 (2015)
  %arXiv:1610.01464v3 [nucl-th]
  % ref 10
 \bibitem{optical1}O. Benhar, Phys. Rev. C87, 024606 (2013); Y. Horikawa~et al., Phys. Rev. C22, 1680 (1980)
 % Cooper et al., Phys. Rev. C47, 297 (1993)
 % ref 11
\bibitem{electron}E. J. Moniz, et al., Phys. Rev. Lett. 26, 445 (1971);  E. J. Moniz, Phys. Rev. 184, 1154 (1969);
       R. R. Whitney et al. Phys, Rev. C9, 2230 (1974)
      %12 
          \bibitem{neutrino} R. A. Smith and E. J. Moniz, Nucl. Phys B43, 605 (1972)
  %13
           \bibitem{T2K} S. Dennis (\textsc{t2k}), talk at  \textsc{nufact}  2018, Virginia Tech, Blacksburg, VA 
             https://indico.phys.vt.edu/event/34/contributions/610/
     %14     
            \bibitem{t2k-impact}   Simon Bienstock (2018). Studying the impact of neutrino cross-section mismodelling on the  \textsc{t2k} oscillation analysis. 
            https://zenodo.org/record/1300504; ibid  \textsc{neutrino}  2018 poster  https://indico.desy.de/indico/event/ 
            18342/session/35/contribution/173    
             %     ?The T2K collaboration finds that, for the statistics of the 2018 data set, a shift of 20 MeV in the binding energy parameter introduces a bias of 20% for sin2(th23) and 40% for dm232 with respect to the size of the systematics errors, assuming maximal sin2(th23).?%\
       % 15     
              \bibitem{gueye} P. Gueye et al.  Phys. Rev. C60, 044308 (1999)
            %16  
              %  K. Abe, et al., Phys.Rev.Lett. 112, 181801 (2014).
            \bibitem{combined}  Christoph Andreas Ternes, talk at \textsc{nufact}  2018, Virginia Tech, Blacksburg, VA   
            https://indico.phys.vt.edu/event/34/contributions/760/ 
            %  2.50 +-0.030
            %17
            \bibitem{Donnelly} C. Maieron, T.W. Donnelly, I. Sick,  Phys.Rev. C65 (2002) 025502; 
 %(nucl-th/0109032);   
  J.E. Amaro, M.B. Barbaro, J.A. Caballero,  T.W. Donnelly, A. Molinari, and I. Sick,
   Phys. Rev. C 71, 015501 (2005)
  % arXiv:nucl-th/0409078.       
        %ref 18
       \bibitem{effective} A. Bodek, M. E. Christy and B, Coppersmith, Eur. Phys. J. C74, 3091 (2014)
       %  (hep-ph/1405.0583).
       %19
        \bibitem{radii} W. M. Seif and Hesham Mansour, Int. J. Mod. Phys. E 24, 1550083 (2015)
        %20
          \bibitem{TUNL}  Jun Chen, Nuclear Data Sheets 140, 1 (2017)\\
           http://www.tunl.duke.edu/nucldata;   http://www.nndc.bnl.gov/nudat2/
   %ref 21
   %*************
\bibitem{nuclear-data} Nuclear Data Tables \\
 www.periodicTable.com/Isotopes/083.208/index.html;
    www-nds.iaea.org/relnsd/vcharthtml/VChartHTML.htm
    %  (nucl-th/1504.08350).
%22
 \bibitem{miller} 
 %Nucleon-Nucleon Correlations, Short-lived Excitations, and the Quarks Within
O. Hen, G.A. Miller, E. Piasetzky, L.B. Weinstein,  Rev. Mod. Phys. 89(4), 045002  (2017) 
%DOI: 10.1103/RevModPhys.89.045002
%e-Print: arXiv:1611.09748 [nucl-ex] 
%23
      \bibitem{brown} B. A. Brown, Prog. Part. Nucl. Phys. 47, 517 (2001)
   % ref 24
 \bibitem{review} {\em Modern Topics in Electron Scattering}, Edited by B. Frois and I. Sick (World Scientific, Singapore, 1991).
 %25
 \bibitem{BodekRitchie} A. Bodek and J. L. Ritchie, Phys.Rev. D23, 1070 (1981)
 % 26
    \bibitem{Koltun}  D. Koltun, %Total Binding Energies of Nuclei, and Particle-Removal Experiments, 
 Phys.Rev.Lett. 28, 182 (1972)
    % ref 27
\bibitem{C12} D. Dutta et al, (Jlab Hall C),  Phys. Rev. C 68, 064603 (2003); D. Dutta, PhD Thesis, Northeastern U. (1999) 
% ref 28
  \bibitem{Saclay}  J. Mougey, M. Bernheim, A. Bussière, A. Gillebert, Phan Xuan Hô, M. Priou, D. Royer, I. Sick, G.J. Wagner,
     Nucl. Phys A262 (1976) 461;  S. Frullani and J. Mougey, Adv. Nucl. Phys. 14, 1(1982)
    % The reactions 6, 7Li, 9Be, 10B (e, e?p)  at 700 MeV and DWIA analysis
    % ref 29
\bibitem{Tokyo1}   K. Nakamura, S. Hiramatsu, T. Kamae, H. Muramatsu,  Y. Watase, Nucl. Phys A268, 381 (1976) 
% The reaction 12C (e, e?p)  at 700 MeV and DWIA analysis
% ref 30
    \bibitem{Tokyo2} 
     K. Nakamura,S. Hiramatsu,T. Kamae,H. Muramatsu, Y. Watas, Nucl. Phys. A296, 431 (1978)
     % ref 31
      \bibitem{Tokyo3} 
    K. Nakamura, S. Hiramatsu, T. Kamae, H. Muramatsu,   Y. Watase, Nucl. Phys. A271,  221 (1976)
  %The 27Al, 40Ca and 51V (e, e?p)  reactions and observation of deep hole states
    % \bibitem{Ca40} O. Benhar et al,  arXivnucl-ex/1406.4080. 
    % ref 29
 %    \bibitem{p2p} G. Jacob and T. A. J. Maria, Rev. Mod. Phys. 43, 6 (1973).
 % ref 32
     \bibitem{huberts} P. K. A. de Witt Huberts, J. Phys, G.: Nucl. Part. Phys. (1990)  507;
        A. E. L. Dieperink and  P. K. A. de Witt Huberts, Annu. Rev, Nucl. Part. Sci. 40, 239 (1990)
 %Kess de Jager "Quasi-elastic proton knockout from $^{16}$O".  (Hall A),
 % Ref 33
  \bibitem{NIKHEFC12} G. van der Steenhoven et al., Nucl. Phys. A480 547 (1988); ibid   A484, 445 (1988) 
 %ref 34
 \bibitem{O16}  N. Liyanage et al, (Jlab Hall A), Phys. Rev. Lett. 86, 5670 (2001);  N. Liyanage, PhD Thesis, MIT (1999); 
 K. G. Fissum et al. Phys.Rev. C70, 034606 (2004)
  %35
% \bibitem{Benharsrc}  % Benhar large Emiss average   Nuclear binding, correlations and the origin of EMC effect  C12=52.2,  NM=70.5
%Omar Benhar and Ingo Sick, Nuclear binding, correlations and the origin of EMC effect  arXiv:1207.4595 (2012)    
   % ref 35
   \bibitem{spectral-theory}A. M. Ankowski and J. T. Sobczyk, Phys. Rev. C 77, 044311 (2008) Xiv:0711.2031 [nucl-th]
        % this has levels of O16,  Ca40 Ar40.  I
 %ref 36
 \bibitem{levels} Vautherin and D. M. Brink, Phys. Rev. C5, 626  (1972).
 %"Energy levels of light nuclei A=11", F. Ajzenberg-Selove and C.L. Busch (2014). $http://www.tunl.duke.edu/nucldata/fas/11/1980.pdf$; 
 %"Energy Levels of Light Nuclei A = 15", F. Ajzenberg-Selove (2015)   $http://www.tunl.duke.edu/nucldata/fas/15/1976.pdf$
  %  \bibitem{spectral-proposal} "Measurement of the spectral function of 40Ar through the (e,$e^{\prime}$ p" reaction
%  A.Ankowski et al,  https://arxiv.org/abs/1406.4080
% Ca calc spectral function
 % \bibitem{Tokyo3} K. Nakamura, S. Hiramatsu, T. Kamae, H. Muramatsu, N. lzutsu and Y. Watase, Nucl. Phys. A271, 221 (1976)
 %37
 \bibitem{archive} Quasielastic Electron Nucleus Scattering Archive: http://faculty.virginia.edu/qes-archive/;
 O. Benhar, D. Day and I. Sick,  Reviews of Modern Physics [Rev. Mod. Phys. 80, 189-224, 2008.
 %38-59
   \bibitem{Heimlich:1973} F.H. Heimlich et al, Nuclear Physics A 231, 509 (1979) (Li6 Heimlich:1973); 
F.H. Heimlich et al. Nucl. Phys. A231, 509  (1974) (Li6 Heimlich:1974rk)
   %39
\bibitem{Whitney:1974hr} R. R. Whitney et al. Phys, Rev. C9, 2230 (1974) (Li6,C12,Ca40,Pb208 Whitney:1974hr)
   %40
 \bibitem{Barreau:1983ht}P. Barreau et al Nucl. Phys. 402A , 515 (1983) (C12  Barreau:1983ht) 
 %41
   \bibitem{Sealock:1989nx}  R. Sealock et al. Phys. Rev. Lett., 62, 1350 (1989) (C12, Fe56 Sealock:1989nx)
   %42
 \bibitem{Baran:1988tw} D. Baran et al, Phys. Rev. Lett., 61, 400 (1988) (C12, Fe56  Baran:1988tw)
%43
\bibitem {Bagdasaryan:1988hp}Bagdasaryan, D. S. and others, YERPHI-1077-40-88  (C12, Fe56 Bagdasaryan:1988hp)
%44
\bibitem {Zeller:1973ge} Diethelm Zeller, DESY Internal Report F23-73/2 (1973);  F.H. Heimlich et al. DESY Report 74/20 (1974) (C12 Zeller:1973ge) 
%45
 \bibitem  {Arrington:1995hs}J. Arrington et al., Phys. Rev C 53 (1996) 2248  ( C12, Fe56 Arrington:1995hs)
 %46     
  \bibitem{Fomin:2010ei} Fomin, N. et al. Phys.Rev.Lett. 105 (2010) 212502 (C12  Fomin:2010ei)
  %47 
  \bibitem{Anghinolfi:1996vm6} M. Anghinolfi et al. Nucl. Phys. A602 (1996) 405 (O16 Anghinolfi:1996vm)
 %48
\bibitem{O'Connell:1987ag}   J.S.O'Connell et al. Phys. Rev. C 35 (1987) 1243 (C12, O16 O'Connell:1987ag) 
%49
\bibitem{E12-14-012}  H. Dai et al   arXiv:1810.10575 [nucl-ex] (Ar40 E12-14-012)
%50  
 \bibitem{Anghinolfi:1995}  M. Anghinolfi et, al., J. Phys. G: Nucl. Pan. Phys. 21 (1995)  L9-LI5. (Ar40 Anghinolfi:1995)
 %51   
  \bibitem{Bosted:1992fy} P.Y. Bosted et al., Phys. Rev. C 46, (1992) 2505 (Al27 Bosted:1992fy),   Steve Rock, private comm. (Al27 Rock-pc)
 %52 
   \bibitem{Williamson:1997}  C.F. Williamson et al. Phys. Rev. C. 56 (1997) 3152M and T.C. Yates et al. Phys. Lett. B 312 (1993) 382 (Ca40 Williamson:1997)
  %53
  \bibitem{Hotta:1994}A. Hotta et al. Phys. Rev. C 30 (1984) 87(Fe Hotta:1994)
  %54
   \bibitem{Meziani:1984is} Z.E. Meziani et al. Phys. Rev. Lett. 52 (1984) 2130 (Ca40 Fe56 Meziani:1984is) 
   %55
   \bibitem{Arrington:1998ps}J. Arrington et al., Phys. Rev. Lett. 82, 2056 (1999) (Fe56  Arrington:1998ps)
%56
\bibitem{Chen:1990kq} J. P. Chen et al. Phys. Rev. Lett. 66, 1283 (1991) (Fe56 Chen:1990kq)  
 %57 
     \bibitem{Day:1993md}D. B. Day,  Phys. Rev. C48, 1849 (1993) (C12, Fe, Al27,  Au Day:1993md)
%58    
    \bibitem{Zghiche:1993xg} A. Zghiche et al., Nucl. Phys. A 572,  513 (1994)(Pb208 Zghiche:1993xg)    
  % ref 59 
       \bibitem{Tsai} Y. S. Tsai, Rev. Mod. Phys. 46, 815, 123 (1974)

\end{thebibliography}
\end{document}